\documentclass[twoside,11pt]{article} 
\usepackage[margin=1.0in]{geometry}

\usepackage{amssymb,amsmath,amsthm}
\usepackage{mathtools}
\usepackage{bm,lscape}
\usepackage{bbm}
\usepackage{mathrsfs,dsfont}
\usepackage{wasysym}
\RequirePackage[
    colorlinks=true,
    linkcolor=blue,
    urlcolor=blue,
    citecolor=blue]{hyperref}
\RequirePackage{natbib}
\usepackage[utf8]{inputenc}
\usepackage[english]{babel}
\usepackage{algorithm}
\usepackage{algpseudocode}

\usepackage{graphicx}
\usepackage{color}
\usepackage{rotating}
\usepackage{authblk}
\usepackage{appendix}

\allowdisplaybreaks

\def\bSig\mathbf{\Sigma}

\newcommand{\E}{\mathbb{E}}

\newcommand{\ddr}{\mathrm{d}}

\def\simiid{\stackrel{\mbox{\scriptsize{iid}}}{\sim}}

\usepackage{booktabs}

\newtheorem{theorem}{Theorem}
\newtheorem{proposition}[theorem]{Proposition}

\providecommand{\keywords}[1]
{
  \small	
  \textbf{\textit{Keywords:}} #1
}

\begin{document}

\title{Bayesian nonparametric estimation of coverage probabilities and distinct counts from sketched data}


\author[1]{Stefano Favaro\thanks{stefano.favaro@unito.it}}
\author[2]{Matteo Sesia\thanks{sesia@marshall.usc.edu}}
\affil[1]{\small{Department of Economics and Statistics, University of Torino and Collegio Carlo Alberto, Italy}}
\affil[2]{\small{Department of Data Sciences and Operations, University of Southern California, Marshall School of Business, Los Angeles, California, USA}}

\maketitle

\begin{abstract}
The estimation of coverage probabilities, and in particular of the missing mass, is a classical statistical problem with applications in numerous scientific fields. In this paper, we study this problem in relation to randomized data compression, or sketching. This is a novel but practically relevant perspective, and it refers to situations in which coverage probabilities must be estimated based on a compressed and imperfect summary, or sketch, of the true data, because neither the full data nor the empirical frequencies of distinct symbols can be observed directly. Our contribution is a Bayesian nonparametric methodology to estimate coverage probabilities from data sketched through random hashing, which also solves the challenging problems of recovering the numbers of distinct counts in the true data and of distinct counts with a specified empirical frequency of interest. The proposed Bayesian estimators are shown to be easily applicable to large-scale analyses in combination with a Dirichlet process prior, although they involve some open computational challenges under the more general Pitman-Yor process prior. The empirical effectiveness of our methodology is demonstrated through numerical experiments and applications to real data sets of Covid DNA sequences, classic English literature, and IP addresses.
\end{abstract}

\keywords{Bayesian nonparametrics; coverage probability; Dirichlet process prior; distinct counts; missing mass; Pitman-Yor process prior; random hashing; sketch.}


\section{Introduction}\label{sec1}

\subsection{Estimation of coverage probabilities}

The estimation of coverage probabilities, and in particular of the missing mass, is a classical statistical problem, dating back to the seminal work of Alan M. Turing and Irving J. Good in 1940s \citep{Goo(53)}. To understand this task, consider a generic population of individuals with values in a (possibly infinite) universe $\mathbb{S}$ of symbols or species labels. In its most common formulation, the problem assumes $n\geq1$ observable data points modeled as random samples from an unknown distribution $p=\sum_{j\geq1}p_{j}\delta_{s_{j}}$, where $p_{j}$ is the probability of symbol $s_{j}\in\mathbb{S}$. Then, denoting by $(N_{j,n})_{j\geq1}$ the empirical frequencies of distinct symbols, the goal is to estimate the {\em coverage probability} of order $r\geq0$:
\begin{align} \label{eq:coverage-prob}
\mathfrak{p}_{r,n}=\sum_{j\geq1}p_{j}I(N_{j,n}=r).
\end{align}
This is the total probability mass of the symbols with empirical frequency $r$. Of special interest is the probability mass of symbols not observed in the sample, namely $\mathfrak{p}_{0,n}$, also known as the {\em missing mass}. The Good-Turing estimator \citep{Goo(53),Rob(56),Rob(68)} is the most popular nonparametric estimator of $\mathfrak{p}_{r,n}$, and it has been the subject of numerous theoretical and methodological studies; e.g., \citet{Efr(03)}, \citet{McA(03)}, \citet{Orl(03)}, \citet{Zha(09)}, \citet{Mos(15)}, \citet{Ben(17)} and \citet{Fad(18)}. A Bayesian nonparametric (BNP) counterpart of the Good-Turing estimator has been proposed by \citet{Fav(12)}, and it relies on the specification of a (nonparametric) prior for the unknown distribution $p$ \citep{Lij(07),Fav(09),Fav(12),Fav(16),Arb(17)}. See \citet{Bal(22)} and references therein for an up-to-date review on frequentist and Bayesian approaches to the estimation of coverage probabilities and generalizations thereof.

The problem of estimating coverage probabilities is encountered in many scientific fields. It first appeared in ecology \citep{Bun(93)}, but its relevance to other areas has grown dramatically over the past three decades, primarily driven by applications in the biological and physical sciences \citep{Mao(02),Gao(07),Dal(13)}. An up-to-date account on the use of coverage probabilities in biology can be found in \citet{Den(19)}. The estimation of the missing mass in biology is typically related to the optimal allocation of resources. In genetic studies, for example, the missing mass may indicate the probability of detecting unobserved genetic variants in a new sample, which is useful to determine how many additional genomes must be sequenced to explain a certain proportion of genetic variation \citep{Ion(09)}. The estimation of coverage probabilities has also found applications in machine learning \citep{Zha(05),Bub(13)}, computer science \citep{Mot(06)}, information theory \citep{Orl(04),Ben(18)}, empirical linguistics and natural language processing \citep{Gal(95), Oha(12)}, and in forensic DNA analysis \citep{Ane(17),Cer(17)}.

\subsection{Data sketching}

While the estimation of coverage probabilities has a long history in the statistics literature, it is not yet a well-studied problem in relation to randomized data {\em compression}, or {\em sketching}. Precisely, this refers to situations in which coverage probabilities must be estimated based on a compressed and imperfect summary, or sketch, of the true data, because neither the full data nor the corresponding empirical frequencies of distinct symbols can be observed directly. In general, sketching is designed to provide compact data structures that can be easily updated and queried in order to estimate meaningful statistics of the true data, the most common being the number of distinct counts and their corresponding empirical frequencies; see the monographs by \citet{Cor(12)} and \citet{Cor(20)} for a comprehensive account on sketching. Sketching is utilized in various applications within biological sciences, to allow time and memory-efficient statistical analyses of large data sets containing many distinct symbols \citep{Zha(14),Sol(16),Ber(18),Mar(19),Leo(20)}, or as a solution to protect sensitive individual information \citep{Koc(20)}. Sketching is increasingly deployed in many fields involving sensitive information and privacy concerns~\citep{Mel(16),Cor(18)}, including in applications seeking to identify most popular websites~\citep{Erl(14)}, count new words typed by users~\citep{Bas(17)}, analyze wireless device locations~\citep{Din(17)}, or track viewership of targeted online advertisements \citep{Heu(13),Cor(17)}. 

The estimation of coverage probabilities from sketched data is a relevant problem in many of the aforementioned domains. For example, in marketing, this task may be useful in order to estimate the likelihood of a targeted advertisement reaching a new user based on anonymized data, allowing the design of more effective online campaigns~\citep{Fah(12)} that respect consumer privacy~\citep{Tou(10)}. In the context of biological sciences, it may lead to principled estimates of the numbers of additional bacterial or viral genomic sequences needed to ensure sufficient genetic diversity in large genomic archives \citep{Bra(19)}. Further, in the context of natural language processing, it may be leveraged to estimate the probability of rare words \citep{Oha(12)} using data that was sketched due to communication bottlenecks or privacy concerns~\citep{Rot(20)}. Despite such potential impacts, we are not aware of any statistical or algorithmic approach to this problem; hence the motivation for this paper.

\subsection{Our contributions}

The contribution of this paper is a BNP methodology to estimate coverage probabilities from data sketched through a single hash function, which also solves the challenging problems of estimating or recovering the number of distinct counts in the true data, as well as the number of distinct counts with empirical frequency $r\geq1$. To the best of our knowledge, this is the first work estimating coverage probabilities from sketched data, as well as the first one to recover the partition structure of the true data. Concretely, we focus on a sketch obtained by means of a random hash function which maps each $\mathbb{S}$-valued data point into the lower dimensional set $\{1,\ldots,J\}$, for a suitable choice of the number of {\em buckets} $J\geq 1$, in such a way that distinct data points tend to populate all buckets uniformly  \citep{Cor(20)}. This is a simplified (single-hashing) version of the popular count-min sketch algorithm of \citet{Cor(05)}. Note that sketching through hash functions compresses the data at the cost of some loss of information due to possible random hash collisions---different symbols may end up in the same bucket.

As a BNP model, here we assume the sketch to be obtained from $n\geq1$ data points randomly sampled from an unknown distribution $p$, which is endowed by a Dirichlet process (DP) prior \citep{Fer(73)}. By relying on a peculiar finite-dimensional projective property of the DP, which leads to a Dirichlet-Multinomial distribution for the sketched data, our main result provides a BNP estimator of $\mathfrak{p}_{r,n}$ with respect to the squared loss. Then, as a corollary, we derive BNP estimators for the number of distinct counts observed, and the number of distinct counts with empirical frequency $r\geq1$. These estimators are simple and can scale to massive data sets. Further, the parameters of the prior distribution can be easily estimated by a computationally efficient empirical Bayes approach, which makes our solution practical in real-world applications. This paper additionally presents an extension of the above results to the Pitman-Yor process (PYP) prior, which is a generalization of the DP prior with a more flexible tail behaviour, ranging from heavy power-law tails to the geometric tails of the DP prior \citep{Pit(97)}. The PYP does not feature an analogous finite-dimensional projective property, and thus it turns out to yield an unwieldy generalization of the Dirichlet-Multinomial distribution for the sketched data. Consequently, the BNP estimator of $\mathfrak{p}_{r,n}$ under the PYP prior admits a closed form expression but quickly becomes intractable to evaluate as the sample size $n$ grows. Although we also derive an alternative representation that in theory enables a Monte Carlo approximation of the estimator, we will explain that the problem under the PYP prior remains computationally challenging even for moderately large $n$, and that is why the applications presented in this paper will focus on the DP prior.

The effectiveness of our BNP methodology is tested via numerical experiments based on synthetic data and with three real-data applications. In particular, we analyze: a data set of $k$-mers within SARS-CoV-2 virus DNA sequences published by the National Center for Biotechnology Information~\citep{hatcher2017virus}, a data set of 2-grams from classic pieces of English literature from the Gutenberg Corpus~\citep{Gutenberg}, and a data set of IP addresses originally utilized to study personalized telecommunication service degradation policies~\citep{rojas2018personalized}.  Although some of these data sets exhibit a power-law tail behaviour that is not well-described by the DP prior, our results demonstrate the proposed methodology often leads to reasonable approximations of the coverage probabilities and missing mass, as long as power-law tails are not too accentuated.

\subsection{Organization of the paper}

The paper is structured as follows. Section \ref{sec2} reviews the relevant BNP model under the PYP prior, of which the DP prior is a special case, and the corresponding methods for estimating coverage probabilities from the true data. Section \ref{sec3} introduces our BNP model for sketched data, obtains the corresponding estimators for the coverage probabilities and the number of distinct counts under the DP prior, and then extends these results to the more general PYP prior. Section \ref{sec4} applies our BNP methodology to synthetic and real data. Section \ref{sec5} concludes by discussing promising directions for future research. All mathematical proofs and additional numerical experiments are deferred to the appendices.


\section{Review of BNP estimation of coverage probabilities}\label{sec2}

\subsection{A BNP model under the PYP and DP prior}

For $n\geq1$, let $(x_{1},\ldots,x_{n})$ be a collection of $\mathbb{S}$-valued data ponts. The BNP approach assumes that data are modeled according to a random sample $\mathbf{X}_{n} = (X_{1},\ldots,X_{n})$ from an unknown discrete distribution $P$, which is endowed with a PYP prior. Formally, we write
\begin{align}\label{eq:exchangeable_model}
\begin{split}
X_1,\ldots,X_{n}\,|\,P &\, \simiid\, P \\
P&\, \sim\,\text{PYP}(\alpha,\theta),
\end{split}
\end{align}
where $\text{PYP}(\alpha,\theta)$ denotes the PYP prior indexed by a discount parameter $\alpha\in[0,1)$ and a scale parameter $\theta>-\alpha$. An intuitive definition of the PYP is given by its stick-breaking construction \citep{Per(92),Pit(95)}. Let: i) $(V_{i})_{i\geq1}$ be independent random variables, with each $V_{i}$ following a Beta$(1-\alpha,\theta+i\alpha)$ distribution; ii) $(S_{j})_{j\geq1}$ be random variables following non-atomic distribution $\nu$ on $\mathbb{S}$ and independent of each other as well as of the $V_{i}$'s. If $P_{1}=V_{1}$ and $P_{j}=V_{j}\prod_{1\leq i\leq j-1}(1-V_{i})$ for $j\geq1$, so that $P_{j}\in(0,1)$ for any $j\geq1$ and $\sum_{j\geq1}P_{j}=1$ almost surely, then the (almost sure) discrete random probability measure $P=\sum_{j\geq1}P_{j}\delta_{S_{j}}$ is a PYP on $\mathbb{S}$ with  parameters $(\alpha,\theta)$. The DP prior corresponds to $\alpha=0$. The parameter $\alpha\in[0,1)$ controls the tail behaviour of $P$. In particular, if $(P_{(j)})_{j\geq1}$ denote the decreasingly ordered random probabilities $P_{j}$'s of $P$, then, for $\alpha\in(0,1)$, as $j\rightarrow+\infty$ the $P_{(j)}$'s follow a power-law distribution of exponent $c=\alpha^{-1}$ \citep{Pit(97)}. That is, $\alpha\in(0,1)$ controls the power-law tail behaviour of the PYP through the small $P_{(j)}$'s: the larger $\alpha$, the heavier the tail of $P$. As a limiting case for $\alpha\rightarrow0$, the DP features geometric tails \citep[Chapter 3 and Chapter 4]{Pit(06)}.

\subsection{Sampling properties of the PYP and DP prior}

Due to the (almost sure) discreteness of the PYP, a random sample $\mathbf{X}_{n}$ from $P\sim\text{PYP}(\alpha,\theta)$ induces a random partition of  $[n]=\{1,\ldots,n\}$ into $K_{n}=k\leq n$ blocks, labelled by $\{S_{1}^{\ast},\ldots,S^{\ast}_{K_{n}}\}$, with corresponding frequencies $(N_{1,n},\ldots,N_{K_{n},n})=(n_{1},\ldots,n_{k})$ such that $n_{i}>0$ and $\sum_{1\leq i\leq k}n_{i}=n$ \citep{Pit(95)}. The distribution of
this partition is determined by the predictive distribution, or generative scheme, of the PYP prior, that is
\begin{displaymath}
\text{Pr}[X_{1}\in\cdot]=\nu(\cdot)
\end{displaymath}
and for $n\geq1$
\begin{equation}\label{eq:pred}
\text{Pr}[X_{n+1}\in\cdot\,|\, \mathbf{X}_{n}]=\frac{\theta+k\alpha}{\theta+n}\nu(\cdot)+\frac{1}{\theta+n}\sum_{i=1}^{k}(n_{i}-\alpha)\delta_{S_{i}^{\ast}}(\cdot).
\end{equation}
The expression in \eqref{eq:pred} provides the conditional distribution of the random partition of $[n+1]$ obtained after sampling one additional data point, given the previous random partition of $[n]$. Note that \eqref{eq:pred} is a linear combination of: i) the probability $(\theta+k\alpha)/(\theta+n)$ that $X_{n+1}$ belongs to a new symbol, i.e., creating a new block in the partition of $[n]$; ii) the probability  $(n_{i}-\alpha)/(\theta+n)$ that $X_{n+1}$ is of symbol $S^{\ast}_{i}$, i.e., increasing by $1$ the size of the block $S^{\ast}_{i}$ in the partition of $[n]$, for $i=1,\ldots,k$ \citep[Chapter 3]{Pit(06)}. The parameter $\alpha\in(0,1)$ controls the rates at which previous symbols are re-observed and new symbols arise. A larger value of $\alpha$ corresponds to a higher probability of observing new symbols. If $\alpha=0$, i.e. under the DP prior, the probabilities in \eqref{eq:pred} become proportional to the frequencies of each symbol, and the probability of generating a new symbol no longer depends on the number of observed symbols \citep{Bac(17)}.

For any $r \in [n]$, let $M_{r,n}$ be the number of distinct symbols with frequency $r$ in a sample $\mathbf{X}_{n}$ from $P\sim\text{PYP}(\alpha,\theta)$; i.e., $M_{r,n}=\sum_{1\leq i\leq K_{n}}I(N_{i,n}=r)$ such that $\sum_{1\leq r\leq n}M_{r,n}=K_{n}$ and $\sum_{1\leq r\leq n}rM_{r,n}=n$. 
The distribution of $\mathbf{M}_{n}=(M_{1,n}, \ldots,M_{n,n})$ is given by
\begin{equation}\label{eq_ewe_py}
\text{Pr}[\mathbf{M}_{n}=(m_{1},\ldots,m_{n})]=n!\frac{\left(\frac{\theta}{\alpha}\right)_{(\sum_{i=1}^{n}m_{i})}}{(\theta)_{(n)}}\prod_{i=1}^{n}\left(\frac{\alpha(1-\alpha)_{(i-1)}}{i!}\right)^{m_{i}}\frac{1}{m_{i}!},
\end{equation}
where $(a)_{(u)}$ is the $u$-th rising factorial of $a$; i.e., $(a)_{(u)}=\prod_{0\leq i\leq u-1}(a+i)$, for $a\geq0$ and $u\in\mathbb{N}_{0}$ \citep{Pit(95)}. The distribution in \eqref{eq_ewe_py} generalizes the Ewens sampling model, which corresponds to $\alpha=0$. The distribution $K_{n}$ follows by a suitable marginalization of \eqref{eq_ewe_py}; i.e.,
\begin{equation}\label{eq_dist_py}
\text{Pr}[K_{n}=k]=\frac{\left(\frac{\theta}{\alpha}\right)_{(k)}}{(\theta)_{(n)}}\mathscr{C}(n,k;\alpha),
\end{equation}
where $\mathscr{C}(u,0,\alpha)=0$ for all $u\geq1$, and $\mathscr{C}(0,0;\alpha)=1$, while for any $a>0$ and $v \leq u \in\mathbb{N}_{0}$,
\begin{equation}\label{gfc_int}
\mathscr{C}(u,v;a)=\frac{1}{v!}\sum_{i=0}^{v}(-1)^{i}{v\choose i}(-ia)_{(u)}.
\end{equation}
See Appendix~\ref{app_combin} for details about the generalized factorial coefficients in \eqref{gfc_int} and their interplay with (signless) Stirling numbers as $\alpha\rightarrow0$. See Appendix~\ref{app_tail} for an account on the tail behaviour of the PYP prior with respect to the large $n$ asymptotic behaviour of  $K_{n}$ and $M_{r,n}$.

\subsection{BNP estimation of coverage probabilities} \label{sec:review-coverage}

Under the BNP model in \eqref{eq:exchangeable_model}, an estimator of $\mathfrak{p}_{r,n}$, with respect to the squared loss, is obtained by a direct application of the predictive distribution in \eqref{eq:pred}. Assume the random sample $\mathbf{X}_{n}$ from $P\sim\text{PYP}(\alpha,\theta)$ features $K_{n}=k$ distinct symbols, labelled by $\{S_{1}^{\ast},\ldots,S^{\ast}_{K_{n}}\}$, with frequencies $(N_{1,n},\ldots,N_{K_{n},n})=(n_{1},\ldots,n_{k})$ such that $M_{r,n}=m_{r}$, and define
\begin{displaymath}
\mathcal{S}_{0}=\mathbb{S}-\{S_{1}^{\ast},\ldots,S^{\ast}_{K_{n}}\},
\end{displaymath}
and for $r\geq1$
\begin{displaymath}
\mathcal{S}_{r}=\bigcup_{i=1}^{K_{n}}\{S^{\ast}_{i}\in\{S_{1}^{\ast},\ldots,S^{\ast}_{K_{n}}\}\text{ : }N_{i,n}=r\}.
\end{displaymath}
That is, $\mathcal{S}_{0}$ denotes the set of symbols not observed in $\mathbf{X}_{n}$, whereas $\mathcal{S}_{r}$ denotes the set of symbols observed in $\mathbf{X}_{n}$ with frequency $r$, for $r=1,\ldots,n$. Then, from \eqref{eq:pred} it holds that
\begin{equation}\label{bnp_est}
\hat{\mathfrak{p}}_{r,n}=\E[P(\mathcal{S}_{r})\,|\,\mathbf{X}_{n}]=\text{Pr}[X_{n+1}\in\mathcal{S}_{r}\,|\,\mathbf{X}_{n}]=\begin{cases} 
 \frac{\theta+k\alpha}{\theta+n}&\mbox{ if } r=0\\[0.4cm] 
\frac{m_{r}(r-\alpha)}{\theta+n}&\mbox{ if } r\geq1
\end{cases}
\end{equation}
is a BNP estimator of $\mathfrak{p}_{r,n}$, with respect to the squared loss, under the PYP prior. The estimator in \eqref{bnp_est} is the BNP counterpart of the Good-Turing estimator. We refer to \citet{Fav(12)} \citet{Fav(16)} and \citet{Arb(17)} for a detailed account on \eqref{bnp_est}, with emphasis on credible intervals, and its interplay with the Good-Turing estimator.


\section{BNP estimation of coverage probabilities from sketched data} \label{sec3}

\subsection{A BNP model for data sketched through random hashing}

We consider a situation in which the $\mathbb{S}$-valued data points $(x_{1},\ldots,x_{n})$ are not directly observable, and instead we have access only to a sketch of them obtained through random hashing \citep[Chapter 5 and Chapter 15]{Mit(17)}. For an integer $J \geq 1$, let $h$ be a (random) hash function of width $J$, which is a random mapping from $\mathbb{S}$ to $[J] = \{1,\ldots,J\}$ chosen from a pairwise independent hash family $\mathcal{H}_{J}$, independently of $\mathbf{X}_{n}$. That is, $h: \mathbb{S} \to [J]$, and, for any $j_1,j_2 \in [J]$ and fixed $x_1,x_2 \in \mathbb{S}$ such that $x_1 \neq x_2$,
\begin{displaymath}
\text{Pr}[h(x_{1})=j_{1},\,h(x_{2})=j_{2}]= \frac{1}{J^{2}}.
\end{displaymath}
Pairwise independence between hash functions is also known as {\em strong universality}, and it implies uniformity, meaning that $\Pr[h(x)=j]=J^{-1}$ for any $j=1,\ldots,J$. Strong universality is known to provide a common setting of mathematical convenience, but it not easy to achieve exactly in practice. However, real-world hash functions effectively perform as if they were perfectly random \citep{Chu(13)}. Hashing $(x_{1},\ldots,x_{n})$  through $h$ produces a vector (sketch) $\mathbf{C}_{n} = (C_{1,n},\ldots,C_{J,n})$, whose $j$-th element (bucket) is
\begin{displaymath}
C_{j,n} = \sum_{i=1}^{n} I( h(x_i) = j),
\end{displaymath}
so that $\sum_{1\leq j\leq J} C_{j,n} = n$.
In general, $\mathbf{C}$ has a smaller (physical) size than $(x_{1},\ldots,x_{n})$ due to the hash collisions \citep[Chapter 3]{Cor(20)}. The above sketch is a special version of the popular count-min sketch of \citet{Cor(05)}, which simultaneously sketches the same data points using several independent hash functions. 

Our BNP model for the sketch $\mathbf{C}_{n}$ relies on two assumptions: i) $(x_{1},\ldots,x_{n})$ are modeled as a random sample $\mathbf{X}_{n}=(X_{1},\ldots,X_{n})$ from $P\sim\text{PYP}(\alpha,\theta)$, i.e., according to the BNP model in \eqref{eq:exchangeable_model}; ii) $h$ is a hash function from a strong universal hash family $\mathcal{H}_{J}$, and it is independent of $\mathbf{X}_{n}$, i.e. independent of $P\sim\text{PYP}(\alpha,\theta)$. Then, we write the BNP model
\begin{align}\label{eq:exchangeable_model_hash} 
\begin{split}
  C_{j,n} &\,=\, \sum_{i=1}^{n} I( h(X_i) = j)\qquad j=1,\ldots,J\\
  h&\,\sim\,\mathcal{H}_{J}\\
  X_1,\ldots,X_{n}\,|\,P &\,\simiid\, P\\
  P&\,\sim\,\text{PYP}(\alpha,\theta).
\end{split}
\end{align}
The BNP model in \eqref{eq:exchangeable_model_hash} may be viewed as a finite-dimensional projection on $[J]$ of the BNP model in \eqref{eq:exchangeable_model}, with the projection determined by $h$. As the random partition $(K_{n},N_{1,n},\ldots,N_{K_{n},n})$ is a sufficient statistic for the $X_{i}$'s under \eqref{eq:exchangeable_model}, the sketch $\mathbf{C}_{n}$ is a sufficient statistics for the $h(X_{i})$'s under \eqref{eq:exchangeable_model_hash}. The distributional properties of $\mathbf{C}_{n}$ are thus critical to make any inferences under the BNP model in \eqref{eq:exchangeable_model_hash}, and their (mathematical) tractability depends on the finite-dimensional projective properties of the prior distribution. We will show how this peculiar feature distinguishes the DP prior from the more general PYP prior.

Under the BNP model in \eqref{eq:exchangeable_model_hash}, we consider the problem of estimating the coverage probability $\mathfrak{p}_{r,n}$ for any $r\geq0$, given only the sketch $\mathbf{C}_{n}$. In particular, for any $r\geq0$, we will compute
\begin{equation}\label{estim_sketc}
\tilde{\mathfrak{p}}_{r,n}=\E[P(\mathcal{S}_{r})\,|\,\mathbf{C}_{n}=(c_{1},\ldots,c_{J})]=\text{Pr}[X_{n+1}\in\mathcal{S}_{r}\,|\,\mathbf{C}_{n}=(c_{1},\ldots,c_{J})],
\end{equation}
which provides a BNP estimator of $\mathfrak{p}_{r,n}$ with respect to the squared loss. Further, we will also present a solution to the problem of estimating or recovering from $\mathbf{C}_{n}$ the number $K_{n}$ of distinct counts in the true data $(x_{1},\ldots,x_{n})$, as well as the number $M_{r,n}$ of distinct counts with empirical frequency $r\geq1$. That is, still under the model \eqref{eq:exchangeable_model_hash}, we will compute
\begin{displaymath}
\tilde{\mathfrak{k}}_{n}=\E[K_{n}\,|\, \mathbf{C}_{n}=(c_{1},\ldots,c_{J}) ],
\end{displaymath}
and, for any $r\geq1$,
\begin{displaymath}
\tilde{\mathfrak{m}}_{r,n}=\E[M_{r,n}\,|\, \mathbf{C}_{n}=(c_{1},\ldots,c_{J}) ],
\end{displaymath}
which are the BNP estimators of $K_{n}$ and $M_{r,n}$, respectively, with respect to the squared loss. In particular, by resorting to the BNP estimator $\hat{\mathfrak{p}}_{r,n}$ displayed in \eqref{bnp_est}, which is a linear function of $M_{r,n}$, it becomes possible to express $\tilde{\mathfrak{k}}_{n}$ and $\tilde{\mathfrak{m}}_{r,n}$ as suitable linear functions of the $\tilde{\mathfrak{p}}_{r,n}$'s. In particular, this result shows how under the BNP approach with a PYP prior, the estimation of coverage probabilities from the sketch $\mathbf{C}_{n}$ also solves the problem of recovering the partition structure $(M_{1,n},\ldots,M_{n,n})$ of the true data $(x_{1},\ldots,x_{n})$. Below, we will first derive $\tilde{\mathfrak{p}}_{r,n}$, $\tilde{\mathfrak{k}}_{n}$ and $\tilde{\mathfrak{m}}_{r,n}$ explicitly for $\alpha=0$, i.e., under the DP prior, and then we shall present their extensions to the more general case of $\alpha\in[0,1)$.

\subsection{Estimation of coverage probabilities under the DP prior} \label{sec:coverage}
The DP features a peculiar finite-dimensional projective property, according to which the finite-dimensional (marginal) distributions of $P\sim\text{PYP}(0,\theta)$ are Dirichlet distribution \citep{Fer(73)}.  Because of the strong universality of the hash family $\mathcal{H}_{J}$, i.e., uniformity, and of the independence between $h\sim\mathcal{H}_{j}$ and $P$, it turns out that $h$ induces a fixed partition $\{B_{1},\ldots,B_{J}\}$ of $\mathbb{S}$, with $B_{j}=\{s\in\mathbb{S}\text{ : }h(s)=j\}$ and $\nu(B_{j})=J^{-1}$ for $j \in [J]$. Accordingly, the finite-dimensional projective property of the DP prior implies that
\begin{equation}\label{dp_fd}
(P(B_{1}),\ldots,P(B_{J}))\sim\text{Dirichlet}\left(\frac{\theta}{J},\ldots,\frac{\theta}{J}\right),
\end{equation}
where $\text{Dirichlet}(\cdot)$ denotes the Dirichlet distribution. See also \citet{Reg(01)} and \citet{Gho(17)} for details. According to \eqref{dp_fd}, under the model in \eqref{eq:exchangeable_model_hash} with $\alpha=0$,
\begin{equation}\label{dp_fdm}
\text{Pr}[\mathbf{C}_{n}=(c_{1},\ldots,c_{J})]={n\choose c_{1},\ldots,c_{J}}\frac{1}{(\theta)_{(n)}}\prod_{j=1}^{J}\left(\frac{\theta}{J}\right)_{(c_{j})},
\end{equation}
which is a Dirichlet-Multinomial distribution with parameter $(n,\theta/J,\ldots,\theta/J)$.
The distribution in \eqref{dp_fdm} leads to an explicit and straightforward expression of $\tilde{\mathfrak{p}}_{r,n}$ and, as we shall discuss later, it also allows estimating the prior parameter $\theta>0$ from the information contained in the sketch $\mathbf{C}_{n}$. The next theorem exploits \eqref{dp_fdm}, in combination with the exchangeability of the $X_{i}$'s and some combinatorial arguments related to $\mathbf{C}_{n}$, to obtain $\tilde{\mathfrak{p}}_{r,n}$. We also provide the estimators $\tilde{\mathfrak{k}}_{n}$ and $\tilde{\mathfrak{m}}_{r,n}$, for $r\geq1$, as functions of the $\tilde{\mathfrak{p}}_{r,n}$'s.

\begin{theorem}\label{teo_dp}
For $n\geq1$, assume the sketch $\mathbf{C}_{n}$ to be modeled according to the BNP model in \eqref{eq:exchangeable_model_hash} with $\alpha=0$, i.e., under the DP prior, and let $\mathbf{C}_{n}=(c_{1},\ldots,c_{J})$. Then, for any $r\geq0$,
\begin{equation}\label{post_dp}
\tilde{\mathfrak{p}}_{r,n}
=\frac{\left(\frac{\theta}{J}\right)r!}{(\theta+n)}\sum_{j=1}^{J}{c_{j}\choose r}\frac{\left(\frac{\theta}{J}\right)_{(c_{j}-r)}}{\left(\frac{\theta}{J} \right)_{(c_{j})}}.
\end{equation}
Moreover, for $r\ge1$
\begin{equation}\label{recov_sketc_unseen_freq_estim_dp}
\tilde{\mathfrak{m}}_{r,n}=\frac{\theta+n}{r}\tilde{\mathfrak{p}}_{r,n},
\end{equation}
and
\begin{equation}\label{recov_sketc_unseen_estim_dp}
\tilde{\mathfrak{k}}_{n}=\sum_{r\geq1}\tilde{\mathfrak{m}}_{r,n}=- \theta \psi\left(1-\frac{\theta}{J}\right) + \frac{\theta}{J} \sum_{j=1}^{J}  \psi\left(1-\frac{\theta}{J}-c_{j}\right),
\end{equation}
where $\psi$ is the digamma function; i.e., $\psi(x) = \frac{d}{dx} \log \Gamma(x)$.
\end{theorem}

See Appendix \ref{app_proof_teo_dp} for the proof of Theorem \ref{teo_dp}. The estimator $\tilde{\mathfrak{p}}_{r,n}$ is the natural counterpart, with respec to sketched data, of the BNP estimator $\hat{\mathfrak{p}}_{r,n}$ in \eqref{bnp_est} with $\alpha=0$, i.e., under the DP prior. In particular, $\hat{\mathfrak{p}}_{r,n}$ is recovered from $\tilde{\mathfrak{p}}_{r,n}$ for a lossless hash function $h:\mathbb{S}\rightarrow\mathbb{S}$.  
As one would expect, it can be seen from \eqref{post_dp} that $\tilde{\mathfrak{p}}_{r,n} =0$ for all $r>\max\{c_{1},\ldots,c_{J}\}$.
Of special interest is the estimator $\tilde{\mathfrak{p}}_{0,n}$ of the missing mass $\mathfrak{p}_{0,n}$, i.e.,
\begin{displaymath}
\tilde{\mathfrak{p}}_{0,n}=\frac{\theta}{\theta+n},
\end{displaymath}
which, interestingly, coincides with $\hat{\mathfrak{p}}_{0,n}$. However, the equivalence between the estimators $\hat{\mathfrak{p}}_{0,n}$ and $\tilde{\mathfrak{p}}_{0,n}$ under the DP prior should not be surprising, because $\hat{\mathfrak{p}}_{0,n}$ depends on $\mathbf{X}_n$ only through the sample size $n$. This is a peculiar property of the DP prior. In fact, according to the Johnson's ``sufficientness" postulate, the DP prior is characterized as the sole (discrete) nonparametric prior for which $\hat{\mathfrak{p}}_{0,n}$ depends on the sampling information only through $n$ \citep{Reg(78),Zab(82)}.  To make Theorem \ref{teo_dp} practically applicable, one must first estimate the prior parameter $\theta$ from the information contained in the sketch $\mathbf{C}_{n}$. This problem can be solved with an empirical Bayes approach that exploits the Dirichlet-Multinomial distribution in \eqref{dp_fdm}. In particular, we estimate the unknown parameter $\theta$ with the value $\tilde{\theta}$ that maximizes the (marginal) likelihood of $\mathbf{C}_{n}$ in \eqref{dp_fdm}.
This likelihood function is log-concave in $\log(\theta)$~\citep{Cai(18)}, which makes it computationally easy to estimate $\theta$ using standard optimization techniques.
Alternatively, one could also estimate $\theta$ by placing a suitable prior distribution on it and following a fully Bayes, or hierarchical Bayes, approach, but that option is not explored in this paper.

\subsection{Estimation of coverage probabilities under the PYP prior} \label{sec:coverage-pyp}

Theorem \ref{teo_dp} is extended here to the general case of the PYP prior, with any $\alpha \in[0,1)$ and $\theta>-\alpha$. As before, because of the strong universality of the hash family $\mathcal{H}_{J}$ and of the independence between $h\sim\mathcal{H}_{j}$ and $P$, it turns out that $h$ induces a fixed partition $\{B_{1},\ldots,B_{J}\}$ of $\mathbb{S}$, with $B_{j}=\{s\in\mathbb{S}\text{ : }h(s)=j\}$ and $\nu(B_{j})=J^{-1}$ for $j\in[J]$. However, the distribution of $(P(B_{1}),\ldots,P(B_{J}))$ is not available in closed-form for any $\alpha\in(0,1)$ because the PYP does not feature a finite-dimensional projective property analogous to that of the DP. Only some moment formulae for this distribution are available \citep{San(06)}, though they have unwieldy expressions whose computation are impractical. See also \citet{Pit(97)} and references therein for further details. In particular, if $I_{(c_{1},\ldots,c_{J})}$ denotes the cartesian product $\times_{1\leq s\leq J}\{0,1,\ldots,c_{s}\}$, with $\boldsymbol{i}=(i_{1},\ldots,i_{J})$ being an element of $I_{(c_{1},\ldots,c_{J})}$ such that $|\boldsymbol{i}|=i_{1}+\cdots+i_{J}$, then \citet[Equation 3.3]{San(06)} shows that
\begin{equation}\label{pyp_fdm}
\text{Pr}[\mathbf{C}_{n}=(c_{1},\ldots,c_{J})]={n\choose c_{1},\ldots,c_{J}}\frac{1}{(\theta)_{(n)}}\sum_{\boldsymbol{i}\in I_{(c_{1},\ldots,c_{J})}}\frac{\left(\frac{\theta}{\alpha}\right)_{(|\boldsymbol{i}|)}}{J^{|\boldsymbol{i}|}}\prod_{j=1}^{J}\mathscr{C}(c_{j},i_{j};\alpha),
\end{equation}
The distribution in \eqref{pyp_fdm} reduces to the Dirichlet-Multinomial distribution in \eqref{dp_fdm} as $\alpha\rightarrow0$; see Appendix \ref{dp_fdm_pyp_fdm} for details. The next theorem exploits \eqref{pyp_fdm}, in combination with the exchangeability of the $X_{i}$'s and some combinatorial arguments related to $\mathbf{C}_{n}$, to obtain $\tilde{\mathfrak{p}}_{r,n}$. We also provide the estimators $\tilde{\mathfrak{k}}_{n}$ and $\tilde{\mathfrak{m}}_{r,n}$, for $r\geq1$, as functions of the $\tilde{\mathfrak{p}}_{r,n}$'s.

\begin{theorem}\label{teo_pyp}
For $n\geq1$, assume the sketch $\mathbf{C}_{n}$ to be modeled according to the BNP model in \eqref{eq:exchangeable_model_hash}, and let $\mathbf{C}_{n}=(c_{1},\ldots,c_{J})$. For any $a\in\mathbb{Z}$ and  $j \in [J]$, let $I_{(c_{1},\ldots,c_{J}),j,a}$ be the cartesian product $\times_{s=1}^{J}\{0,1,\ldots,c_{s}+a\delta_{s,j}\}$, where $\delta_{s,j}$ is the Kronecker delta, and let $\boldsymbol{i}=(i_{1},\ldots,i_{J})$ be an element of $I_{(c_{1},\ldots,c_{J}),j,a}$, with $|\boldsymbol{i}|=i_{1}+\cdots+i_{J}$. Then, for any $r\geq0$,
\begin{align}\label{post_pyp}
\tilde{\mathfrak{p}}_{r,n}
  &= \frac{\left(\frac{\theta}{J}\right)(1-\alpha)_{(r)}}{(\theta+n)}\sum_{j=1}^{J}{c_{j}\choose r}\frac{\sum_{\boldsymbol{i}\in I_{(c_{1},\ldots,c_{J}),j,-r}} \frac{\left(\frac{\theta+\alpha}{\alpha}\right)_{|\boldsymbol{i}|}}{J^{|\boldsymbol{i}|}} \prod_{s=1}^{J}\mathscr{C}(c_{s}-r\delta_{s,j},i_{s};\alpha)}{\sum_{\boldsymbol{i}\in I_{(c_{1},\ldots,c_{J})}}\frac{\left(\frac{\theta}{\alpha}\right)_{|\boldsymbol{i}|}}{J^{|\boldsymbol{i}|}} \prod_{s=1}^{J}\mathscr{C}(c_{s},i_{s};\alpha)}.
\end{align}
Moreover, for $r\ge1$
\begin{equation}\label{recov_sketc_unseen_freq_estim_pyp}
\tilde{\mathfrak{m}}_{r,n}=\frac{\theta+n}{r-\alpha}\tilde{\mathfrak{p}}_{r,n}
\end{equation}
and 
\begin{equation}\label{recov_sketc_unseen_estim_pyp}
\tilde{\mathfrak{k}}_{n}=\frac{\theta+n}{\alpha}\tilde{\mathfrak{p}}_{0,n}-\frac{\theta}{\alpha}.
\end{equation}
\end{theorem}

See Appendix \ref{app_proof_teo_pyp} for the proof of Theorem \ref{teo_pyp}. Note that Theorem \ref{teo_dp} follows directly from Theorem \ref{teo_pyp} by taking the limit of $\alpha\rightarrow0$; see Appendix \ref{app_dp_pyp} for details. The estimator $\tilde{\mathfrak{p}}_{r,n}$ in Theorem \ref{teo_pyp} is the natural counterpart, with respect to sketched data, of the BNP estimator $\hat{\mathfrak{p}}_{r,n}$ in \eqref{bnp_est}. As for Theorem \ref{teo_dp}, it follows from \eqref{post_pyp} that $\tilde{\mathfrak{p}}_{r,n} =0$ and $\tilde{\mathfrak{m}}_{r,n}=0$ for any $r>\max\{c_{1},\ldots,c_{J}\}$.   Of special interest is the missing mass estimator $\tilde{\mathfrak{p}}_{0,n}$,
\begin{equation}\label{post_pyp_missing}
\tilde{\mathfrak{p}}_{0,n}=\frac{\theta}{\theta+n}\frac{\sum_{\boldsymbol{i}\in I_{(c_{1},\ldots,c_{J})}}\frac{\left(\frac{\theta+\alpha}{\alpha}\right)_{|\boldsymbol{i}|}}{J^{|\boldsymbol{i}|}} \prod_{s=1}^{J}\mathscr{C}(c_{s},i_{s};\alpha)}{\sum_{\boldsymbol{i}\in I_{(c_{1},\ldots,c_{J})}} \frac{\left(\frac{\theta}{\alpha}\right)_{|\boldsymbol{i}|}}{J^{|\boldsymbol{i}|}} \prod_{s=1}^{J}\mathscr{C}(c_{s},i_{s};\alpha)}.
\end{equation} 
Under suitable assumptions on the large-$n$ behaviour of the $c_{j}$'s, one can obtain a simple large $n$ asymptotic approximation of \eqref{post_pyp_missing}. In this respect, it is useful to observe that $\lim_{n\rightarrow+\infty}\E[n^{-1}C_{j,n}]=J^{-1}$ for any $j\in[J]$; this follows from \eqref{pyp_fdm} by de Finetti's theorem, and it is a consequence of the strong universality of the hash family $\mathcal{H}_{J}$ and of its independence of $\mathbf{X}_{n}$. Now, for $\alpha\in(0,1)$, if we assume that $c_{j}=nJ^{-1}$ for any $j\in[J]$, then 
\begin{equation}\label{asymp_pyp}
\lim_{n\rightarrow+\infty}n^{1-\alpha}\tilde{\mathfrak{p}}_{0,n}=J^{1-\alpha}\frac{\Gamma(\theta+J\alpha-\alpha+1)}{\Gamma(\theta+J\alpha)}.
\end{equation}
See Appendix \ref{app_proofs_3} for the proof of Equation \eqref{asymp_pyp}. The result in \eqref{asymp_pyp} leads to a large $n$ approximation of \eqref{post_pyp_missing}, though this is only of a qualitative nature because we cannot quantify the approximation error. In particular, because of the assumption on the $c_{j}$'s, we expect that a very large $n$ is required in order to make this asymptotic approximation accurate.

Unfortunately, the estimator $\tilde{\mathfrak{p}}_{r,n}$ in \eqref{post_pyp} is impractical to compute because it involves summing a potentially very large number of generalized factorial coefficients, depending on the number of buckets $J$. 
In general, generalized factorial coefficients can be computed recursively as
\begin{equation}\label{triangular}
\mathscr{C}(u+1,v;a)=(u-va)\mathscr{C}(u,v;a)+a\mathscr{C}(u,v+1;a),
\end{equation}
for any $a>0$, $v\leq u\in\mathbb{N}_{0}$, with the proviso $\mathscr{C}(u,0,\alpha)=0$ for $u\geq1$ and $\mathscr{C}(0,0;\alpha)=1$ \citep[Theorem 2.18]{Cha(05)}. The recursion follows directly from \eqref{gfc_int}, which does not admits a closed-form solution; see Appendix \ref{app_combin} for details on generalized factorial coefficients. Therefore, as the sample size $n$ grows, the computational cost of evaluating $\tilde{\mathfrak{p}}_{r,n}$ in \eqref{post_pyp} becomes overwhelming, thus preventing the implementation of our BNP estimator in concrete applications for $\alpha\in(0,1)$. This computational challenge motivates the results in next section, which show that the estimator $\tilde{\mathfrak{p}}_{r,n}$ can be equivalently rewritten in terms of expected values of random variables that can be sampled exactly, thereby opening a path for a Monte Carlo evaluation of \eqref{post_pyp}. 
Note that the practical problem of estimating empirically the parameters of the PYP prior is postponed until Section~\ref{sec:eb-pyp}.

\subsection{Monte Carlo approximation of $\tilde{\mathfrak{p}}_{r,n}$ under the PYP prior} \label{sec:coverage-pyp-mc}

The following result (proved in Appendix \ref{app_main_post_mc}) shows that the estimator $\tilde{\mathfrak{p}}_{r,n}$ can be written as a suitable functional of the number of distinct symbols in a random sample from $P\sim\text{PYP}(\alpha,\theta)$, for any $\alpha\in(0,1)$. 
This will be useful to enable a Monte Carlo evaluation of \eqref{post_pyp}, in combination with the exact and approximate Monte Carlo techniques for sampling the number of distinct symbols under our BNP model with PYP prior.

\begin{proposition}\label{main_post_mc}
For $n\geq1$, assume the sketch $\mathbf{C}_{n}$ to be modeled according to the BNP model in \eqref{eq:exchangeable_model_hash}, and let $\mathbf{C}_{n}=(c_{1},\ldots,c_{J})$. For any $s \in [J]$, let $K_{c_{s}}$ be the number of distinct symbols in a random sample $\mathbf{X}_{c_{s}}$ from $P\sim\text{PYP}(\alpha,\theta)$, which is independent of $\mathbf{C}_{n}$. Note that $K_{c_{i}}$ is assumed to be independent of $K_{c_{j}}$ for all $i \neq j$. Then, for any $r\geq0$,
\begin{align}\label{post_pyp_mc}
  &\tilde{\mathfrak{p}}_{r,n}
    =\frac{\left(\frac{\theta}{J}\right)(1-\alpha)_{(r)}}{(\theta+n)}\sum_{j=1}^{J}{c_{j}\choose r}\frac{(\theta)_{(c_{j}-r)} \E[Z_{r,j}]}{(\theta)_{(c_{j})}\E[Z'_{r,j}]},
\end{align}
where  
\begin{align} \label{post_pyp_mc-Z}
\begin{split}
  Z_{r,j} & = \frac{\left(1 + \frac{\theta}{\alpha}\right)_{(\sum_{s=1}^{J}K_{c_{s}-r \delta_{s,j}})}}{J^{\sum_{s=1}^{J}K_{c_{s}-r \delta_{s,j}}} \prod_{s=1}^{J}\left(\frac{\theta}{\alpha}\right)_{(K_{c_{s}}-r \delta_{s,j})}}, \\
  Z'_{r,j} & = \frac{\left(\frac{\theta}{\alpha}\right)_{(\sum_{s=1}^{J}K_{c_{s}-r \delta_{s,j}})}}{J^{\sum_{s=1}^{J}K_{c_{s}-r \delta_{s,j}}}  \prod_{s=1}^{J}\left(\frac{\theta}{\alpha}\right)_{(K_{c_{s}}-r\delta_{s,j})}}.
\end{split}
\end{align}
\end{proposition}

The expression in \eqref{post_pyp_mc} suggests a Monte Carlo approximation of the estimator $\tilde{\mathfrak{p}}_{r,n}$ through independent random sampling of $K_{c_{s}-r}$. Sampling $K_{c_{s}-r}$ reduces to sampling  $c_{s}-r-1$ independent Bernoulli random variables, due to form of the predictive distribution in \eqref{eq:pred}; see Algorithm \ref{sampling_k}. This operation can be made computationally faster through large-sample approximations in the limit of large $c_s$, for $r\ll c_{s}$. For $\alpha\in(0,1)$, a first-order approximation of $\tilde{\mathfrak{p}}_{r,n}$ for large values of $c_{s}$ is obtained from \eqref{eq:sigma_diversity} by replacing each $K_{c_{s}}$ with $c_{s}^{\alpha}S_{\alpha,\theta}$. Random sampling of  $S_{\alpha,\theta}$ can then be carried out efficiently through adaptive rejection sampling \citep{Dev(09)}. Unfortunately, the random variables $Z_{r,n}$ and $Z'_{r,n}$ in~\eqref{post_pyp_mc-Z} tend to have a highly skewed distribution for all but very small values of $n$, and this makes it difficult to obtain even approximately unbiased estimates of the ratio of expected values in~\eqref{post_pyp_mc}; see \citet{quenouille1956notes}. Although solutions to mitigate the bias of ratio estimators have been proposed \citep{tin1965comparison}, the specific problem we face in~\eqref{post_pyp_mc} is especially challenging because $Z_{r,n}$ and $Z'_{r,n}$ involve a ratio of rising factorials, which makes the distribution of their ratio extremely skewed. Therefore, the number of Monte Carlo samples required for an accurate estimate of~\eqref{post_pyp_mc} tends to be prohibitively large even for moderate values of $n$, as demonstrated empirically in Section \ref{sec4}. This is a practical limitation of Theorem~\ref{main_post_mc}, which we have not yet been able to overcome.

\begin{algorithm}[H] \label{sampling_k}
\caption{Sampling $K_{c-r}$ for any $r \in \{0,\dots,c\}$}
\begin{algorithmic}
\State $K[0] \gets 0$;
\State $K[1] \gets 1$;
\State $i\gets 1$;
\While{$i \leq c$}
    \State $Ber\gets\text{random sample from Bernoulli}\left(\frac{\theta+\alpha K[i-1]}{\theta+i}\right)$;
    \State $K[i]\gets K[i-1] + Ber$;
    \State $i\gets i+1$;
\EndWhile
\State \Return reverse of $K$
\end{algorithmic}
\end{algorithm}

\subsection{Empirical Bayes estimation of $(\alpha,\theta)$ from sketched data} \label{sec:eb-pyp}

To make the results of Theorem \ref{teo_pyp} directly applicable, one must first estimate the PYP prior parameters $\alpha\in(0,1)$ and $\theta>-\alpha$ from the sketch $\mathbf{C}_{n}$. Differently from the special case of the DP prior, for which $\mathbf{C}_{n}$ has the Dirichlet-Multinomial distribution in \eqref{dp_fdm}, the distribution of $\mathbf{C}_{n}$ generally takes the cumbersome expression in \eqref{pyp_fdm}, whose evaluation involves the same computational issues as the estimator in \eqref{post_pyp}. This difficulty prevents a maximum marginal likelihood strategy in the spirit of that described in Section~\ref{sec:coverage} for the special case of the DP prior. An alternative route is offered by the likelihood-free Wasserstein distance approach~\citep{Ber(19)} first proposed by \citet{Dol(22)} in the context of BNP empirical frequency estimation from a sketch obtained from multiple independent hash functions. This solution is still quite computationally expensive if the sample size $n$ is very large, but it is easy to explain and implement.The key idea is to simulate independent data sets $\mathbf{X}^{\prime}_n$ from the BNP model in \eqref{eq:exchangeable_model_hash} using different values of the prior parameters, namely $(\alpha^{\prime},\theta^{\prime})$. The simulated data are sketched into $\mathbf{C}^{\prime}_{n}$ with the same hash function $h$ as in \eqref{eq:exchangeable_model_hash}. Then, the sorted entries of the vector $\mathbf{C}^{\prime}_{n}$ are compared, with respect to the Wasserstein distance, to the sorted entries of the original sketch $\mathbf{C}_{n}$. The empirical estimates for $(\alpha,\theta)$ are defined as those approximately minimizing the Wasserstein distance between the empirical distributions of $\mathbf{C}_{n}$ and $\mathbf{C}^{\prime}_{n}$. If $n$ is very large, the computational cost of this procedure can be greatly reduced, at the cost of some additional approximations, by simulating a smaller data set $\mathbf{X}'_{n'}$ with $n' \ll n$, and comparing the resulting $\mathbf{C}^{\prime}_{n}$ with a down-scaled version of $\mathbf{C}_{n}$, in which all entries of the latter are multiplied by $n^{\prime}/n$. We refer to \citet{Dol(22)} for further details.


\section{Numerical experiments} \label{sec4}

We begin by investigating the empirical performance on simulated data of the BNP coverage probability estimators derived in Section~\ref{sec:coverage}, for the special case of the DP prior, and in Section~\ref{sec:coverage-pyp}, for the general PYP prior.
Synthetic data are generated from the BNP model in~\eqref{eq:exchangeable_model_hash} using different values of the parameters $(\alpha,\theta)$, and then they are sketched as in~\eqref{eq:exchangeable_model_hash} with a hash function of width 128.
Our BNP estimates are compared to the true coverage probabilities in~\eqref{eq:coverage-prob}, which are available in these experiments because we know the prior parameters of the true data-generating model and have access to the non-sketched data.
All experiments are repeated 20 times and the results averaged, utilizing independent data sets and independent hash functions.

\begin{figure}[!htb]
\centering
\includegraphics[width=0.9\linewidth]{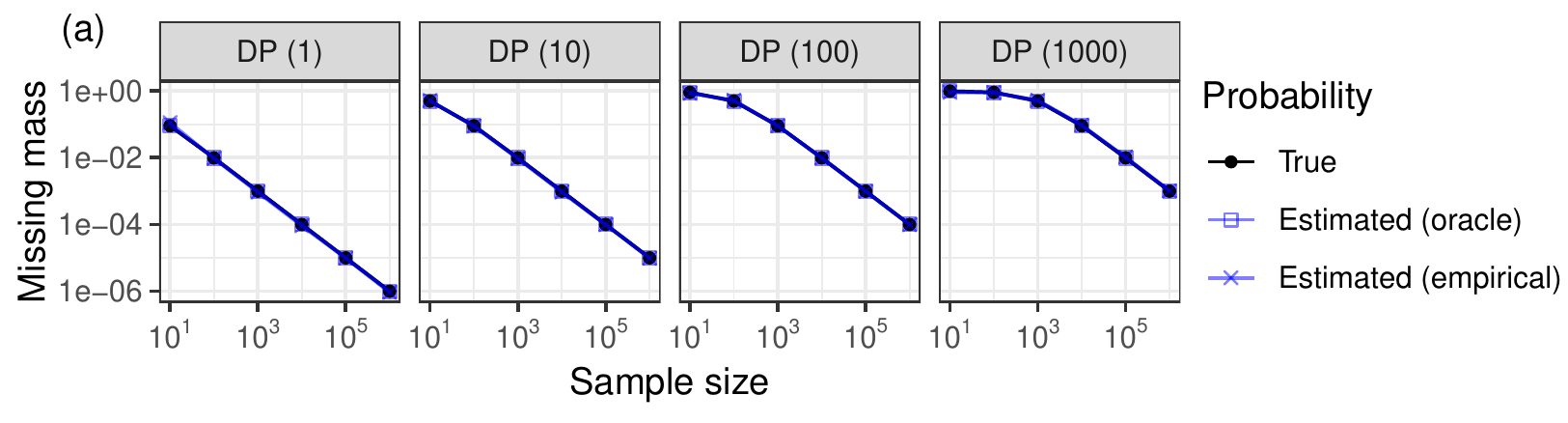}
\includegraphics[width=0.9\linewidth]{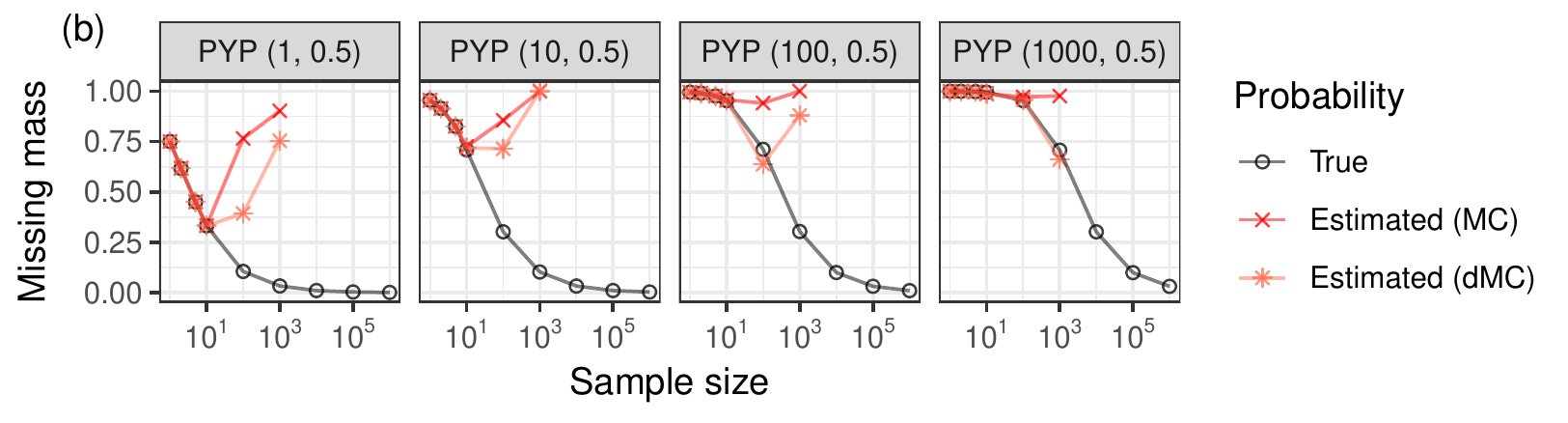}
\caption{True and estimated missing mass for sketched data simulated from a DP prior model (a) with different parameters $\theta$, and from a PYP prior model (b) with different parameters $(\alpha,\theta)$. In (b), the Monte Carlo estimated probabilities are not shown for $n>1000$.}
\label{fig:exp-dp}
\end{figure}

\begin{figure}[!htb]
\centering
\includegraphics[width=0.8\linewidth]{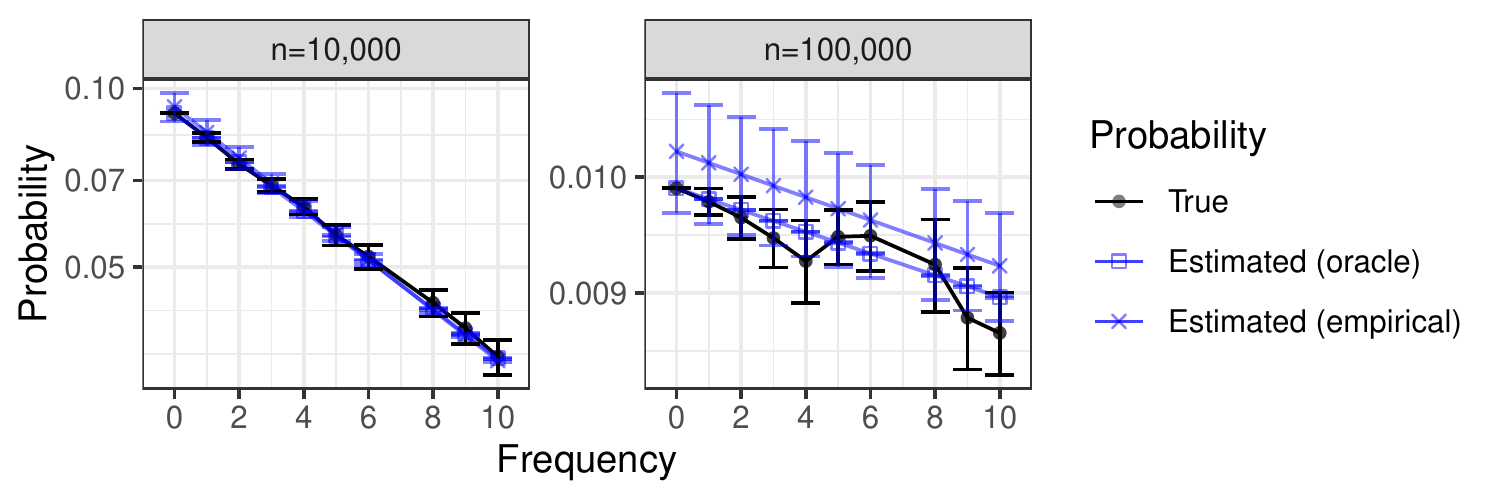}
\caption{True and estimated coverage probabilities $\mathfrak{p}_{r,n}$ for synthetic data sets of different size generated from a DP prior with $\theta=100$, for different values of $r$. The height of the error bars is equal to two standard errors. Other details are as in Figure~\ref{fig:exp-dp}.}
\label{fig:exp-dp-coverage}
\end{figure}

Figure~\ref{fig:exp-dp} compares the true and estimated missing-mass probabilities as a function of the sample size, separately for data generated from DP prior models with $\alpha=0$ and different values of $\theta$, and for PYP models with $\alpha=0.5$ and different values of $\theta$. 
In Figure~\ref{fig:exp-dp}~(a), where $\alpha=0$, we compare estimated probabilities calculated with perfect ({\em oracle}) knowledge of the true data-generating parameter $\theta$ to those based on an empirical Bayes estimate of $\theta$, as detailed in Section~\ref{sec:coverage}.
Both estimates coincide almost exactly with the true probabilities.
In Figure~\ref{fig:exp-dp}~(b), where $\alpha=0.5$, we compare two versions of our estimators from~\eqref{post_pyp_missing} approximated via the Monte Carlo approach detailed in Section~\ref{sec:coverage-pyp-mc} utilizing 100,000 independent realizations of $Z_{r,n}$ and $Z'_{r,n}$, with and without approximately debiasing the ratio of expected values using Tin's method~\citep{tin1965comparison}.
For simplicity, we do not estimate $\theta$ empirically here, relying instead on the true value of this parameter.
These results show our BNP estimates coincide almost exactly with the true missing mass when the sample size $n$ is very small. By contrast, if $n$ is larger, our Monte Carlo BNP method overestimates the missing mass due to the skewness of $Z_{r,n}$ and $Z'_{r,n}$ in~\eqref{post_pyp_mc}; see \cite{quenouille1956notes}.
Unfortunately, even the approximate debiasing technique of~\citet{tin1965comparison} mitigates but does not eliminate this problem. 
Figure~\ref{fig:exp-dp-coverage} compares the true and estimated coverage probabilities for different values of the frequency $r$, in the special case of the DP prior with $\theta=1000$, and for two different values of the sample size. These results confirm our BNP estimates from Section~\ref{sec:coverage} are accurate, regardless of whether the prior parameter $\theta$ is known or estimated empirically from the sketched data. Additional results from these experiments are in Appendix~\ref{app_experiments}.

Figure~\ref{fig:exp-pyp-alpha} in Appendix~\ref{app_experiments} reports on results analogous to those in Figure~\ref{fig:exp-dp}~(b), but varying the data-generating parameter $\alpha$ instead of $\theta$.
Figures~\ref{fig:exp-dp-pyp} and~\ref{fig:exp-dp-pyp-theta} in Appendix~\ref{app_experiments} show our BNP estimates of the missing mass calculated under a possibly mis-specified assumption that $\alpha=0$, for synthetic data generated from a PYP prior model with different values of $(\alpha,\theta)$. In these experiments, the parameter $\theta$ is estimated empirically via maximum marginal likelihood. Intuitively, we see our estimates  may not be accurate if the prior model is mis-specified, especially if the sample size is large. In particular, the estimates computed with the assumption that $\alpha=0$ tend to underestimate the missing mass if the data-generating model exhibits power-law tail behaviour, as it is the case for the PYP with larger $\alpha$.
Analogous conclusions can be drawn from Figure~\ref{fig:exp-dp-zipf}, which reports on similar experiments based on data simulated from a Zipf distribution.

\begin{figure}[!htb]
\centering
\includegraphics[width=0.9\linewidth]{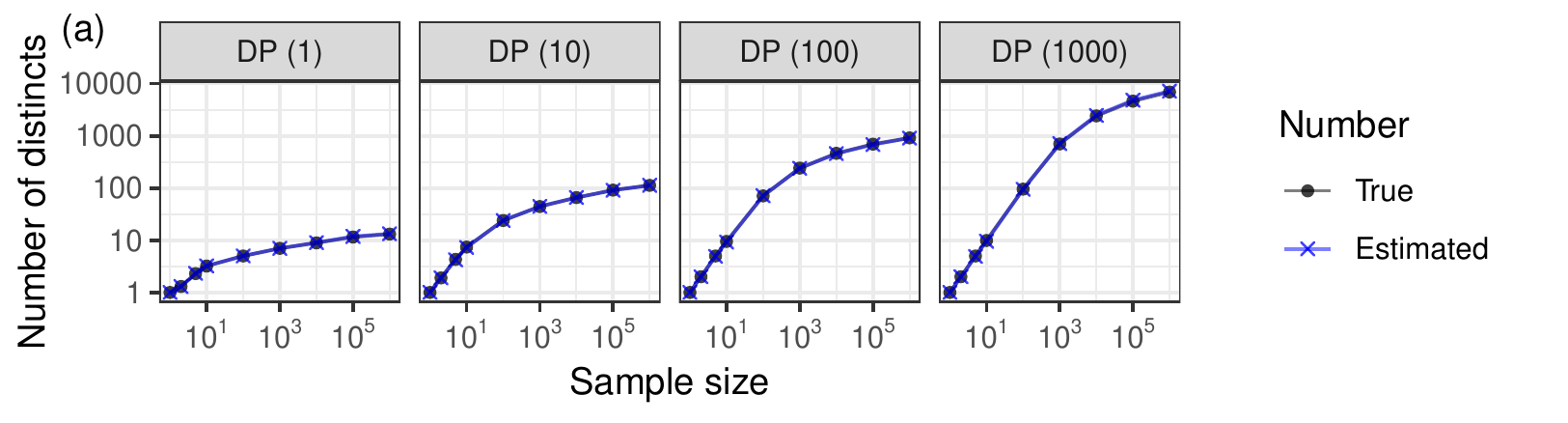}
\includegraphics[width=0.9\linewidth]{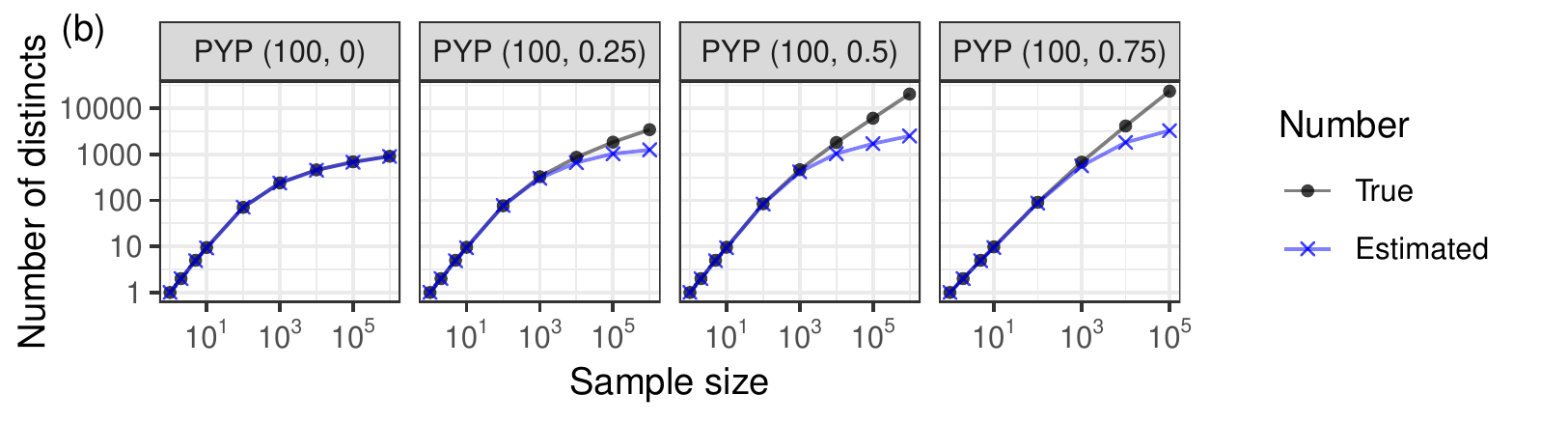}
\caption{True and estimated number of distinct species for synthetic data generated from a PYP prior model with different parameters $(\alpha,\theta)$. The estimated probabilities are computed under the simplifying assumption that $\alpha=0$.
Other details are as in Figure~\ref{fig:exp-dp}.}
\label{fig:exp-dp-dp-distinct}
\end{figure}

Next, we turn our attention to the estimation of $K_{n}$, the number of distinct species in the sample.
Figure~\ref{fig:exp-dp-dp-distinct} reports on experiments similar to those in Figure~\ref{fig:exp-dp}: the true and estimated numbers of distinct species are shown as a function of the sample size $n$, separately for data generated from DP prior models with $\alpha=0$ and different values of $\theta$, and for PYP models with $\theta=100$ and different values of $\alpha$. Note that all estimates are computed under the (possibly mis-specified) assumption that $\alpha=0$, and estimating $\theta$ empirically via maximum marginal likelihood, to avoid the computational issues related to the Monte Carlo estimation experienced in Figure~\ref{fig:exp-dp}~(b).
As predicted by the theory and anticipated by our previous experiments, the results in Figure~\ref{fig:exp-dp-dp-distinct} confirm our estimates are accurate when the DP prior is well-specified, while otherwise they tend to underestimate the number of distinct species, especially if $n$ is large.
Analogous conclusions can be drawn from Figure~\ref{fig:exp-dp-zipf-distinct}, which reports on similar experiments based on synthetic data generated from a Zipf distribution.

\subsection{Experiments with real data}

Finally, we apply the proposed methods to three real data sets. 
The first data set was made publicly available by the National Center for Biotechnology Information~\citep{hatcher2017virus} and contains 43,196 sequences of approximately 30,000 nucleotides each, collected from SARS-CoV-2 viruses.
For each nucleotide sequence, we extract a list of all contiguous DNA sub-sequences of length $16$ (i.e., {\em 16-mers}), and then we sketch the resulting data set with a random hash function.
The second data set consists of 18 open-domain classic pieces of English literature from the Gutenberg Corpus~\citep{Gutenberg}, downloaded using the NLTK Python package~\citep{bird2009natural}. These data are pre-processed with the same approach of \citet{sesia2022conformalized}: after removing punctuation and unusual words (keeping only words contained in a English dictionary of size 25,487), we extract 1,700,000 consecutive pairs of words, or {\em 2-grams}. Then, the 2-grams are sketched with a random hash function, as usual.
The third data set is discussed in \citet{rojas2018personalized} and contains a list of 3,577,296 IP addresses, which we sketch directly without pre-processing; these data were made publicly available through the Kaggle machine-learning competition website.

\begin{figure}[!htb]
\centering
\includegraphics[width=0.8\linewidth]{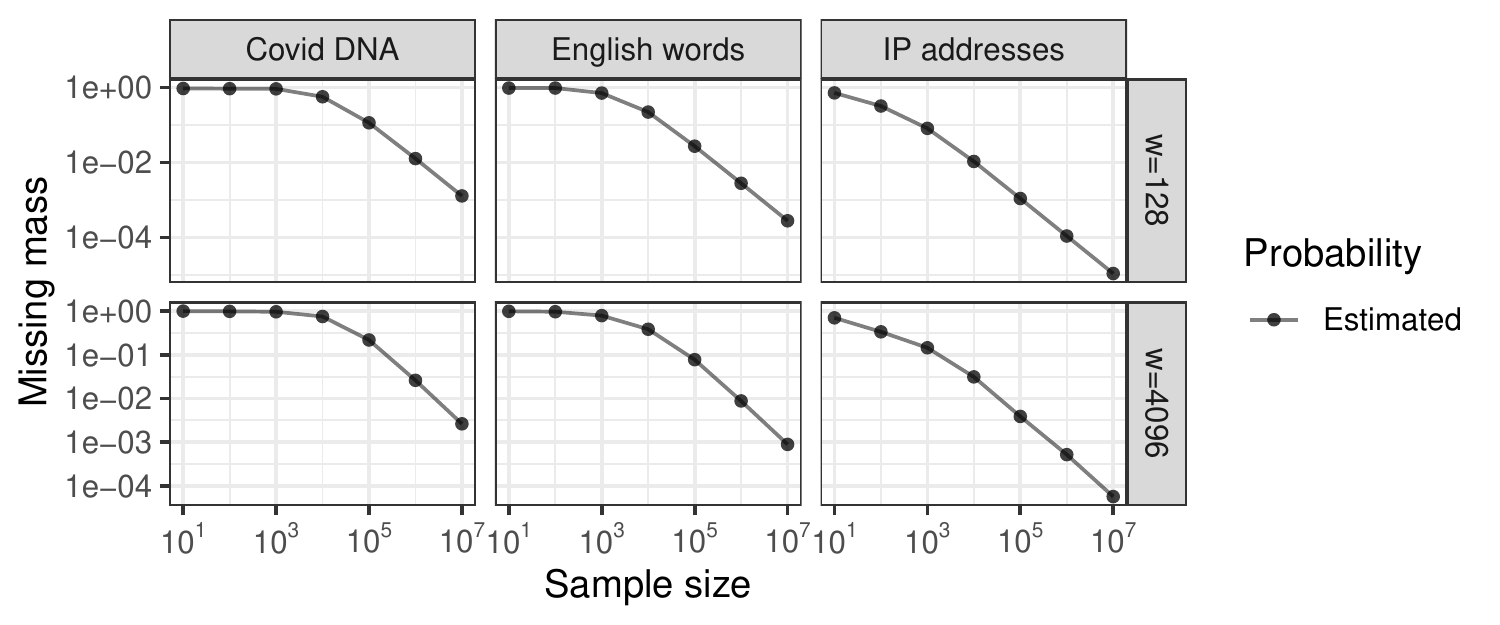}
\caption{True and estimated missing mass for three data sets sketched by a hash function of width 128 or 4096. Other details are as in Figure~\ref{fig:exp-dp}.}
\label{fig:exp-dp-real}
\end{figure}

\begin{figure}[!htb]
\centering
\includegraphics[width=0.8\linewidth]{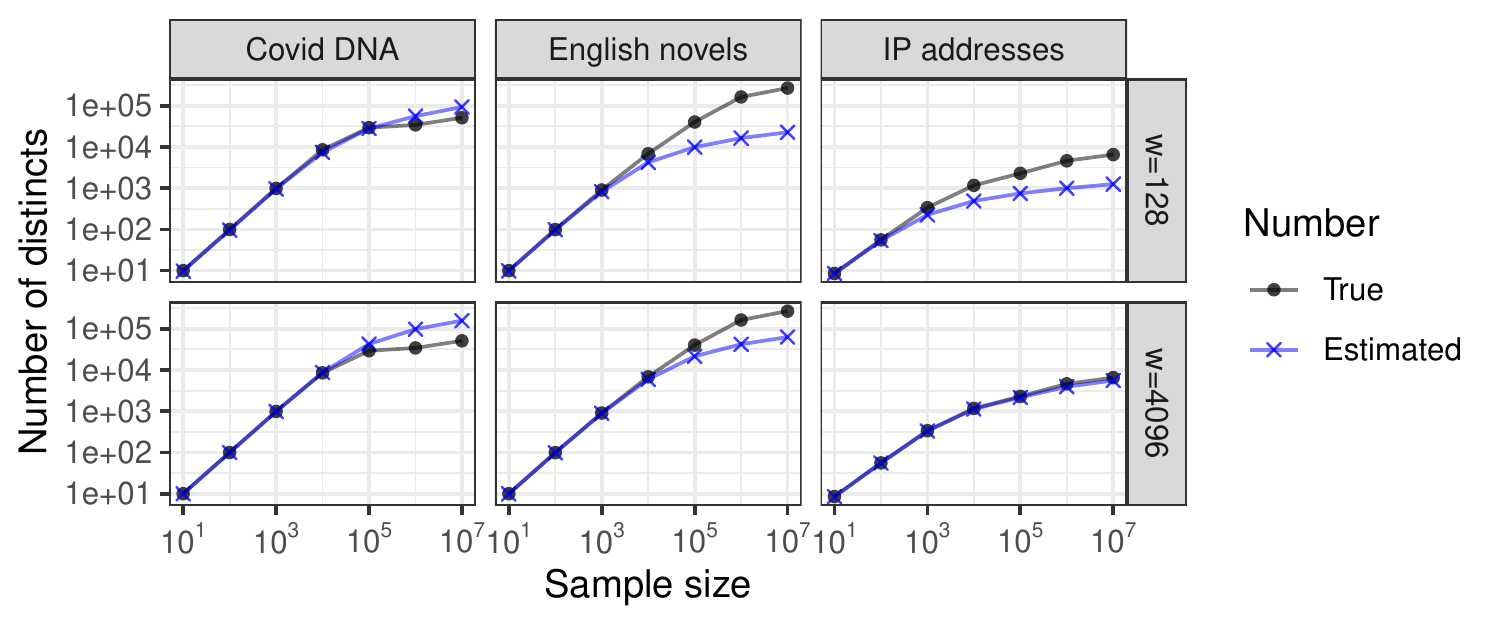}
\caption{Numbers of distinct species for sketched data. Other details are as in Figure~\ref{fig:exp-dp-real}.}
\label{fig:exp-dp-data-distinct}
\end{figure}

Figure~\ref{fig:exp-dp-real} compares the true and estimated missing-mass probabilities for random subsets of the three aforementioned data sets, as a function of the sample size and for two different values of the hash function width. Our BNP estimated probabilities are calculated assuming $\alpha=0$, and estimating $\theta$ empirically via maximum marginal likelihood. As in the previous section, all results are averaged over 20 independent experiments with different hash functions.
The results show our estimated probabilities are relatively accurate for the DNA data set, which does not exhibit power-law tail behaviour \citep{sesia2022conformalized}, but tend to underestimate the true missing mass in the other cases, especially if the hash function width is small.
Figure~\ref{fig:exp-dp-data-distinct} reports analogous results corresponding to the estimation of the number of distinct species.

\clearpage

\section{Discussion} \label{sec5}

This paper studied the classical problem of estimating coverage probabilities from a novel perspective, addressing the increasingly relevant situations in which one only has access to a compressed and imperfect summary, or sketch, of the true data. Despite the clear practical nature of this problem in the age of big data, cloud computing and privacy concerns, we are not aware of other existing statistical or algorithmic approaches for estimating coverage probabilities in relation to randomized data compression, or sketching. This paper began to fill this gap by developing a BNP methodology to estimate coverage probabilities from data sketched through random hashing, assuming a PYP prior. 

The PYP has been widely used in BNP inference for species sampling problems (assuming the true data to be available), mostly because of its uniqueness in combining mathematical tractability with a flexible tail behaviour driven by an easily interpretable parameterization \citep{Bal(22)}. In particular, the mathematical tractability of the PYP prior has proved before to be a desirable feature in order to achieve posterior inferences that are tractable to evaluate, computationally efficient and scalable to massive data sets. However, our paper has shown that the BNP estimation of coverage probabilities from sketched data poses additional challenges regarding the choice of the prior distribution. In particular, the lack of mathematical tractability in the finite-dimensional projection of the PYP prior leads to numerically impractical estimators in the context of sketched data, except for the special case of the DP prior.
Therefore, our work paves the way to investigate more computationally efficient approximations of BNP estimators under the general PYP prior, starting from numerical approximations of the estimator in \eqref{post_pyp} or alternative representations of the Monte Carlo results in Proposition~\ref{main_post_mc}. Theorem \ref{teo_pyp} also motivates further study of the large $n$ asymptotic behaviour of the estimator in \eqref{post_pyp}, with the aim of obtaining simpler large $n$ approximations with reliable error bounds. 

As this is the first work on the estimation of coverage probabilities from sketched data, it opens several new avenues for future research. First of all, our work focused on providing BNP estimators, leaving open the question of how to quantify the uncertainty of such estimators. This is a challenging problem, as assessing the uncertainty would require computing the posterior distribution of the coverage probabilities given the sketched data, or equivalently the posterior distribution of the model given the sketched data. Beyond coverage probabilities, one may consider more general species sampling problems, with the most relevant being the estimation of the number of unseen symbols \citep{Goo(56),Efr(76),Orl(17)}. In particular, based on a sketch of $n$ data points modeled as a random sample from an unknown distribution $p$, which is endowed with a suitable prior, how to estimate the number of hitherto unseen symbols that would be observed if $m$ additional samples were collected from the same distribution? This is an $m$-steps ahead generalization of the estimation of the missing mass considered in this paper. Finally, this paper may inspire future studies of such problems from a frequentist perspective, which is a well-developed area in the context of full data availability, but has yet not been investigated in relation with sketching.

\appendix

\appendixtitleon

\begin{appendices}

\section{Generalized factorial coefficients}\label{app_combin}
We recall the definitions of Stirling number of the first type and generalized factorial coefficient, as well as some of their key properties needed to prove our results \citep[Chapter 2]{Cha(05)}. For $t>0$, the $(u,v)$-th signless Stirling number of the first type, denoted by $|s(u,v)|$, is the defined as the $v$-th coefficient in the expansion of $(t)_{(u)}$ into powers, i.e., 
\begin{equation}\label{defi_stir}
(t)_{(u)}=\sum_{v=0}^{u}|s(u,v)|t^{v},
\end{equation}
with the proviso that  $|s(0,0)|=1$, $|s(u,0)|=0$ for $u>0$ and $|s(u,v)|=0$ for $v>u$. Generalized factorial coefficients provide a generalization of signless Stirling number of the first type. In particular, the $(u,v)$-th (centered) generalized factorial coefficient, namely $\mathscr{C}(u,v;a)$, is defined as the  $v$-th coefficient in the expansion of $(at)_{(u)}$ into rising factorials, i.e.,
\begin{equation}\label{defi_gfc}
(at)_{(u)}=\sum_{v=0}^{u}\mathscr{C}(u,v;a)(t)_{(v)},
\end{equation}
with $\mathscr{C}(0,0;a)=1$, $\mathscr{C}(u,0;a)=0$ for $u>0$, $\mathscr{C}(u,v;a)=0$ for $v>u$. An explicit expression for the generalized factorial coefficient is in \citet[Theorem 2.14]{Cha(05)}, i.e., 
\begin{displaymath}
\mathscr{C}(u,v;\alpha)=\frac{1}{v!}\sum_{i=0}^{v}(-1)^{i}{u\choose v}(-i\alpha)_{(u)}, 
\end{displaymath}
from which
\begin{equation}\label{triangular}
\mathscr{C}(u+1,v;\alpha)=(u-va)\mathscr{C}(u,v;\alpha)+\alpha\mathscr{C}(u,v+1;\alpha).
\end{equation}
See \citet[Theorem 2.18]{Cha(05)}. Stirling numbers of the first type arise from the generalized factorial coefficients as $\alpha\rightarrow0$. From \citet[Theorem 2.16]{Cha(05)} 
\begin{equation}\label{asym_gfc_alpha}
\lim_{\alpha\rightarrow0}\frac{1}{\alpha^{k}}\mathscr{C}(u,v;\alpha)=|s(u,v)|.
\end{equation}
We refer to \citet[Chapter 2 and Chapter 3]{Cha(05)} for further details on Stirling numbers of the first type, generalized factorial coefficients and some generalizations thereof.

\section{Tail behaviour of the PYP prior} \label{app_tail}

\begin{figure}[!htb]
    \centering
    \includegraphics[width=\linewidth]{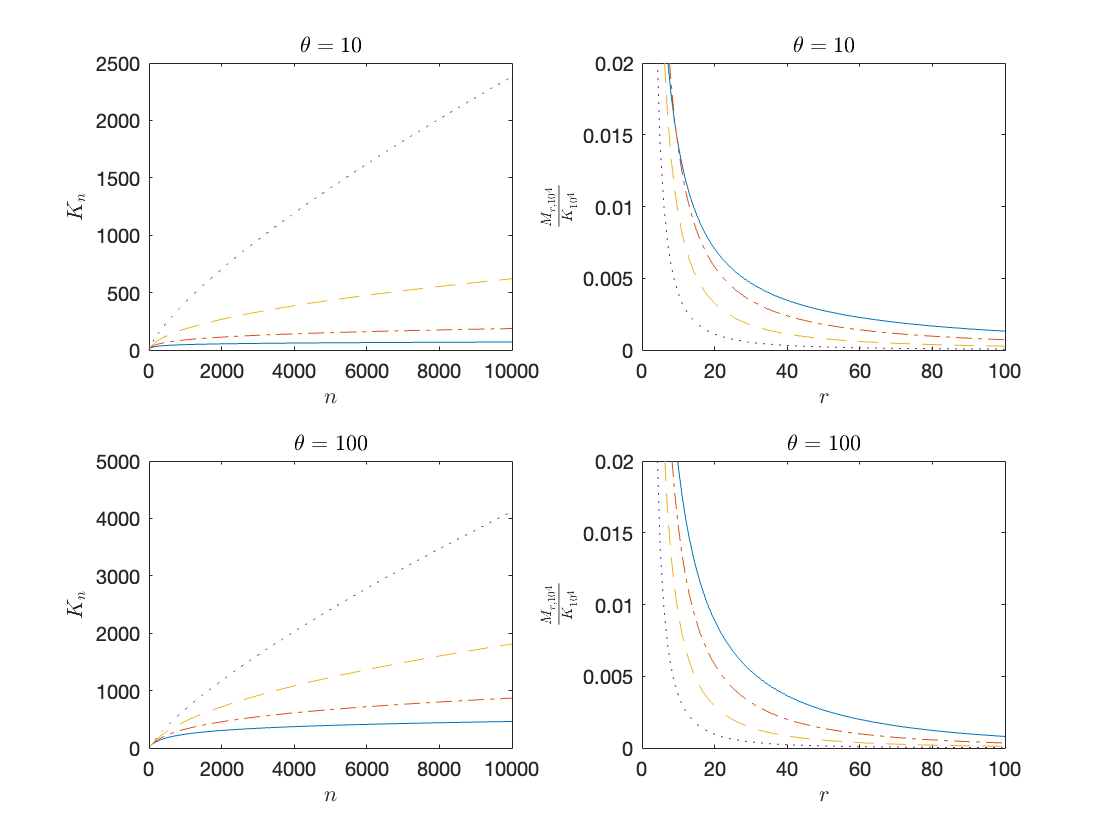}
    \caption{\small{Behaviours in the sample size $n\geq1$ of the statistic $K_{n}$ and the statistic $M_{r,n}/K_{n}$, for $1\leq n\leq 10^{4}$ under $P\sim\text{PYP}(\alpha,\theta)$: $\alpha=0$ (blue -), $\alpha=.25$ (red -.), $\alpha=.5$ (yellow --) and $\alpha=.75$ (purple :).}}
    \label{fig:draws}
\end{figure}

The PYP prior features a power-law tail behaviour, in contrast with the geometric tail behaviour of the DP prior, and such a behaviour emerges from the large-$n$ asymptotics of $K_{n}$ and $M_{r,n}$. For any $\alpha\in(0,1)$ let $f_{\alpha}$ be the positive $\alpha$-stable density function, and for $\theta>-\alpha$ let $S_{\alpha,\theta}$ be a positive random variable whose distribution has a density function
\begin{equation}\label{adivers}
f_{S_{\alpha,\theta}}(s)\propto s^{\frac{\theta-1}{\alpha}-1}f_{\alpha}(s^{-1/\alpha}),
\end{equation}
which is a Mittag-Leffler density function. \citet[Theorem 3.8]{Pit(06)} shows that, as $n\rightarrow+\infty$,
\begin{align} \label{eq:sigma_diversity}
n^{-\alpha}K_{n}\rightarrow S_{\alpha,\theta}
\end{align}
and
\begin{align} \label{eq:sigma_diversity_m}
n^{-\alpha}M_{r,n}\rightarrow \frac{\alpha(1-\alpha)_{(r-1)}}{r!}S_{\alpha,\theta}
\end{align}
almost surely. Equation \eqref{eq:sigma_diversity} shows that $K_{n}$, for large $n$, grows as $n^{\alpha}$; this is the growth of the number of distinct symbols in $n\geq1$ samples from a power-law distribution of exponent $c=\alpha^{-1}$. By combining \eqref{eq:sigma_diversity} and \eqref{eq:sigma_diversity_m}, it holds that $p_{\alpha,r}=\alpha(1-\alpha)_{(r-1)}/r!$ is the large-$n$ asymptotic proportion of the number of distinct symbols with frequency $r$. Then $ p_{\alpha,r}\approx c_{\alpha}r^{-\alpha -1}$ for large $r$, with $c_{\alpha}$ being a constant; this is the distribution of the number of distinct symbols with frequency $r$  in $n\geq1$ samples from a power-law distribution of exponent $c=\alpha^{-1}$. The parameter $\alpha\in(0,1)$ then controls the power-law tail behaviour of $P\sim\text{PYP}(\alpha,\theta)$: the larger $\alpha$ the heavier the tail of $P$. See Figure \ref{fig:draws}. As a limiting case for $\alpha\rightarrow0$, the DP features geometric tails \citep[Chapter 3 and Chapter 4]{Pit(06)}.

\section{Proofs}
\subsection{Proof of Theorem \ref{teo_dp}}\label{app_proof_teo_dp}
The proof relies on the finite-dimensional projective property of the DP, in combination with the exchangeability of $X_{i}$'s and some combinatorial arguments on the sketch $\mathbf{C}_{n}$. The independence between $h$ and $\mathbf{X}_{n}$ allows us to invoke the ``freezing lemma" \citep[Lemma 4.1]{Bal(17)}, according to which we can treat $h$ as fixed, i.e., non-random. Then, from \eqref{estim_sketc},
\begin{align}\label{main_term_dp}
\tilde{\mathfrak{p}}_{r,n}=\sum_{j=1}^{J}\text{Pr}[h(X_{n+1})=j\,|\, \mathbf{C}_{n}=\mathbf{c}]\text{Pr}[X_{n+1}\in \mathcal{S}_{r}\,|\, h(X_{n+1})=j,\mathbf{C}_{n}=\mathbf{c}],
\end{align}
where $\mathbf{c}=(c_{1},\ldots,c_{J})$. First, we consider the evaluation of $\text{Pr}[X_{n+1}\in \mathcal{S}_{r}\,|\, h(X_{n+1})=j,\mathbf{C}_{n}=\mathbf{c}]$, i.e.,
\begin{align}\label{second_term_dp}
&\text{Pr}[X_{n+1}\in \mathcal{S}_{r}\,|\, h(X_{n+1})=j,\mathbf{C}_{n}=\mathbf{c}]=\frac{\text{Pr}[X_{n+1}\in \mathcal{S}_{r},h(X_{n+1})=j,\mathbf{C}_{n}=\mathbf{c}]}{\text{Pr}[h(X_{n+1})=j,\mathbf{C}_{n}=\mathbf{c}]}.
\end{align}
Consider the denominator of \eqref{second_term_dp}. Uniformity of $h$ implies that $h$ induces a partition $\{B_{1},\ldots,B_{J}\}$ of $\mathbb{S}$ such that $B_{j}=\{s\in\mathbb{S}\text{ : }h(s)=j\}$ and $\nu(B_{j})=J^{-1}$ for $j=1,\ldots,J$. The finite-dimensional projective property of the DP implies that $(P(B_{1}),\ldots,P(B_{J}))$ is distributed as a Dirichlet distribution with parameter $(\theta/J,\ldots,\theta/J)$. Therefore, we write
\begin{align}\label{denomin_dp}
&\text{Pr}[h(X_{n+1})=j,\mathbf{C}_{n}=\mathbf{c}]\\
&\notag\quad={n\choose c_{1},\ldots,c_{J}}\E\left[(P(B_{j}))^{c_{j}+1}\prod_{1\leq s\neq j\leq J} (P(B_{s}))^{c_{s}}\right]\\
&\notag\quad={n\choose c_{1},\ldots,c_{J}}\int_{\Delta_{J}}p_{1}^{c_{1}}\cdots p_{j}^{c_{j}+1}\cdots p_{J}^{c_{J}}\frac{\Gamma(\theta)}{\Gamma(\theta/J)\cdots\Gamma(\theta/J)}p_{1}^{\theta/J-1}\cdots p_{J}^{\theta/J-1}\ddr p_{1}\cdots \ddr p_{J}\\
&\notag\quad={n\choose c_{1},\ldots,c_{J}}\frac{\Gamma(\theta)\Gamma(\theta/J+c_{1})\cdots\Gamma(\theta/J+c_{j}+1)\cdots\Gamma(\theta/J+c_{J})}{\Gamma(\theta/J)\cdots\Gamma(\theta/J)\Gamma(\theta+n+1)}.
\end{align}
Now, consider the numerator of \eqref{second_term_dp}. To evaluate the numerator of \eqref{second_term_dp}, we define the event $B(n,r)=\{X_{1}=\cdots=X_{r}=X_{m+1},\{X_{r+1},\ldots,X_{n}\}\cap\{X_{n+1}\}=\emptyset\}$, such that we write
\begin{align*}
&\text{Pr}[X_{n+1}\in \mathcal{S}_{r},h(X_{n+1})=j,\mathbf{C}_{n}=\mathbf{c}]\\
&\quad=\frac{1}{J}{m\choose r}\text{Pr}[B(m,r),h(X_{n+1})=j,\mathbf{C}_{n}=\mathbf{c}]\\
&\quad=\frac{1}{J}{m\choose r}\text{Pr}\left[B(m,r),h(X_{n+1})=j,C_{1,n}=c_{1},\ldots\right.\\
&\quad\quad\quad\quad\quad\quad\quad\quad\quad\quad\quad\quad\quad\quad\quad\left.\ldots,\sum_{i=r+1}^{n}\mathbbm{1}_{\{h(X_{i})\}}(h(X_{n+1}))=c_{j}-r,\dots,C_{J,n}=c_{J}\right].
\end{align*}
The distribution of the random variable $(X_{n+1},h(X_{n+1}),\mathbf{C}_{n})$ is determined by the distribution of the random variable $(X_{1},\ldots,X_{n},X_{n+1})$. In particular, let $\Pi(s,k)$ be the set of all partitions of ${1,\ldots,s}$ into $k$ disjoint subsets $\pi_{1},\ldots,\pi_{k}$ such that $n_{i}$ is the cardinality of $\pi_{i}$. From \citet[Equation 3.5]{San(06)}, for any measurable $A_{1},\ldots,A_{n},A_{n+1}$ we can write
\begin{align*}
&\text{Pr}[X_{1}\in A_{1},\ldots,X_{n}\in A_{n},X_{n+1}\in A_{n+1}]\\
&\quad=\sum_{k=1}^{n+1}\frac{\theta^{k}}{(\theta)_{(n+1)}}\sum_{(\pi_{1},\ldots,\pi_{k})\in\Pi_{n+1,k}}\prod_{i=1}^{k}(n_{i}-1)!\nu(\cap_{n\in\pi_{i}}A_{n})
\end{align*}
for any $n\geq1$. Now, let $\mathcal{S}$ be the Borel $\sigma$-algebra of $\mathbb{S}$, and let $\nu_{\pi_{1},\ldots,\pi_{k}}$ be a probability measure on $(\mathbb{S}^{n+1},\mathcal{S}^{n+1})$ defined as $\nu_{\pi_{1},\ldots,\pi_{k}}=\prod_{1\leq i\leq k}\nu(\cap_{n\in\pi_{i}}A_{n})$ and attaching to $B(n,r)$ a value that is either $0$ or $1$. In particular, $\nu_{\pi_{1},\ldots,\pi_{k}}(B(n,r))=1$ if and only if one of the $\pi_{i}$'s is equal to the set $\{1,\ldots,r,n+1\}$. Then, based on $\nu_{\pi_{1},\ldots,\pi_{k}}$ we can write that
\begin{align*}
&\text{Pr}\left[B(n,r),h(X_{n+1})=j,C_{1,n}=c_{1},\ldots,\sum_{i=r+1}^{n}\mathbbm{1}_{\{h(X_{i})\}}(h(X_{n+1}))=c_{j}-r,\dots,C_{J,n}=c_{J}\right]\\
&\quad=\sum_{k=2}^{n-r+1}\frac{\theta^{k}}{(\theta)_{(n+1)}}\sum_{(\pi_{1},\ldots,\pi_{k-1})\in\Pi_{n-r,k-1}}r!\prod_{i=1}^{k}(n_{i}-1)!\\
&\quad\quad\times\nu_{\pi_{1},\ldots,\pi_{k}}\left(C_{1,n}=c_{1},\ldots,\sum_{i=r+1}^{n}\mathbbm{1}_{\{h(X_{i})\}}(h(X_{n+1}))=c_{j}-r,\dots,C_{J,n}=c_{J}\right)\\
&\quad=\theta\frac{(\theta)_{(n-r)}}{(\theta)_{(n+1)}}r!\sum_{l=1}^{n-r}\frac{\theta^{l}}{(\theta)_{(n-r)}}\sum_{(\pi_{1},\ldots,\pi_{l})\in\Pi_{n-r,l}}r!\prod_{i=1}^{l}(n_{i}-1)!\\
&\quad\quad\times\nu_{\pi_{1},\ldots,\pi_{k}}\left(C_{1,n}=c_{1},\ldots,\sum_{i=r+1}^{n}\mathbbm{1}_{\{h(X_{i})\}}(h(X_{n+1}))=c_{j}-r,\dots,C_{J,n}=c_{J}\right),
\end{align*}
where
\begin{align*}
&\sum_{l=1}^{n-r}\frac{\theta^{l}}{(\theta)_{(n-r)}}\sum_{(\pi_{1},\ldots,\pi_{l})\in\Pi_{n-r,l}}r!\prod_{i=1}^{l}(n_{i}-1)!\\
&\quad\times\nu_{\pi_{1},\ldots,\pi_{k}}\left(C_{1,n}=c_{1},\ldots,\sum_{i=r+1}^{n}\mathbbm{1}_{\{h(X_{i})\}}(h(X_{n+1}))=c_{j}-r,\dots,C_{J,n}=c_{J}\right)
\end{align*}
is the distribution of $(X_{1},\ldots,X_{n-r})$ from a DP with scale $\theta$, which is provided by \citet[Equation 3.5]{San(06)} by relying on the finite-dimensional projective property of the DP, i.e.
\begin{align*}
&\theta\frac{(\theta)_{(n-r)}}{(\theta)_{(n+1)}}r!\sum_{l=1}^{n-r}\frac{\theta^{l}}{(\theta)_{(n-r)}}\sum_{(\pi_{1},\ldots,\pi_{r})\in\Pi_{n-r,l}}r!\prod_{i=1}^{l}(n_{i}-1)!\\
&\quad\quad\times\nu_{\pi_{1},\ldots,\pi_{k}}\left(C_{1,n}=c_{1},\ldots,\sum_{i=r+1}^{n}\mathbbm{1}_{\{h(X_{i})\}}(h(X_{n+1}))=c_{j}-r,\dots,C_{J,n}=c_{J}\right)\\
&\quad=\theta\frac{(\theta)_{(n-r)}}{(\theta)_{(n+1)}}r!{n-r\choose c_{1},\ldots,c_{j}-r,\ldots,c_{J}}\E\left[(P(B_{j}))^{c_{j}-r}\prod_{1\leq s\neq j\leq J} (P(B_{s}))^{c_{s}}\right]\\
&\quad=\theta\frac{(\theta)_{(n-r)}}{(\theta)_{(n+1)}}r!{n-r\choose c_{1},\ldots,c_{j}-r,\ldots,c_{J}}\\
&\quad\quad\times\int_{\Delta_{J}}p_{1}^{c_{1}}\cdots p_{j}^{c_{j}-r}\cdots p_{J}^{c_{J}}\frac{\Gamma(\theta)}{\Gamma(\theta/J)\cdots\Gamma(\theta/J)}p_{1}^{\theta/J-1}\cdots p_{J}^{\theta/J-1}\ddr p_{1}\cdots \ddr p_{J}\\
&\quad=\theta\frac{(\theta)_{(n-r)}}{(\theta)_{(n+1)}}r!{n-r\choose c_{1},\ldots,c_{j}-r,\ldots,c_{J}}\\
&\quad\quad\times\frac{\Gamma(\theta)\Gamma(\theta/J+c_{1})\cdots\Gamma(\theta/J+c_{j}-r)\cdots\Gamma(\theta/J+c_{J})}{\Gamma(\theta/J)\cdots\Gamma(\theta/J)\Gamma(\theta+m-l)},
\end{align*}
where the first identity follows from \citet[Proposition 3.1]{San(06)} under the DP prior; see also the formule displayed at page 469 of \citet{San(06)}. Therefore, we can write that
\begin{align}\label{num_dp}
&\text{Pr}[X_{n+1}\in \mathcal{S}_{r},h(X_{n+1})=j,\mathbf{C}_{n}=\mathbf{c}]\\
&\notag\quad=\frac{1}{J}{n\choose r}\theta\frac{(\theta)_{(n-r)}}{(\theta)_{(n+1)}}r!{n-r\choose c_{1},\ldots,c_{j}-r,\ldots,c_{J}}\\
&\notag\quad\quad\times\frac{\Gamma(\theta)\Gamma(\theta/J+c_{1})\cdots\Gamma(\theta/J+c_{j}-r)\cdots\Gamma(\theta/J+c_{J})}{\Gamma(\theta/J)\cdots\Gamma(\theta/J)\Gamma(\theta+n-r)}.
\end{align}
Now, according to \eqref{second_term_dp}, we combine \eqref{denomin_dp} with \eqref{num_dp} in order to obtain the conditional probability
\begin{align}\label{final_second_term}
&\text{Pr}[X_{n+1}\in \mathcal{S}_{r}\,|\,h(X_{n+1})=j,\mathbf{C}_{n}=\mathbf{c}]=\frac{\theta/J}{\theta/J+c_{j}}\frac{(c_{j}-r+1)_{(l)}}{(\theta/J+c_{j}-l)_{(r)}}
\end{align}
This result completes the evaluation of $\text{Pr}[X_{n+1}\in \mathcal{S}_{r}\,|\,h(X_{n+1})=j,\mathbf{C}_{n}=\mathbf{c}]$. Now, we consider the evaluation of $\text{Pr}[h(X_{n+1})=j\,|\,\mathbf{C}_{n}=\mathbf{c}]$, which corresponds to the following ratio
\begin{align}\label{first_term_dp}
&\text{Pr}[h(X_{n+1})=j\,|\,\mathbf{C}_{n}=\mathbf{c}]=\frac{\text{Pr}[h(X_{n+1})=j,\mathbf{C}_{n}=\mathbf{c}]}{\text{Pr}[\mathbf{C}_{n}=\mathbf{c}]}.
\end{align}
The numerator of \eqref{first_term_dp} is given by \eqref{denomin_dp}, whereas the denominator of \eqref{first_term_dp} follows directly from the finite-dimensional projective property of the DP. In particular, we write
\begin{align}\label{denomin_dp_extra}
&\text{Pr}[\mathbf{C}_{n}=\mathbf{c}]\\
&\notag\quad={n\choose c_{1},\ldots,c_{J}}\E\left[\prod_{s=1}^{J} (P(B_{s}))^{c_{s}}\right]\\
&\notag\quad={n\choose c_{1},\ldots,c_{J}}\int_{\Delta_{J}}p_{1}^{c_{1}}\cdots p_{J}^{c_{J}}\frac{\Gamma(\theta)}{\Gamma(\theta/J)\cdots\Gamma(\theta/J)}p_{1}^{\theta/J-1}\cdots p_{J}^{\theta/J-1}\ddr p_{1}\cdots \ddr p_{J}\\
&\notag\quad={n\choose c_{1},\ldots,c_{J}}\frac{\Gamma(\theta)\Gamma(\theta/J+c_{1})\cdots\Gamma(\theta/J+c_{J})}{\Gamma(\theta/J)\cdots\Gamma(\theta/J)\Gamma(\theta+n)}.
\end{align}
Now, according to \eqref{first_term_dp}, we combine \eqref{denomin_dp} with \eqref{denomin_dp_extra} to obtain the conditional probability
\begin{align}\label{final_primo_term}
&\text{Pr}[h(X_{n+1})=j\,|\,\mathbf{C}_{n}=\mathbf{c}]=\frac{\theta/J+c_{j}}{\theta+n}.
\end{align}
According to \eqref{main_term_dp} the proof of Equation \eqref{post_dp} is completed by combining \eqref{final_primo_term} with \eqref{final_second_term}. With regards to the proof of Equation \eqref{recov_sketc_unseen_freq_estim_dp}, we define the partition set $\mathcal{M}_{n,k}=\{(m_{1},\ldots,m_{n})\text{ : }m_{i}\geq0,\,\sum_{1\leq i\leq n}m_{i}=k\text{ and }\sum_{1\leq i\leq n}im_{i}=n\}$. Then, we can write that
\begin{align*}
\tilde{\mathfrak{p}}_{r,n}&=\sum_{k=1}^{n}\sum_{(m_{1},\ldots,m_{n})\in\mathcal{M}_{n,k}}\text{Pr}[X_{n+1}\in\mathcal{S}_{r}\,|\,\mathbf{C}_{n}=\mathbf{c},\mathbf{M}_{n}=(m_{1},\ldots,m_{n})]\\
&\quad\times\text{Pr}[\mathbf{M}_{n}=(m_{1},\ldots,m_{n})\,|\,\mathbf{C}_{n}=\mathbf{c}]\\
&=\sum_{k=1}^{n}\sum_{(m_{1},\ldots,m_{n})\in\mathcal{M}_{n,k}}\hat{\mathfrak{p}}_{r,n}\text{Pr}[\mathbf{M}_{n}=(m_{1},\ldots,m_{n})\,|\,\mathbf{C}_{n}=\mathbf{c}]\\
&\quad\text{[by Equation \eqref{bnp_est} with $\alpha=0$]}\\
&=\sum_{k=1}^{n}\sum_{(m_{1},\ldots,m_{n})\in\mathcal{M}_{n,k}}\frac{rm_{r}}{\theta+n}\text{Pr}[\mathbf{M}_{n}=(m_{1},\ldots,m_{n})\,|\,\mathbf{C}_{n}=\mathbf{c}]\\
&=\frac{r}{\theta+n}\E[M_{r,n}\,|\,\mathbf{C}_{n}=\mathbf{c}],
\end{align*}
i.e., 
\begin{displaymath}
\tilde{\mathfrak{p}}_{r,n}=\frac{r}{\theta+n}\tilde{\mathfrak{m}}_{r,n}
\end{displaymath}
and
\begin{displaymath}
\tilde{\mathfrak{m}}_{r,n}=\frac{\theta+n}{r}\tilde{\mathfrak{p}}_{r,n},
\end{displaymath}
which completes the proof of Equation \eqref{recov_sketc_unseen_freq_estim_dp}. Finally, with regards to Equation \eqref{recov_sketc_unseen_estim_dp}, we write
\begin{align*}
\tilde{\mathfrak{k}}_{n}&=\sum_{r\geq1}\tilde{\mathfrak{m}}_{r,n}\\
&=\sum_{r\geq1}\frac{\theta+n}{r}\tilde{\mathfrak{p}}_{r,n}\\
    & = \sum_{r\geq1}\frac{\theta+n}{r} \frac{\frac{\theta}{J}r!}{(\theta+n)}\sum_{j=1}^{J}{c_{j}\choose r}\frac{\left(\frac{\theta}{J}\right)_{(c_{j}-r)}}{\left(\frac{\theta}{J}\right)_{(c_{j})}} \\
    & = \frac{\theta}{J} \sum_{j=1}^{J} \frac{1}{\left(\frac{\theta}{J}\right)_{(c_{j})}} \sum_{r=1}^{c_{j}} {c_{j}\choose r} (r-1)! \left(\frac{\theta}{J}\right)_{(c_{j}-r)} \\
    & = \frac{\theta}{J} \sum_{j=1}^{J} \frac{\Gamma(c_{j}+1)}{\Gamma(c_{j}+\theta/J)} \sum_{r=1}^{c_{j}} \frac{\Gamma(c_{j}-r+\theta/J)}{r \Gamma(c_{j}-r+1)}   \\
    & = \frac{\theta}{J} \sum_{j=1}^{J} \frac{\Gamma(c_{j}+1)}{\Gamma(c_{j}+\theta/J)} \frac{\Gamma(\theta/J+c_{j})}{\Gamma(1+c_{j})} \left[ \psi(1-\theta/J-c_{j}) - \psi(1-\theta/J)  \right]\\
    & =  - \theta \psi(1-\theta/J) + \frac{\theta}{J} \sum_{j=1}^{J}  \psi(1-\theta/J-c_{j}),
\end{align*}
where $\psi$ denotes the derivative of the log-Gamma function (digamma function), i.e., $\psi(x) = \frac{d}{dx} \log \Gamma(x)$. This completes the proof of Equation \eqref{recov_sketc_unseen_freq_estim_dp}, and the proof of the theorem

\subsection{Equation \eqref{dp_fdm} from Equation \eqref{pyp_fdm}}\label{dp_fdm_pyp_fdm}
The proof relies on the use Equation \eqref{asym_gfc_alpha}, which characterizes the behaviour of generalized factorial coefficients as $\alpha\rightarrow0$. In particular, by means of Equation \eqref{pyp_fdm} we write
\begin{align*}
&\lim_{\alpha\rightarrow0}\text{Pr}[\mathbf{C}_{n}=(c_{1},\ldots,c_{J})]\\
&\quad=\lim_{\alpha\rightarrow0}{n\choose c_{1},\ldots,c_{J}}\sum_{\boldsymbol{i}\in I_{(c_{1},\ldots,c_{J})}}\frac{\frac{\left(\frac{\theta}{\alpha}\right)_{(|\boldsymbol{i}|)}}{J^{|\boldsymbol{i}|}}}{(\theta)_{(n)}}\prod_{j=1}^{J}\mathscr{C}(c_{j},i_{j};\alpha)\\
&\quad=\lim_{\alpha\rightarrow0}{n\choose c_{1},\ldots,c_{J}}\frac{1}{(\theta)_{(n)}}\sum_{\boldsymbol{i}\in I_{(c_{1},\ldots,c_{J})}}J^{-|\boldsymbol{i}|}\left(\prod_{j=0}^{|\boldsymbol{i}|-1}(\theta+j\alpha)\right)\prod_{j=1}^{J}\frac{\mathscr{C}(c_{j},i_{j};\alpha)}{i_{j}^{\alpha}}\\
&\quad\text{[by Equation \eqref{asym_gfc_alpha}]}\\
&\quad={n\choose c_{1},\ldots,c_{J}}\frac{1}{(\theta)_{(n)}}\sum_{\boldsymbol{i}\in I_{(c_{1},\ldots,c_{J})}}\left(\frac{\theta}{J}\right)^{-|\boldsymbol{i}|}\prod_{j=1}^{J}|s(c_{j},i_{j})|\\
&\quad\text{[by Equation \eqref{defi_stir}]}\\
&\quad={n\choose c_{1},\ldots,c_{J}}\frac{1}{(\theta)_{(n)}}\prod_{j=1}^{J}\left(\frac{\theta}{J}\right)_{(c_{j})},
\end{align*}
i.e., a Dirichlet-Multinomial distribution with parameter $(n,\theta/J,\ldots,\theta/J)$. The proof is completed.

\subsection{Proof of Theorem \ref{teo_pyp}}\label{app_proof_teo_pyp}
The proof is along lines similar to the proof of Theorem \ref{teo_dp}. Differently from the proof of Theorem \ref{teo_dp}, which relies on the finite-dimensional projective property of the DP, this proof relies on marginal properties of the PYP prior that are available from \citet{San(06)}, in combination with the exchangeability of the  $X_{i}$'s \citep[Chapter 3]{Pit(06)} and some combinatorial arguments on the sketch $\mathbf{C}_{n}$. As in the proof of Theorem \ref{teo_dp}, the independence between $h$ and $\mathbf{X}_{n}$ allows us to treat $h$ as fixed, i.e., non-random. Then, from \eqref{estim_sketc},
\begin{align}\label{main_term_pyp}
\tilde{\mathfrak{p}}_{r,n}=\sum_{j=1}^{J}\text{Pr}[h(X_{n+1})=j\,|\, \mathbf{C}_{n}=\mathbf{c}]\text{Pr}[X_{n+1}\in \mathcal{S}_{r}\,|\, h(X_{n+1})=j,\mathbf{C}_{n}=\mathbf{c}],
\end{align}
where $\mathbf{c}=(c_{1},\ldots,c_{J})$. First, we consider the evaluation of $\text{Pr}[X_{n+1}\in \mathcal{S}_{r}\,|\, h(X_{n+1})=j,\mathbf{C}_{n}=\mathbf{c}]$, i.e.,
\begin{align}\label{second_term_pyp}
&\text{Pr}[X_{n+1}\in \mathcal{S}_{r}\,|\, h(X_{n+1})=j,\mathbf{C}_{n}=\mathbf{c}]=\frac{\text{Pr}[X_{n+1}\in \mathcal{S}_{r},h(X_{n+1})=j,\mathbf{C}_{n}=\mathbf{c}]}{\text{Pr}[h(X_{n+1})=j,\mathbf{C}_{n}=\mathbf{c}]}.
\end{align}
Consider the denominator of \eqref{second_term_pyp}. Uniformity of $h$ implies that $h$ induces a $J$-partition $\{B_{1},\ldots,B_{J}\}$ of $\mathbb{S}$ such that $B_{j}=\{v\in\mathbb{S}\text{ : }h(v)=j\}$ and $\nu(B_{j})=J^{-1}$ for any $j=1,\ldots,J$. Then, by a direct application of \citet[Equation 3.3]{San(06)}, we can write that
\begin{align}\label{denomin_pyp}
&\text{Pr}[h(X_{n+1})=j,\mathbf{C}_{n}=\mathbf{c}]\\
&\notag\quad={n\choose c_{1},\ldots,c_{J}}\E\left[(P(B_{j}))^{c_{j}+1}\prod_{1\leq s\neq j\leq J} (P(B_{s}))^{c_{s}}\right]\\
&\notag\quad={n\choose c_{1},\ldots,c_{J}}\sum_{\boldsymbol{i}\in I_{\mathbf{c},j,+1}}\frac{\frac{\left(\frac{\theta}{\alpha}\right)_{(|\boldsymbol{i}|)}}{J^{|\boldsymbol{i}|}}}{(\theta)_{(n+1)}}\mathscr{C}(c_{j}+1,i_{j};\alpha)\prod_{1\leq s\neq j\leq J}\mathscr{C}(c_{s};i_{s};\alpha)
\end{align}
Now, consider the numerator of \eqref{second_term_pyp}. To evaluate the numerator of \eqref{second_term_pyp}, we define the event $B(n,r)=\{X_{1}=\cdots=X_{r}=X_{n+1},\{X_{r+1},\ldots,X_{n}\}\cap\{X_{n+1}\}=\emptyset\}$, such that we write
\begin{align*}
&\text{Pr}[X_{n+1}\in \mathcal{S}_{r},h(X_{n+1})=j,\mathbf{C}_{n}=\mathbf{c}]\\
&\quad=\frac{1}{J}{n\choose r}\text{Pr}[B(n,r),h(X_{n+1})=j,\mathbf{C}_{n}=\mathbf{c}]\\
&\quad=\frac{1}{J}{n\choose r}\text{Pr} \Bigg[B(n,r),h(X_{n+1})=j,C_{1,n}=c_{1},\ldots \\
&\quad\quad\quad\quad\quad\quad\quad\quad\quad\quad\quad\quad\quad\quad\quad\ldots,\sum_{i=r+1}^{n}\mathbbm{1}_{\{h(X_{i})\}}(h(X_{n+1}))=c_{j}-r,\dots,C_{J,n}=c_{J}\Bigg].
\end{align*}
The distribution of the random variable $(X_{n+1},h(X_{n+1}),\mathbf{C}_{n})$ is determined by the distribution of the random variable $(X_{1},\ldots,X_{n},X_{n+1})$. In particular, let $\Pi(s,k)$ be the set of all partitions of ${1,\ldots,s}$ into $k$ disjoint subsets $\pi_{1},\ldots,\pi_{k}$ such that $n_{i}$ is the cardinality of $\pi_{i}$. From \citet[Equation 3.5]{San(06)}, for any measurable $A_{1},\ldots,A_{n},A_{n+1}$ we can write
\begin{align*}
&\text{Pr}[X_{1}\in A_{1},\ldots,X_{n}\in A_{n},X_{n+1}\in A_{n+1}]\\
&\quad=\sum_{k=1}^{n+1}\frac{\prod_{i=0}^{k-1}(\theta+i\alpha)}{(\theta)_{(n+1)}}\sum_{(\pi_{1},\ldots,\pi_{k})\in\Pi_{n+1,k}}\prod_{i=1}^{k}(1-\alpha)_{(n_{i}-1)}\nu(\cap_{n\in\pi_{i}}A_{n})
\end{align*}
for any $n\geq1$. Now, let $\mathscr{S}$ be the Borel $\sigma$-algebra of $\mathbb{S}$, and let $\nu_{\pi_{1},\ldots,\pi_{k}}$ be a probability measure on $(\mathbb{S}^{n+1},\mathscr{S}^{n+1})$ defined as $\nu_{\pi_{1},\ldots,\pi_{k}}=\prod_{1\leq i\leq k}\nu(\cap_{n\in\pi_{i}}A_{n})$ and attaching to $B(n,r)$ a value that is either $0$ or $1$. In particular, $\nu_{\pi_{1},\ldots,\pi_{k}}(B(n,r))=1$ if and only if one of the $\pi_{i}$'s is equal to the set $\{1,\ldots,r,n+1\}$. Then, based on the measure $\nu_{\pi_{1},\ldots,\pi_{k}}$ we write
\begin{align*}
&\text{Pr}\left[B(n,r),h(X_{n+1})=j,C_{1,n}=c_{1},\ldots,\sum_{i=r+1}^{m}\mathbbm{1}_{\{h(X_{i})\}}(h(X_{n+1}))=c_{j}-r,\dots,C_{J,n}=c_{J}\right]\\
&\quad=\sum_{k=2}^{n-r+1}\frac{\prod_{i=0}^{k-1}(\theta+i\alpha)}{(\theta)_{(n+1)}}\sum_{(\pi_{1},\ldots,\pi_{k-1})\in\Pi(n-r,k-1)}(1-\alpha)_{(r)}\prod_{i=1}^{k-1}(1-\alpha)_{(n_{i}-1)}\\
&\quad\quad\times\nu_{\pi_{1},\ldots,\pi_{k}}\left(C_{1,n}=c_{1},\ldots,\sum_{i=r+1}^{m}\mathbbm{1}_{\{h(X_{i})\}}(h(X_{n+1}))=c_{j}-l,\dots,C_{J,n}=c_{J}\right)\\
&\quad=\theta\frac{(\theta+\alpha)_{(n-r)}}{(\theta)_{(n+1)}}(1-\alpha)_{(l)}\sum_{t=1}^{n-r}\frac{\prod_{i=0}^{t-1}(\theta+\alpha+i\alpha)}{(\theta+\alpha)_{(n-r)}}\sum_{(\pi_{1},\ldots,\pi_{t})\in\Pi(n-r,t)}\prod_{i=1}^{t}(1-\alpha)_{(n_{i}-1)}\\
&\quad\quad\times\nu_{\pi_{1},\ldots,\pi_{k}}\left(C_{1,n}=c_{1},\ldots,\sum_{i=r+1}^{m}\mathbbm{1}_{\{h(X_{i})\}}(h(X_{n+1}))=c_{j}-l,\dots,C_{J,n}=c_{J}\right),
\end{align*}
where
\begin{align*}
&\sum_{t=1}^{n-r}\frac{\prod_{i=0}^{t-1}(\theta+\alpha+i\alpha)}{(\theta+\alpha)_{(n-r)}}\sum_{(\pi_{1},\ldots,\pi_{t})\in\Pi(n-r,t)}\prod_{i=1}^{t}(1-\alpha)_{(n_{i}-1)}\\
&\quad\quad\times\nu_{\pi_{1},\ldots,\pi_{k}}\left(C_{1,n}=c_{1},\ldots,\sum_{i=r+1}^{m}\mathbbm{1}_{\{h(X_{i})\}}(h(X_{n+1}))=c_{j}-l,\dots,C_{J,n}=c_{J}\right)
\end{align*}
is the distribution of a random sample $(X_{1},\ldots,X_{n-r})$ from a PYP with discount parameter $\alpha$ and scale parameter $\theta+\alpha$, which is provided by \citet[Equation 3.5]{San(06)}; note that the scale parameter is updated. Then, we evaluate the above probability as follow
\begin{align*}
&\theta\frac{(\theta+\alpha)_{(n-r)}}{(\theta)_{(n+1)}}(1-\alpha)_{(l)}\sum_{t=1}^{n-r}\frac{\prod_{i=0}^{t-1}(\theta+\alpha+i\alpha)}{(\theta+\alpha)_{(n-r)}}\sum_{(\pi_{1},\ldots,\pi_{t})\in\Pi(n-r,t)}\prod_{i=1}^{t}(1-\alpha)_{(n_{i}-1)}\\
&\quad\quad\times\nu_{\pi_{1},\ldots,\pi_{k}}\left(C_{1,n}=c_{1},\ldots,\sum_{i=r+1}^{m}\mathbbm{1}_{\{h(X_{i})\}}(h(X_{n+1}))=c_{j}-r,\dots,C_{J,n}=c_{J}\right)\\
&\quad=\theta\frac{(\theta+\alpha)_{(n-r)}}{(\theta)_{(n+1)}}(1-\alpha)_{(r)}{n-r\choose c_{1},\ldots,c_{j}-l,\ldots,c_{J}}\E\left[(P(B_{j}))^{c_{j}-r}\prod_{1\leq s\neq j\leq J}(P(B_{s}))^{c_{s}}\right]\\
&\quad=\theta\frac{(\theta+\alpha)_{(n-r)}}{(\theta)_{(n+1)}}(1-\alpha)_{(l)}{n-r\choose c_{1},\ldots,c_{j}-r,\ldots,c_{J}}\\
&\quad\quad\times\sum_{\boldsymbol{i}\in I_{\mathbf{c},j,-r}}\frac{\frac{\left(\frac{\theta+\alpha}{\alpha}\right)_{(|\boldsymbol{i}|)}}{J^{|\boldsymbol{i}|}}}{(\theta+\alpha)_{(n-r)}}\mathscr{C}(c_{j}-r,i_{j};\alpha)\prod_{1\leq s\neq j\leq J}\mathscr{C}(c_{s},i_{s};\alpha),
\end{align*}
where the first identity and the second identity follow from \citet[Proposition 3.1]{San(06)} and \citet[Equation 3.3]{San(06)}, respectively, under the PYP prior; see also formule displayed at page 469 of \citet{San(06)}. Hence, we can write the above probability as
\begin{align}\label{num_pyp}
&\text{Pr}[X_{n+1}\in \mathcal{S}_{r},h(X_{n+1})=j,\mathbf{C}_{n}=\mathbf{c}]\\
&\notag\quad=\frac{1}{J}{n\choose r}\theta\frac{(\theta+\alpha)_{(n-r)}}{(\theta)_{(n+1)}}(1-\alpha)_{(r)}{n-r\choose c_{1},\ldots,c_{j}-r,\ldots,c_{J}}\\
&\notag\quad\quad\times\sum_{\boldsymbol{i}\in I_{\mathbf{c},j,-r}}\frac{\frac{\left(\frac{\theta+\alpha}{\alpha}\right)_{(|\boldsymbol{i}|)}}{J^{|\boldsymbol{i}|}}}{(\theta+\alpha)_{(n-r)}}\mathscr{C}(c_{j}-r,i_{j};\alpha)\prod_{1\leq s\neq j\leq J}\mathscr{C}(c_{s},i_{s};\alpha).
\end{align}
Now, according to \eqref{second_term_pyp}, we combine \eqref{denomin_pyp} with \eqref{num_pyp} to obtain the conditional probability
\begin{align}\label{final_second_term_pyp}
&\text{Pr}[X_{n+1}\in \mathcal{S}_{r}\,|\, h(X_{n+1})=j,\mathbf{C}_{n}=\mathbf{c}]\\
&\notag\quad=\frac{\theta}{J}{c_{j}\choose r}\frac{(\theta+\alpha)_{(n-r)}}{(\theta)_{(n+1)}}(1-\alpha)_{(r)}\\
&\notag\quad\quad\times\frac{\sum_{\boldsymbol{i}\in I_{\mathbf{c},j,-r}}\frac{\frac{\left(\frac{\theta+\alpha}{\alpha}\right)_{(|\boldsymbol{i}|)}}{J^{|\boldsymbol{i}|}}}{(\theta+\alpha)_{(n-r)}}\mathscr{C}(c_{j}-r,i_{j};\alpha)\prod_{1\leq s\neq j\leq J}\mathscr{C}(c_{s},i_{s};\alpha)}{\sum_{\boldsymbol{i}\in I_{\mathbf{c},j,+1}}\frac{\frac{\left(\frac{\theta}{\alpha}\right)_{(|\boldsymbol{i}|)}}{J^{|\boldsymbol{i}|}}}{(\theta)_{(n+1)}}\mathscr{C}(c_{j}+1,i_{j};\alpha)\prod_{1\leq s\neq j\leq J}\mathscr{C}(c_{s},i_{s};\alpha)}\\
&\notag\quad=\frac{\theta}{J}{c_{j}\choose l}(1-\alpha)_{(r)}\\
&\notag\quad\quad\times\frac{\sum_{\boldsymbol{i}\in I_{\mathbf{c},j,-r}}\frac{\left(\frac{\theta+\alpha}{\alpha}\right)_{(|\boldsymbol{i}|)}}{J^{|\boldsymbol{i}|}}\mathscr{C}(c_{j}-r,i_{j};\alpha)\prod_{1\leq s\neq j\leq J}\mathscr{C}(c_{s},i_{s};\alpha)}{\sum_{\boldsymbol{i}\in I_{\mathbf{c},j,+1}}\frac{\left(\frac{\theta}{\alpha}\right)_{(|\boldsymbol{i}|)}}{J^{|\boldsymbol{i}|}}\mathscr{C}(c_{j}+1,i_{j};\alpha)\prod_{1\leq s\neq j\leq J}\mathscr{C}(c_{s},i_{s};\alpha)}.
\end{align}
This completes the evaluation of $\text{Pr}[X_{n+1}\in \mathcal{S}_{r}\,|\, h(X_{n+1})=j,\mathbf{C}_{n}=\mathbf{c}]$, and now we consider the evaluation of the conditional probability $\text{Pr}[h(X_{n+1})=j\,|\, \mathbf{C}_{n}=\mathbf{c}]$. That is, we compute
\begin{align}\label{first_term_pyp}
&\text{Pr}[h(X_{n+1})=j\,|\, \mathbf{C}_{n}=\mathbf{c}]=\frac{\text{Pr}[h(X_{n+1})=j,\mathbf{C}_{n}=\mathbf{c}]}{\text{Pr}[\mathbf{C}_{n}=\mathbf{c}]}.
\end{align}
The numerator of \eqref{first_term_pyp} is given by Equation \eqref{denomin_pyp}, whereas the denominator of \eqref{first_term_pyp} follows 
\begin{align}\label{denomin_pyp_extra}
\text{Pr}[\mathbf{C}_{n}=\mathbf{c}]&={n\choose c_{1},\ldots,c_{J}}\E\left[\prod_{s=1}^{J}(P(B_{s}))^{c_{s}}\right]\\
&\notag={n\choose c_{1},\ldots,c_{J}}\sum_{\boldsymbol{i}\in I_{\mathbf{c},j,0}}\frac{\frac{\left(\frac{\theta}{\alpha}\right)_{(|\boldsymbol{i}|)}}{J^{|\boldsymbol{i}|}}}{(\theta)_{(n)}}\prod_{s=1}^{J}\mathscr{C}(c_{s},i_{s};\alpha).
\end{align}
Now, according to \eqref{first_term_pyp}, we combine \eqref{denomin_pyp} with \eqref{denomin_pyp_extra} to obtain the conditional probability
\begin{align}\label{final_primo_term_pyp}
&\text{Pr}[h(X_{n+1})=j\,|\, \mathbf{C}_{n}=\mathbf{c}]\\
&\notag\quad=\frac{\sum_{\boldsymbol{i}\in I_{\mathbf{c},j,+1}}\frac{\frac{\left(\frac{\theta}{\alpha}\right)_{(|\boldsymbol{i}|)}}{J^{|\boldsymbol{i}|}}}{(\theta)_{(n+1)}}\mathscr{C}(c_{j}+1,i_{j};\alpha)\prod_{1\leq s\neq j\leq J}\mathscr{C}(c_{s},i_{s};\alpha)}{\sum_{\boldsymbol{i}\in I_{\mathbf{c},j,0}}\frac{\frac{\left(\frac{\theta}{\alpha}\right)_{(|\boldsymbol{i}|)}}{J^{|\boldsymbol{i}|}}}{(\theta)_{(n)}}\prod_{s=1}^{J}\mathscr{C}(c_{s},i_{s};\alpha)}\\
&\notag\quad=\frac{1}{\theta+n}\frac{\sum_{\boldsymbol{i}\in I_{\mathbf{c},j,+1}}\frac{\left(\frac{\theta}{\alpha}\right)_{(|\boldsymbol{i}|)}}{J^{|\boldsymbol{i}|}}\mathscr{C}(c_{j}+1,i_{j};\alpha)\prod_{1\leq s\neq j\leq J}J^{-i_{s}}\mathscr{C}(c_{s},i_{s};\alpha)}{\sum_{\boldsymbol{i}\in I_{\mathbf{c},j,0}}\frac{\left(\frac{\theta}{\alpha}\right)_{(|\boldsymbol{i}|)}}{J^{|\boldsymbol{i}|}}\prod_{s=1}^{J}\mathscr{C}(c_{s},i_{s};\alpha)}.
\end{align}
According to \eqref{main_term_pyp}, the proof of Equation \eqref{post_pyp} is completed by combining \eqref{final_primo_term_pyp} with \eqref{final_second_term_pyp}. With regards to the proof of Equation \eqref{recov_sketc_unseen_freq_estim_pyp}, we define the partition set $\mathcal{M}_{n,k}=\{(m_{1},\ldots,m_{n})\text{ : }m_{i}\geq0,\,\sum_{1\leq i\leq n}m_{i}=k\text{ and }\sum_{1\leq i\leq n}im_{i}=n\}$. Then, we can write that
\begin{align*}
\tilde{\mathfrak{p}}_{r,n}&=\sum_{k=1}^{n}\sum_{(m_{1},\ldots,m_{n})\in\mathcal{M}_{n,k}}\text{Pr}[X_{n+1}\in\mathcal{S}_{r}\,|\,\mathbf{C}_{n}=\mathbf{c},\mathbf{M}_{n}=(m_{1},\ldots,m_{n})]\\
&\quad\times\text{Pr}[\mathbf{M}_{n}=(m_{1},\ldots,m_{n})\,|\,\mathbf{C}_{n}=\mathbf{c}]\\
&=\sum_{k=1}^{n}\sum_{(m_{1},\ldots,m_{n})\in\mathcal{M}_{n,k}}\hat{\mathfrak{p}}_{r,n}\text{Pr}[\mathbf{M}_{n}=(m_{1},\ldots,m_{n})\,|\,\mathbf{C}_{n}=\mathbf{c}]\\
&\quad\text{[by Equation \eqref{bnp_est}]}\\
&=\sum_{k=1}^{n}\sum_{(m_{1},\ldots,m_{n})\in\mathcal{M}_{n,k}}\frac{(r-\alpha)m_{r}}{\theta+n}\text{Pr}[\mathbf{M}_{n}=(m_{1},\ldots,m_{n})\,|\,\mathbf{C}_{n}=\mathbf{c}]\\
&=\frac{r-\alpha}{\theta+n}\E[M_{r,n}\,|\,\mathbf{C}_{n}=\mathbf{c}],
\end{align*}
i.e., 
\begin{displaymath}
\tilde{\mathfrak{p}}_{r,n}=\frac{r-\alpha}{\theta+n}\tilde{\mathfrak{m}}_{r,n}
\end{displaymath}
and
\begin{displaymath}
\tilde{\mathfrak{m}}_{r,n}=\frac{\theta+n}{r-\alpha}\tilde{\mathfrak{p}}_{r,n},
\end{displaymath}
which completes the proof of Equation \eqref{recov_sketc_unseen_freq_estim_pyp}. Finally, with regards to Equation \eqref{recov_sketc_unseen_estim_pyp}, we write
\begin{align*}
\tilde{\mathfrak{p}}_{0,n}&=\sum_{k=1}^{n}\sum_{(m_{1},\ldots,m_{n})\in\mathcal{M}_{n,k}}\text{Pr}[X_{n+1}\in\mathcal{S}_{0}\,|\,\mathbf{C}_{n}=\mathbf{c},\mathbf{M}_{n}=(m_{1},\ldots,m_{n})]\\
&\quad\times\text{Pr}[\mathbf{M}_{n}=(m_{1},\ldots,m_{n})\,|\,\mathbf{C}_{n}=\mathbf{c}]\\
&=\sum_{k=1}^{n}\sum_{(m_{1},\ldots,m_{n})\in\mathcal{M}_{n,k}}\hat{\mathfrak{p}}_{0,n}\text{Pr}[\mathbf{M}_{n}=(m_{1},\ldots,m_{n})\,|\,\mathbf{C}_{n}=\mathbf{c}]\\
&\quad\text{[by Equation \eqref{bnp_est}]}\\
&=\sum_{k=1}^{n}\sum_{(m_{1},\ldots,m_{n})\in\mathcal{M}_{n,k}}\frac{\theta+k\alpha}{\theta+n}\text{Pr}[\mathbf{M}_{n}=(m_{1},\ldots,m_{n})\,|\,\mathbf{C}_{n}=\mathbf{c}]\\
&=\frac{\theta}{\theta+n}+\frac{\alpha}{\theta+n}\E[K_{n}\,|\,\mathbf{C}_{n}=\mathbf{c}],
\end{align*}
i.e., 
\begin{displaymath}
\tilde{\mathfrak{p}}_{0,n}=\frac{\theta}{\theta+n}+\frac{\alpha}{\theta+n}\tilde{\mathfrak{k}}_{n}
\end{displaymath}
and 
\begin{displaymath}
\tilde{\mathfrak{k}}_{n}=\frac{\theta+n}{\alpha}\tilde{p}_{0,n}-\frac{\theta}{\alpha}.
\end{displaymath}
This completes the proof of Equation \eqref{recov_sketc_unseen_freq_estim_pyp}, and the proof of the theorem.

\subsection{Equation \eqref{post_dp} from Equation \eqref{post_pyp}}\label{app_dp_pyp}
The proof relies on the use Equation \eqref{asym_gfc_alpha}, which characterizes the behaviour of generalized factorial coefficients as $\alpha\rightarrow0$. In particular, by means of Theorem \ref{teo_pyp} we can write that
\begin{align*}
&\lim_{\alpha\rightarrow0}\tilde{\mathfrak{p}}_{r,n}\\
&\quad=\lim_{\alpha\rightarrow0}\frac{(\theta/J)(1-\alpha)_{(r)}}{(\theta+n)}\sum_{j=1}^{J}{c_{j}\choose r}\frac{\sum_{\boldsymbol{i}\in I_{(c_{1},\ldots,c_{J}),j,-r}}\frac{\left(\frac{\theta+\alpha}{\alpha}\right)_{|\boldsymbol{i}|}}{J^{|\boldsymbol{i}|}}\prod_{s=1}^{J}\mathscr{C}(c_{s}-r\delta_{s,j},i_{s};\alpha)}{\sum_{\boldsymbol{i}\in I_{(c_{1},\ldots,c_{J})}}\frac{\left(\frac{\theta}{\alpha}\right)_{|\boldsymbol{i}|}}{J^{|\boldsymbol{i}|}}\prod_{s=1}^{J}\mathscr{C}(c_{s},i_{s};\alpha)}\\
&\quad=\lim_{\alpha\rightarrow0}\frac{(\theta/J)(1-\alpha)_{(r)}}{\theta+n}\\
&\quad\quad\times\sum_{j=1}^{J}{c_{j}\choose r}\frac{\sum_{\boldsymbol{i}\in I_{(c_{1},\ldots,c_{J}),j,-r}}J^{-|\boldsymbol{i}|}\left(\prod_{j=0}^{|\boldsymbol{i}|-1}(\theta+\alpha+j\alpha)\right)\prod_{s=1}^{J}\frac{\mathscr{C}(c_{s}-r\delta_{s,j},i_{s};\alpha)}{i_{s}^{\alpha}}}{\sum_{\boldsymbol{i}\in I_{(c_{1},\ldots,c_{J})}}J^{-|\boldsymbol{i}|}\left(\prod_{j=0}^{|\boldsymbol{i}|-1}(\theta+j\alpha)\right)\prod_{s=1}^{J}\frac{\mathscr{C}(c_{s},i_{s};\alpha)}{i_{s}^{\alpha}}}\\
&\quad\text{[by Equation \eqref{asym_gfc_alpha}]}\\
&\quad=\frac{(\theta/J)r!}{\theta+n}\sum_{j=1}^{J}{c_{j}\choose r}\frac{\sum_{\boldsymbol{i}\in I_{(c_{1},\ldots,c_{J}),j,-r}}\left(\frac{\theta}{J}\right)^{|\boldsymbol{i}|}\prod_{s=1}^{J}|s(c_{s}-r\delta_{s,j},i_{s})|}{\sum_{\boldsymbol{i}\in I_{\mathbf{c}_{m},j,0}}\left(\frac{\theta}{J}\right)^{|\boldsymbol{i}|}\prod_{s=1}^{J}|s(c_{s},i_{s})|}\\
&\quad\text{[by Equation \eqref{defi_stir}]}\\
&\quad=\frac{(\theta/J)r!}{\theta+n}\sum_{j=1}^{J}{c_{j}\choose r}\frac{\prod_{s=1}^{J}\left(\frac{\theta}{J}\right)_{(c_{s}-r\delta_{s,j})}}{\prod_{s=1}^{J}\left(\frac{\theta}{J}\right)_{(c_{s})}}\\
&\quad=\frac{(\theta/J)r!}{\theta+n}\sum_{j=1}^{J}{c_{j}\choose r}\frac{\Gamma(\theta/J+c_{j}-r)}{\Gamma(\theta/J+c_{j})}.
\end{align*}
Equation \eqref{post_dp} follows from the definition of rising factorial numbers in terms of the ratio of Gamma functions, i.e., $(a)_{(n)}=\Gamma(a+n)/\Gamma(a)$ for $a>0$ and $n\in\mathbb{N}_{0}$. The proof is completed.

\subsection{Proof of Equation \eqref{asymp_pyp}}\label{app_proofs_3}
The proof relies on the use of a Poisson asymptotic property for (normalized) generalized factorial coefficients. In particular, according to \citet[Lemma 2]{DF(20)}, for $z>0$
\begin{equation}\label{asym_gfc_n}
\lim_{n\rightarrow+\infty}\frac{z^{k}\mathscr{C}(n,k;\alpha)}{\sum_{k=0}^{n}z^{k}\mathscr{C}(n,k;\alpha)}=\text{e}^{-z}\frac{z^{k-1}}{(k-1)!}.
\end{equation}
First, we rewrite the numerator and the denominator of \eqref{post_pyp_missing}. We write the numerator of \eqref{post_pyp_missing} as
\begin{align}\label{num_posterior}
&n^{1-\alpha}\frac{\theta}{\theta+n}\sum_{\boldsymbol{i}\in I_{(c_{1},\ldots,c_{J})}}\frac{\left(\frac{\theta+\alpha}{\alpha}\right)_{(|\boldsymbol{i}|)}}{J^{|\boldsymbol{i}|}}\prod_{s=1}^{J}\mathscr{C}(c_{s},i_{s};\alpha)\\
&\notag\quad=n^{1-\alpha}\frac{\theta}{\theta+n}\left(\prod_{s=2}^{J}\sum_{i_{s}=0}^{c_{s}}\frac{1}{J^{i_{s}}}\mathscr{C}(c_{s},i_{s};\alpha)\right)\\
&\notag\quad\quad\times\sum_{i_{1}=0}^{c_{1}}\left(\frac{\theta+\alpha}{\alpha}\right)_{(i_{1})}\left(\frac{1}{J}\right)^{i_{1}}\mathscr{C}(c_{1},i_{1};\alpha)\cdots\\
&\notag\quad\quad\quad\cdots\times\sum_{i_{j}=0}^{c_{j}}\left(\frac{\theta+\alpha}{\alpha}+i_{1}+\cdots+i_{j-1}\right)_{(i_{j})}\left(\frac{1}{J}\right)^{i_{j}}\mathscr{C}(c_{j},i_{j};\alpha)\cdots\\
&\notag\quad\quad\quad\quad\cdots\times\sum_{i_{J}=0}^{c_{J}}\left(\frac{\theta+\alpha}{\alpha}+i_{1}+\cdots+i_{J-1}\right)_{(i_{J})}\frac{\left(\frac{1}{J}\right)^{i_{J}}\mathscr{C}(c_{J},i_{J};\alpha)}{\sum_{i_{J}=0}^{c_{J}}\left(\frac{1}{J}\right)^{i_{J}}\mathscr{C}(c_{J},i_{J};\alpha)}.
\end{align}
Then, we apply the same arguments to rewrite the denominator of \eqref{post_pyp}. In particular, we have 
\begin{align}\label{den_posterior}
&\sum_{\boldsymbol{i}\in I_{(c_{1},\ldots,c_{J})}}\frac{\left(\frac{\theta}{\alpha}\right)_{(\boldsymbol{i})}}{J^{|\boldsymbol{i}|}}\prod_{s=1}^{J}\mathscr{C}(c_{s},i_{s};\alpha)\\
&\notag\quad=\left(\prod_{s=2}^{J}\sum_{i_{s}=0}^{c_{s}}\frac{1}{J^{i_{s}}}\mathscr{C}(c_{s},i_{s};\alpha)\right)\\
&\notag\quad\quad\times\sum_{i_{1}=0}^{c_{1}}\left(\frac{\theta}{\alpha}\right)_{(i_{1})}\left(\frac{1}{J}\right)^{i_{1}}\mathscr{C}(c_{1},i_{1};\alpha)\cdots\\
&\notag\quad\quad\quad\cdots\times\sum_{i_{j}=0}^{c_{j}}\left(\frac{\theta}{\alpha}+i_{1}+\cdots+i_{j-1}\right)_{(i_{j})}\left(\frac{1}{J}\right)^{i_{j}}\mathscr{C}(c_{j},i_{j};\alpha)\cdots\\
&\notag\quad\quad\quad\quad\cdots\times\sum_{i_{J}=0}^{c_{J}}\left(\frac{\theta}{\alpha}+i_{1}+\cdots+i_{J-1}\right)_{(i_{J})}\frac{\left(\frac{1}{J}\right)^{i_{J}}\mathscr{C}(c_{J},i_{J};\alpha)}{\sum_{i_{J}=0}^{c_{J}}\left(\frac{1}{J}\right)^{i_{J}}\mathscr{C}(c_{J},i_{J};\alpha)}.
\end{align}
Now, under the assumption that $c_{j}=nJ^{-1}$ for any $j\in[J]$, we consider \eqref{num_posterior} and apply \eqref{asym_gfc_n} to obtain an approximation of it. For any $h\in\{2,\ldots,J\}$, by a direct application of \eqref{asym_gfc_n} we write
\begin{align}\label{asym_num}
&\lim_{n\rightarrow+\infty}\sum_{i_{h}=0}^{nJ^{-1}}\left(\frac{\theta+\alpha}{\alpha}+i_{1}+\cdots+i_{h-1}\right)_{(i_{h})}\frac{\left(\frac{1}{J}\right)^{i_{h}}\mathscr{C}(nJ^{-1},i_{h};\alpha)}{\sum_{i_{h}=0}^{nJ^{-1}}\left(\frac{1}{J}\right)^{i_{h}}\mathscr{C}(nJ^{-1},i_{h};\alpha)}\\
&\notag\quad=\sum_{i_{h}\geq1}\left(\frac{\theta+\alpha}{\alpha}+i_{1}+\cdots+i_{h-1}\right)_{(i_{h})}\text{e}^{-\frac{1}{J}}\frac{\left(\frac{1}{J}\right)^{i_{h}-1}}{(i_{h}-1)!}.
\end{align}
Similarly, under the assumption that $c_{j}=nJ^{-1}$ for any $j\in[J]$, we consider \eqref{den_posterior} and apply \eqref{asym_gfc_n} to obtain an approximation of it. For any $h\in\{2,\ldots,J\}$, by an application of \eqref{asym_gfc_n} we write
\begin{align}\label{asym_den}
&\lim_{n\rightarrow+\infty}\sum_{i_{h}=0}^{nJ^{-1}}\left(\frac{\theta}{\alpha}+i_{1}+\cdots+i_{h-1}\right)_{(i_{h})}\frac{\left(\frac{1}{J}\right)^{i_{h}}\mathscr{C}(nJ^{-1},i_{h};\alpha)}{\sum_{i_{h}=0}^{nJ^{-1}}\left(\frac{1}{J}\right)^{i_{h}}\mathscr{C}(nJ^{-1},i_{h};\alpha)}\\
&\notag\quad=\sum_{i_{h}\geq1}\left(\frac{\theta}{\alpha}+i_{1}+\cdots+i_{h-1}\right)_{(i_{h})}\text{e}^{-\frac{1}{J}}\frac{\left(\frac{1}{J}\right)^{i_{h}-1}}{(i_{h}-1)!}.
\end{align}
Now, for $h\in\{2,\ldots,J\}$, we make use of  \eqref{asym_num} and \eqref{asym_den} in \eqref{num_posterior} and \eqref{den_posterior}, respectively, to find
\begin{align}\label{limit_j_final_num}
&\lim_{n\rightarrow+\infty}\sum_{\boldsymbol{i}\in I_{(nJ^{-1},\ldots,nJ^{-1})}}\frac{\left(\frac{\theta+\alpha}{\alpha}\right)_{(|\boldsymbol{i}|)}}{J^{|\boldsymbol{i}|}}\prod_{s=1}^{J}\mathscr{C}(nJ^{-1},i_{s};\alpha)
\end{align}
and
\begin{align}\label{limit_j_final_den}
&\lim_{n\rightarrow+\infty}\sum_{\boldsymbol{i}\in I_{(nJ^{-1},\ldots,nJ^{-1})}}\left(\frac{\theta}{\alpha}\right)_{(|\boldsymbol{i}|)}\prod_{s=1}^{J}\frac{\mathscr{C}(nJ^{-1},i_{s};\alpha)}{J^{i_{s}}}.
\end{align}
In order to apply \eqref{asym_num} and \eqref{asym_den} to get \eqref{limit_j_final_num} and \eqref{limit_j_final_den}, respectively, we proceed iteratively from $h=J$ to $h=2$ on the numerator \eqref{num_posterior} and on the denominator \eqref{den_posterior}, i.e.
\begin{itemize}
\item[i)] for $h=J$, 
\begin{align*}
&\lim_{n\rightarrow+\infty}\sum_{i_{J}=0}^{nJ^{-1}}\left(\frac{\theta+\alpha}{\alpha}+i_{1}+\cdots+i_{J-1}\right)_{(i_{J})}\frac{\left(\frac{1}{J}\right)^{i_{J}}\mathscr{C}(nJ^{-1},i_{J};\alpha)}{\sum_{i_{J}=0}^{nJ^{-1}}\left(\frac{1}{J}\right)^{i_{J}}\mathscr{C}(nJ^{-1},i_{J};\alpha)}\\
&\quad=\sum_{i_{J}\geq1}\left(\frac{\theta+\alpha}{\alpha}+i_{1}+\cdots+i_{J-1}\right)_{(i_{J})}\text{e}^{-\frac{1}{J}}\frac{\left(\frac{1}{J}\right)^{i_{J}-1}}{(i_{J}-1)!}=\text{e}^{-\frac{1}{J}}\frac{\frac{\theta+\alpha}{\alpha}+i_{1}+\cdots+i_{J-1}}{\left(1-\frac{1}{J}\right)^{\frac{\theta+\alpha}{\alpha}+i_{1}+\cdots+i_{J-1}+1}}
\end{align*}
and
\begin{align*}
&\lim_{n\rightarrow+\infty}\sum_{i_{J}=0}^{nJ^{-1}}\left(\frac{\theta}{\alpha}+i_{1}+\cdots+i_{J-1}\right)_{(i_{J})}\frac{\left(\frac{1}{J}\right)^{i_{J}}\mathscr{C}(nJ^{-1},i_{J};\alpha)}{\sum_{i_{J}=0}^{nJ^{-1}}\left(\frac{1}{J}\right)^{i_{J}}\mathscr{C}(nJ^{-1},i_{J};\alpha)}\\
&\quad=\sum_{i_{J}\geq1}\left(\frac{\theta}{\alpha}+i_{1}+\cdots+i_{J-1}\right)_{(i_{J})}\text{e}^{-\frac{1}{J}}\frac{\left(\frac{1}{J}\right)^{i_{J}-1}}{(i_{J}-1)!}=\text{e}^{-\frac{1}{J}}\frac{\frac{\theta}{\alpha}+i_{1}+\cdots+i_{J-1}}{\left(1-\frac{1}{J}\right)^{\frac{\theta}{\alpha}+i_{1}+\cdots+i_{J-1}+1}};
\end{align*}
\item[ii)] for $h=J-1$, 
\begin{align*}
&\lim_{n\rightarrow+\infty}\sum_{i_{J-1}=0}^{nJ^{-1}}\frac{\left(\frac{\theta+\alpha}{\alpha}+i_{1}+\cdots+i_{J-2}\right)_{(i_{J-1})}}{\text{e}^{\frac{1}{J}}\frac{\left(1-\frac{1}{J}\right)^{\frac{\theta+\alpha}{\alpha}+i_{1}+\cdots+i_{J-1}+1}}{\frac{\theta+\alpha}{\alpha}+i_{1}+\cdots+i_{J-1}}}\frac{\left(\frac{1}{J}\right)^{i_{J-1}}\mathscr{C}(nJ^{-1},i_{J-1};\alpha)}{\sum_{i_{J-1}=0}^{nJ^{-1}}\left(\frac{1}{J}\right)^{i_{J-1}}\mathscr{C}(nJ^{-1},i_{J-1};\alpha)}\\
&\quad=\sum_{i_{J-1}\geq1}\left(\frac{\theta+\alpha}{\alpha}+i_{1}+\cdots+i_{J-2}\right)_{(i_{J-1})}\text{e}^{-\frac{1}{J}}\frac{\left(\frac{1}{J}\right)^{i_{J-1}-1}}{(i_{J-1}-1)!}\text{e}^{-\frac{1}{J}}\frac{\frac{\theta+\alpha}{\alpha}+i_{1}+\cdots+i_{J-1}}{\left(1-\frac{1}{J}\right)^{\frac{\theta+\alpha}{\alpha}+i_{1}+\cdots+i_{J-1}+1}}\\
&\quad=\text{e}^{-\frac{2}{J}}\frac{1}{\left(1-\frac{1}{J}\right)^{\frac{\theta+\alpha}{\alpha}+i_{1}+\cdots+i_{J-2}+2}}\sum_{i_{J-1}\geq1}\left(\frac{\theta+\alpha}{\alpha}+i_{1}+\cdots+i_{J-2}\right)_{(i_{J-1}+1)}\frac{\left(\frac{1}{J-1}\right)^{i_{J-1}-1}}{(i_{J-1}-1)!}\\
&\quad=\text{e}^{-\frac{2}{J}}\frac{1}{\left(1-\frac{1}{J}\right)^{\frac{\theta+\alpha}{\alpha}+i_{1}+\cdots+i_{J-2}+2}}\frac{\left(\frac{\theta+\alpha}{\alpha}+i_{1}+\cdots+i_{J-2}\right)_{(2)}}{\left(1-\frac{1}{J-1}\right)^{\frac{\theta+\alpha}{\alpha}+i_{1}+\cdots+i_{J-2}+2}}\\
&\quad=\text{e}^{-\frac{2}{J}}\frac{\left(\frac{\theta+\alpha}{\alpha}+i_{1}+\cdots+i_{J-2}\right)_{(2)}}{\left(1-\frac{2}{J}\right)^{\frac{\theta+\alpha}{\alpha}+i_{1}+\cdots+i_{J-2}+2}}
\end{align*}
and
\begin{align*}
&\lim_{n\rightarrow+\infty}\sum_{i_{J-1}=0}^{nJ^{-1}}\frac{\left(\frac{\theta}{\alpha}+i_{1}+\cdots+i_{J-2}\right)_{(i_{J-1})}}{\text{e}^{\frac{1}{J}}\frac{\left(1-\frac{1}{J}\right)^{\frac{\theta}{\alpha}+i_{1}+\cdots+i_{J-1}+1}}{\frac{\theta}{\alpha}+i_{1}+\cdots+i_{J-1}}}\frac{\left(\frac{1}{J}\right)^{i_{J-1}}\mathscr{C}(nJ^{-1},i_{J-1};\alpha)}{\sum_{i_{J-1}=0}^{nJ^{-1}}\left(\frac{1}{J}\right)^{i_{J-1}}\mathscr{C}(nJ^{-1},i_{J-1};\alpha)}\\
&\quad=\sum_{i_{J-1}\geq1}\left(\frac{\theta}{\alpha}+i_{1}+\cdots+i_{J-2}\right)_{(i_{J-1})}\text{e}^{-\frac{1}{J}}\frac{\left(\frac{1}{J}\right)^{i_{J-1}-1}}{(i_{J-1}-1)!}\text{e}^{-\frac{1}{J}}\frac{\frac{\theta}{\alpha}+i_{1}+\cdots+i_{J-1}}{\left(1-\frac{1}{J}\right)^{\frac{\theta}{\alpha}+i_{1}+\cdots+i_{J-1}+1}}\\
&\quad=\text{e}^{-\frac{2}{J}}\frac{1}{\left(1-\frac{1}{J}\right)^{\frac{\theta}{\alpha}+i_{1}+\cdots+i_{J-2}+2}}\sum_{i_{J-1}\geq1}\left(\frac{\theta}{\alpha}+i_{1}+\cdots+i_{J-2}\right)_{(i_{J-1}+1)}\frac{\left(\frac{1}{J-1}\right)^{i_{J-1}-1}}{(i_{J-1}-1)!}\\
&\quad=\text{e}^{-\frac{2}{J}}\frac{1}{\left(1-\frac{1}{J}\right)^{\frac{\theta}{\alpha}+i_{1}+\cdots+i_{J-2}+2}}\frac{\left(\frac{\theta}{\alpha}+i_{1}+\cdots+i_{J-2}\right)_{(2)}}{\left(1-\frac{1}{J-1}\right)^{\frac{\theta}{\alpha}+i_{1}+\cdots+i_{J-2}+2}}\\
&\quad=\text{e}^{-\frac{2}{J}}\frac{\left(\frac{\theta}{\alpha}+i_{1}+\cdots+i_{J-2}\right)_{(2)}}{\left(1-\frac{2}{J}\right)^{\frac{\theta}{\alpha}+i_{1}+\cdots+i_{J-2}+2}};
\end{align*}
\item[iii)] for $h=j$,
\begin{align*}
&\lim_{n\rightarrow+\infty}\sum_{i_{j}=0}^{nJ^{-1}}\frac{\left(\frac{\theta+\alpha}{\alpha}+i_{1}+\cdots+i_{j-1}\right)_{(i_{j})}}{\text{e}^{\frac{J-j}{J}}\frac{\left(1-\frac{J-j}{J}\right)^{\frac{\theta+\alpha}{\alpha}+i_{1}+\cdots+i_{j}+J-j}}{\left(\frac{\theta+\alpha}{\alpha}+i_{1}+\cdots+i_{j}\right)_{(J-j)}}}\frac{\left(\frac{1}{J}\right)^{i_{j}}\mathscr{C}(nJ^{-1},i_{j};\alpha)}{\sum_{i_{j}=0}^{nJ^{-1}}\left(\frac{1}{J}\right)^{i_{j}}\mathscr{C}(nJ^{-1},i_{j};\alpha)}\\
&\quad=\sum_{i_{j}\geq1}\left(\frac{\theta+\alpha}{\alpha}+i_{1}+\cdots+i_{j-1}\right)_{(i_{j})}\text{e}^{-\frac{1}{J}}\frac{\left(\frac{1}{J}\right)^{i_{j}-1}}{(i_{j}-1)!}\text{e}^{-\frac{J-j}{J}}\frac{\left(\frac{\theta+\alpha}{\alpha}+i_{1}+\cdots+i_{j}\right)_{(J-j)}}{\left(1-\frac{J-j}{J}\right)^{\frac{\theta+\alpha}{\alpha}+i_{1}+\cdots+i_{j}+J-j}}\\
&\quad=\text{e}^{-\frac{J-j+1}{J}}\frac{1}{\left(1-\frac{J-j}{J}\right)^{\frac{\theta+\alpha}{\alpha}+i_{1}+\cdots+i_{j-1}+J-j+1}}\sum_{i_{j}\geq1}\left(\frac{\theta+\alpha}{\alpha}+i_{1}+\cdots+i_{j-1}\right)_{(i_{j}+J-j)} \cdot \\
  & \qquad \qquad \cdot \frac{\left(\frac{1}{j}\right)^{i_{j}-1}}{(i_{j}-1)!}\\
&\quad=\text{e}^{-\frac{J-j+1}{J}}\frac{1}{\left(1-\frac{J-j}{J}\right)^{\frac{\theta+\alpha}{\alpha}+i_{1}+\cdots+i_{j-1}+J-j+1}}\frac{\left(\frac{\theta+\alpha}{\alpha}+i_{1}+\cdots+i_{j-1}\right)_{(J-j+1)}}{\left(1-\frac{1}{j}\right)^{\frac{\theta+\alpha}{\alpha}+i_{1}+\cdots+i_{j-1}+J-j+1}}\\
&\quad=\text{e}^{-\frac{J-j+1}{J}}\frac{\left(\frac{\theta+\alpha}{\alpha}+i_{1}+\cdots+i_{j-1}\right)_{(J-j+1)}}{\left(1-\frac{J-j+1}{J}\right)^{\frac{\theta+\alpha}{\alpha}+i_{1}+\cdots+i_{j-1}+J-j+1}}
\end{align*}
and
\begin{align*}
&\lim_{n\rightarrow+\infty}\sum_{i_{j}=0}^{nJ^{-1}}\frac{\left(\frac{\theta}{\alpha}+i_{1}+\cdots+i_{j-1}\right)_{(i_{j})}}{\text{e}^{\frac{J-j}{J}}\frac{\left(1-\frac{J-j}{J}\right)^{\frac{\theta}{\alpha}+i_{1}+\cdots+i_{j}+J-j}}{\left(\frac{\theta}{\alpha}+i_{1}+\cdots+i_{j}\right)_{(J-j)}}}\frac{\left(\frac{1}{J}\right)^{i_{j}}\mathscr{C}(nJ^{-1},i_{j};\alpha)}{\sum_{i_{j}=0}^{nJ^{-1}}\left(\frac{1}{J}\right)^{i_{j}}\mathscr{C}(nJ^{-1},i_{j};\alpha)}\\
&\quad=\sum_{i_{j}\geq1}\left(\frac{\theta}{\alpha}+i_{1}+\cdots+i_{j-1}\right)_{(i_{j})}\text{e}^{-\frac{1}{J}}\frac{\left(\frac{1}{J}\right)^{i_{j}-1}}{(i_{j}-1)!}\text{e}^{-\frac{J-j}{J}}\frac{\left(\frac{\theta}{\alpha}+i_{1}+\cdots+i_{j}\right)_{(J-j)}}{\left(1-\frac{J-j}{J}\right)^{\frac{\theta}{\alpha}+i_{1}+\cdots+i_{j}+J-j}}\\
&\quad=\text{e}^{-\frac{J-j+1}{J}}\frac{1}{\left(1-\frac{J-j}{J}\right)^{\frac{\theta}{\alpha}+i_{1}+\cdots+i_{j-1}+J-j+1}}\sum_{i_{j}\geq1}\left(\frac{\theta}{\alpha}+i_{1}+\cdots+i_{j-1}\right)_{(i_{j}+J-j)}\frac{\left(\frac{1}{j}\right)^{i_{j}-1}}{(i_{j}-1)!}\\
&\quad=\text{e}^{-\frac{J-j+1}{J}}\frac{1}{\left(1-\frac{J-j}{J}\right)^{\frac{\theta}{\alpha}+i_{1}+\cdots+i_{j-1}+J-j+1}}\frac{\left(\frac{\theta}{\alpha}+i_{1}+\cdots+i_{j-1}\right)_{(J-j+1)}}{\left(1-\frac{1}{j}\right)^{\frac{\theta}{\alpha}+i_{1}+\cdots+i_{j-1}+J-j+1}}\\
&\quad=\text{e}^{-\frac{J-j+1}{J}}\frac{\left(\frac{\theta}{\alpha}+i_{1}+\cdots+i_{j-1}\right)_{(J-j+1)}}{\left(1-\frac{J-j+1}{J}\right)^{\frac{\theta}{\alpha}+i_{1}+\cdots+i_{j-1}+J-j+1}};
\end{align*}
\item[iv)] for $h=2$,
\begin{align*}
&\lim_{n\rightarrow+\infty}\sum_{i_{2}=0}^{nJ^{-1}}\frac{\left(\frac{\theta+\alpha}{\alpha}+i_{1}\right)_{(i_{2})}}{\text{e}^{\frac{J-2}{J}}\frac{\left(1-\frac{J-2}{J}\right)^{\frac{\theta+\alpha}{\alpha}+i_{1}+i_{2}+J-2}}{\left(\frac{\theta+\alpha}{\alpha}+i_{1}+i_{2}\right)_{(J-2)}}}\frac{\left(\frac{1}{J}\right)^{i_{2}}\mathscr{C}(nJ^{-1},i_{2};\alpha)}{\sum_{i_{2}=0}^{nJ^{-1}}\left(\frac{1}{J}\right)^{i_{2}}\mathscr{C}(nJ^{-1},i_{2};\alpha)}\\
&\quad=\sum_{i_{2}\geq1}\left(\frac{\theta+\alpha}{\alpha}+i_{1}\right)_{(i_{2})}\text{e}^{-\frac{1}{J}}\frac{\left(\frac{1}{J}\right)^{i_{2}-1}}{(i_{2}-1)!}\text{e}^{-\frac{J-2}{J}}\frac{\left(\frac{\theta+\alpha}{\alpha}+i_{1}+i_{2}\right)_{(J-2)}}{\left(1-\frac{J-2}{J}\right)^{\frac{\theta+\alpha}{\alpha}+i_{1}+i_{2}+J-2}}\\
&\quad=\text{e}^{-\frac{J-1}{J}}\frac{1}{\left(1-\frac{J-2}{J}\right)^{\frac{\theta+\alpha}{\alpha}+i_{1}+J-1}}\sum_{i_{2}\geq1}\left(\frac{\theta+\alpha}{\alpha}+i_{1}\right)_{(i_{2}+J-2)}\frac{\left(\frac{1}{2}\right)^{i_{2}-1}}{(i_{2}-1)!}\\
&\quad=\text{e}^{-\frac{J-1}{J}}\frac{\left(\frac{\theta+\alpha}{\alpha}+i_{1}\right)_{(J-1)}}{\left(1-\frac{J-1}{J}\right)^{\frac{\theta+\alpha}{\alpha}+i_{1}+J-1}}
\end{align*}
and
\begin{align*}
&\lim_{n\rightarrow+\infty}\sum_{i_{2}=0}^{nJ^{-1}}\frac{\left(\frac{\theta}{\alpha}+i_{1}\right)_{(i_{2})}}{\text{e}^{\frac{J-2}{J}}\frac{\left(1-\frac{J-2}{J}\right)^{\frac{\theta}{\alpha}+i_{1}+i_{2}+J-2}}{\left(\frac{\theta}{\alpha}+i_{1}+i_{2}\right)_{(J-2)}}}\frac{\left(\frac{1}{J}\right)^{i_{2}}\mathscr{C}(nJ^{-1},i_{2};\alpha)}{\sum_{i_{2}=0}^{nJ^{-1}}\left(\frac{1}{J}\right)^{i_{2}}\mathscr{C}(nJ^{-1},i_{2};\alpha)}\\
&\quad=\sum_{i_{2}\geq1}\left(\frac{\theta}{\alpha}+i_{1}\right)_{(i_{2})}\text{e}^{-\frac{1}{J}}\frac{\left(\frac{1}{J}\right)^{i_{2}-1}}{(i_{2}-1)!}\text{e}^{-\frac{J-2}{J}}\frac{\left(\frac{\theta}{\alpha}+i_{1}+i_{2}\right)_{(J-2)}}{\left(1-\frac{J-2}{J}\right)^{\frac{\theta}{\alpha}+i_{1}+i_{2}+J-2}}\\
&\quad=\text{e}^{-\frac{J-1}{J}}\frac{1}{\left(1-\frac{J-2}{J}\right)^{\frac{\theta}{\alpha}+i_{1}+J-1}}\sum_{i_{2}\geq1}\left(\frac{\theta}{\alpha}+i_{1}\right)_{(i_{2}+J-2)}\frac{\left(\frac{1}{2}\right)^{i_{2}-1}}{(i_{2}-1)!}\\
&\quad=\text{e}^{-\frac{J-1}{J}}\frac{\left(\frac{\theta}{\alpha}+i_{1}\right)_{(J-1)}}{\left(1-\frac{J-1}{J}\right)^{\frac{\theta}{\alpha}+i_{1}+J-1}}.
\end{align*}
\end{itemize}
Then, we complete by considering the last term involving $nJ^{-1}$. In particular, we can write
\begin{align*}
&\lim_{n\rightarrow+\infty}n^{1-\alpha}\frac{\theta}{\theta+n}\frac{\sum_{i_{1}=0}^{nJ^{-1}}\frac{\left(\frac{\theta+\alpha}{\alpha}\right)_{(i_{1})}}{\text{e}^{\frac{J-1}{J}}\frac{\left(1-\frac{J-1}{J}\right)^{\frac{\theta+\alpha}{\alpha}+i_{1}+J-1}}{\left(\frac{\theta+\alpha}{\alpha}+i_{1}\right)_{(J-1)}}}\left(\frac{1}{J}\right)^{i_{1}}\mathscr{C}(nJ^{-1},i_{1};\alpha)}{\sum_{i_{1}=0}^{nJ^{-1}}\frac{\left(\frac{\theta}{\alpha}\right)_{(i_{1})}}{\text{e}^{\frac{J-1}{J}}\frac{\left(1-\frac{J-1}{J}\right)^{\frac{\theta}{\alpha}+i_{1}+J-1}}{\left(\frac{\theta}{\alpha}+i_{1}\right)_{(J-1)}}}\left(\frac{1}{J}\right)^{i_{1}}\mathscr{C}(nJ^{-1},i_{1};\alpha)}\\
&\quad=\lim_{n\rightarrow+\infty}n^{1-\alpha}\frac{\theta}{\theta+n}\frac{\quad\frac{\text{e}^{-\frac{J-1}{J}}}{\left(\frac{1}{J}\right)^{\frac{\theta+\alpha}{\alpha}+J}}\left(\frac{\theta+\alpha}{\alpha}\right)_{(J-1)}\sum_{i_{1}=1}^{nJ^{-1}}\left(\frac{\theta+\alpha}{\alpha}+J-1\right)_{(i_{i})}\mathscr{C}(nJ^{-1},i_{1};\alpha)}{\frac{\text{e}^{-\frac{J-1}{J}}}{\left(\frac{1}{J}\right)^{\frac{\theta}{\alpha}+J}}\left(\frac{\theta}{\alpha}\right)_{(J-1)}\sum_{i_{1}=1}^{nJ^{-1}}\left(\frac{\theta}{\alpha}+J-1\right)_{(i_{i})}\mathscr{C}(nJ^{-1},i_{1};\alpha)}\\
&\quad=\lim_{n\rightarrow+\infty}n^{1-\alpha}\frac{\theta}{\theta+n}\frac{\frac{\text{e}^{-\frac{J-1}{J}}}{\left(\frac{1}{J}\right)^{\frac{\theta+\alpha}{\alpha}+J}}\left(\frac{\theta+\alpha}{\alpha}\right)_{(J-1)}\left(\theta+J\alpha\right)_{(nJ^{-1})}}{\frac{\text{e}^{-\frac{J-1}{J}}}{\left(\frac{1}{J}\right)^{\frac{\theta}{\alpha}+J}}\left(\frac{\theta}{\alpha}\right)_{(J-1)}\left(\theta+J\alpha-\alpha\right)_{(nJ^{-1})}}\\
&\quad=\lim_{n\rightarrow+\infty}n^{1-\alpha}\frac{\theta}{\theta+n}J\frac{(\theta+J\alpha-\alpha)\left(\theta+J\alpha\right)_{(nJ^{-1})}}{\theta\left(\theta+J\alpha -\alpha\right)_{(nJ^{-1})}}\\
&\quad=J^{1-\alpha}\frac{\Gamma(\theta+J\alpha-\alpha+1)}{\Gamma(\theta+J\alpha)},
\end{align*}
where the last equality follows by means of Stirling approximation for the ration of Gamma functions, i.e., $\Gamma(a+n)/\Gamma(n)\approx a^{n}$ as $n\rightarrow+\infty$. This complete the proof of Equation \eqref{asymp_pyp}.

\subsection{Proof of Proposition \ref{main_post_mc} }\label{app_main_post_mc}

The proof relies on the use of the distribution in \eqref{eq_dist_py} of the number of distinct blocks in the random partition induced by a random sample from $P\sim\text{PYP}(\alpha,\theta)$. From Theorem \ref{teo_pyp}, we write
\begin{align*}
  \tilde{\mathfrak{p}}_{r,n}&=\frac{(\theta/J)(1-\alpha)_{(r)}}{(\theta+n)}\sum_{j=1}^{J}{c_{j}\choose r}\frac{\sum_{\boldsymbol{i}\in I_{(c_{1},\ldots,c_{J}),j,-r}}\frac{\left(\frac{\theta+\alpha}{\alpha}\right)_{|\boldsymbol{i}|}}{J^{|\boldsymbol{i}|}}\prod_{s=1}^{J}\mathscr{C}(c_{s}-r\delta_{s,j},i_{s};\alpha)}{\sum_{\boldsymbol{i}\in I_{(c_{1},\ldots,c_{J})}}\frac{\left(\frac{\theta}{\alpha}\right)_{|\boldsymbol{i}|}}{J^{|\boldsymbol{i}|}}\prod_{s=1}^{J}\mathscr{C}(c_{s},i_{s};\alpha)}\\
&=\frac{(\theta/J)\theta(1-\alpha)_{(r)}}{(\theta+n)}\\
&\quad\times\sum_{j=1}^{J}{c_{j}\choose r}\frac{(\theta)_{(c_{j}-r)}}{(\theta)_{(c_{j})}}\frac{\sum_{\boldsymbol{i}\in I_{(c_{1},\ldots,c_{J}),j,-r}}\frac{\left(\frac{\theta+\alpha}{\alpha}\right)_{|\boldsymbol{i}|}}{J^{|\boldsymbol{i}|}\prod_{s=1}^{J}\left(\frac{\theta}{\alpha}\right)_{(i_{s})}}\prod_{s=1}^{J}\frac{\left(\frac{\theta}{\alpha}\right)_{(i_{s})}}{(\theta)_{(c_{s}-r\delta_{s,j})}}\mathscr{C}(c_{s}-r\delta_{s,j},i_{s};\alpha)}{\sum_{\boldsymbol{i}\in I_{(c_{1},\ldots,c_{J})}}\frac{\left(\frac{\theta}{\alpha}\right)_{|\boldsymbol{i}|}}{J^{|\boldsymbol{i}|}\prod_{s=1}^{J}\left(\frac{\theta}{\alpha}\right)_{(i_{s})}}\prod_{s=1}^{J}\frac{\left(\frac{\theta}{\alpha}\right)_{(i_{s})}}{(\theta)_{(c_{s})}}\mathscr{C}(c_{s},i_{s};\alpha)}\\
&\text{[by Equation \eqref{eq_dist_py}]}\\
&=\frac{(\theta/J)(1-\alpha)_{(r)}}{(\theta+n)}\sum_{j=1}^{J}{c_{j}\choose r}\frac{(\theta)_{(c_{j}-r)}}{(\theta)_{(c_{j})}}\frac{\sum_{\boldsymbol{i}\in I_{(c_{1},\ldots,c_{J}),j,-r}}\frac{\left(\frac{\theta+\alpha}{\alpha}\right)_{|\boldsymbol{i}|}}{J^{|\boldsymbol{i}|}\prod_{s=1}^{J}\left(\frac{\theta}{\alpha}\right)_{(i_{s})}}\prod_{s=1}^{J}\text{Pr}[K_{c_{s}-r\delta_{s,j}}=i_{s}]}{\sum_{\boldsymbol{i}\in I_{(c_{1},\ldots,c_{J})}}\frac{\left(\frac{\theta}{\alpha}\right)_{|\boldsymbol{i}|}}{J^{|\boldsymbol{i}|}\prod_{s=1}^{J}\left(\frac{\theta}{\alpha}\right)_{(i_{s})}}\prod_{s=1}^{J}\text{Pr}[K_{c_{s}}=i_{s}]}\\
&=\frac{(\theta/J)(1-\alpha)_{(r)}}{(\theta+n)}\sum_{j=1}^{J}{c_{j}\choose l}\frac{(\theta)_{(c_{j}-r)}\E\left[\frac{\frac{\left(\frac{\theta+\alpha}{\alpha}\right)_{(\sum_{s=1}^{J}K_{c_{s}-r\delta_{s,j}})}}{J^{\sum_{s=1}^{J}K_{c_{s}-r\delta_{s,j}}}}}{\prod_{s=1}^{J}\left(\frac{\theta}{\alpha}\right)_{(K_{c_{s}}-r\delta_{s,j})}}\right]}{(\theta)_{(c_{j})}\E\left[\frac{\frac{\left(\frac{\theta}{\alpha}\right)_{(\sum_{s=1}^{J}K_{c_{s}})}}{J^{\sum_{s=1}^{J}K_{c_{s}}}}}{\prod_{s=1}^{J}\left(\frac{\theta}{\alpha}\right)_{(K_{c_{s}})}}\right]},
\end{align*}
where $K_{c_{s}}$ is the number of distinct blocks in the random partition induced by a random sample $\mathbf{X}_{c_{s}}$ from $P\sim\text{PYP}(\alpha,\theta)$ for any $s=1,\ldots,J$, with the random variable $K_{c_{r}}$ independent of the random variable $K_{c_{s}}$ for any $s\neq r$ with $r,s=1,\ldots,J$. The proof is completed.

\section{Additional results from numerical experiments} \label{app_experiments}



\begin{figure}[H]
\centering
\includegraphics[width=\linewidth]{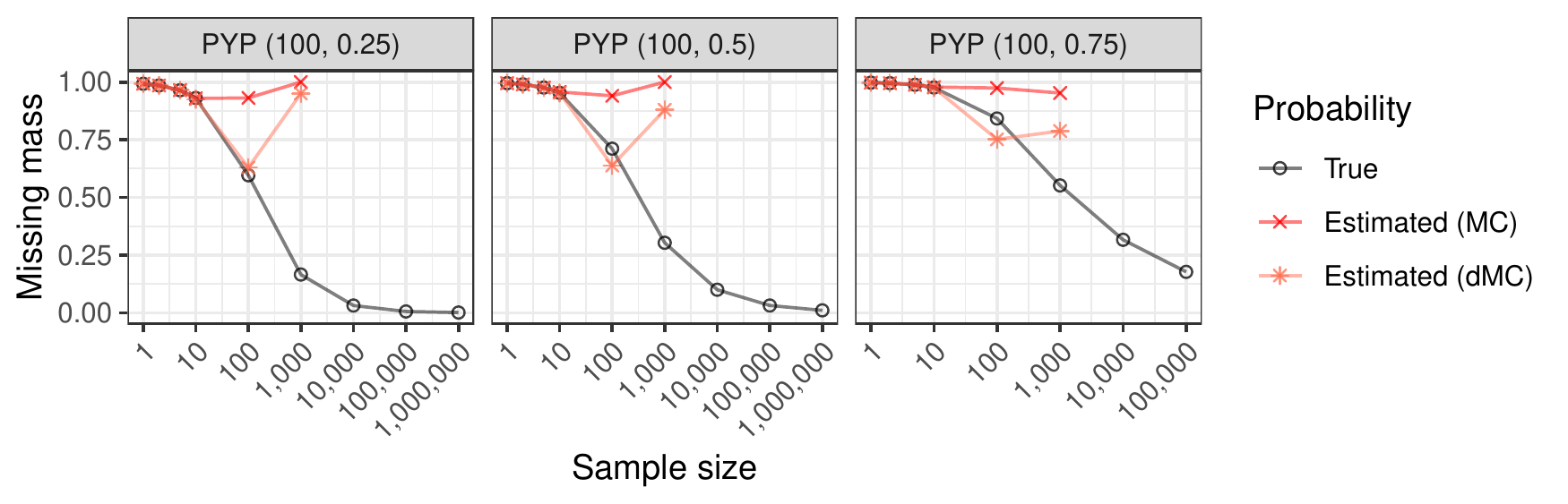}
\caption{True and estimated coverage probabilities for data sampled from a PYP prior model with different parameters $(\alpha,\theta)$.  Other details are as in Figure~\ref{fig:exp-dp}~(b).}
\label{fig:exp-pyp-alpha}
\end{figure}

\begin{figure}[H]
\centering
\includegraphics[width=0.9\linewidth]{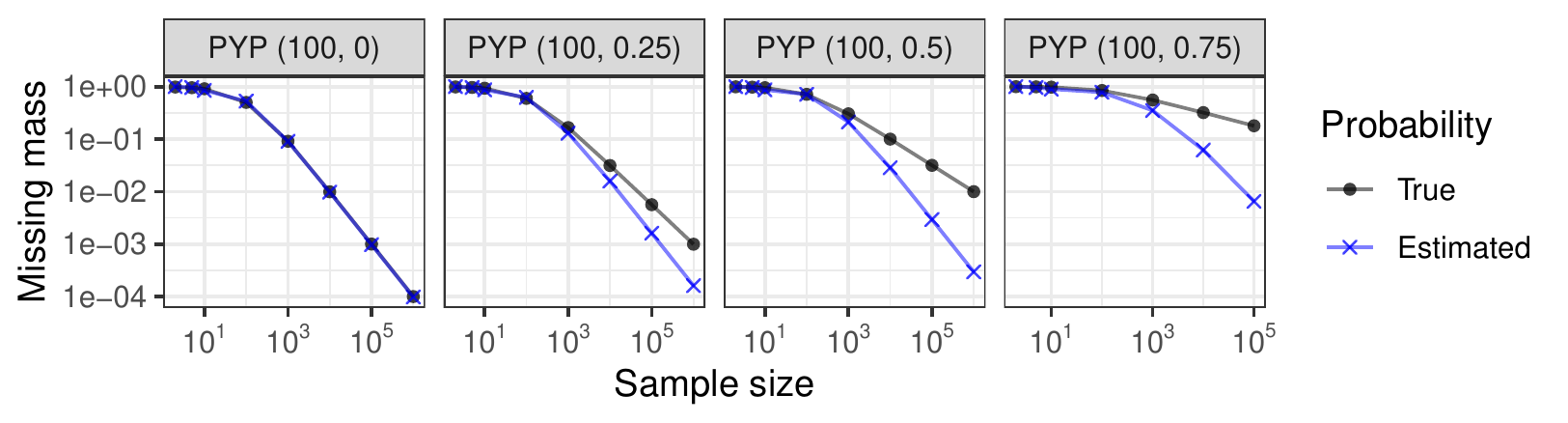}
\caption{True and estimated missing mass for sketched data simulated from a PYP prior model with different parameters $(\alpha,\theta)$. The estimated is estimated assuming a mis-specified Dirichlet process prior with parameter estimated via maximum marginal likelihood. Other details are as in Figure~\ref{fig:exp-dp}.}
\label{fig:exp-dp-pyp}
\end{figure}

\begin{figure}[H]
\centering
\includegraphics[width=0.7\linewidth]{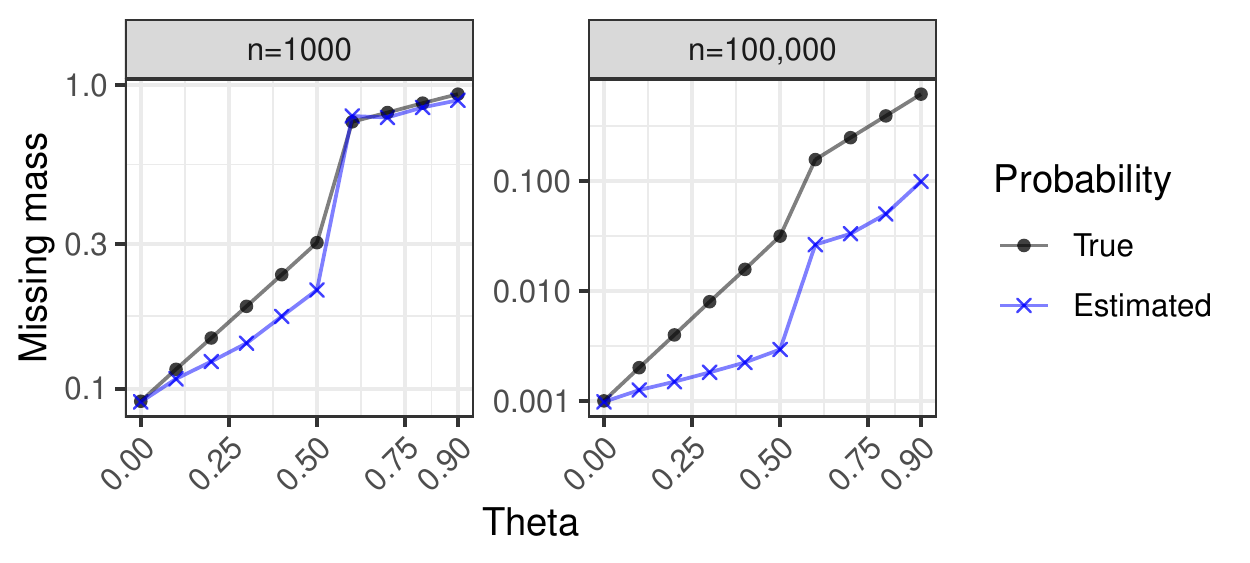}
\caption{True and estimated missing mass for sketched data simulated from a PYP prior model with parameter $\alpha=0.5$ and different $\theta$. Other details are as in Figure~\ref{fig:exp-dp-pyp}.}
\label{fig:exp-dp-pyp-theta}
\end{figure}

\begin{figure}[H]
\centering
\includegraphics[width=0.9\linewidth]{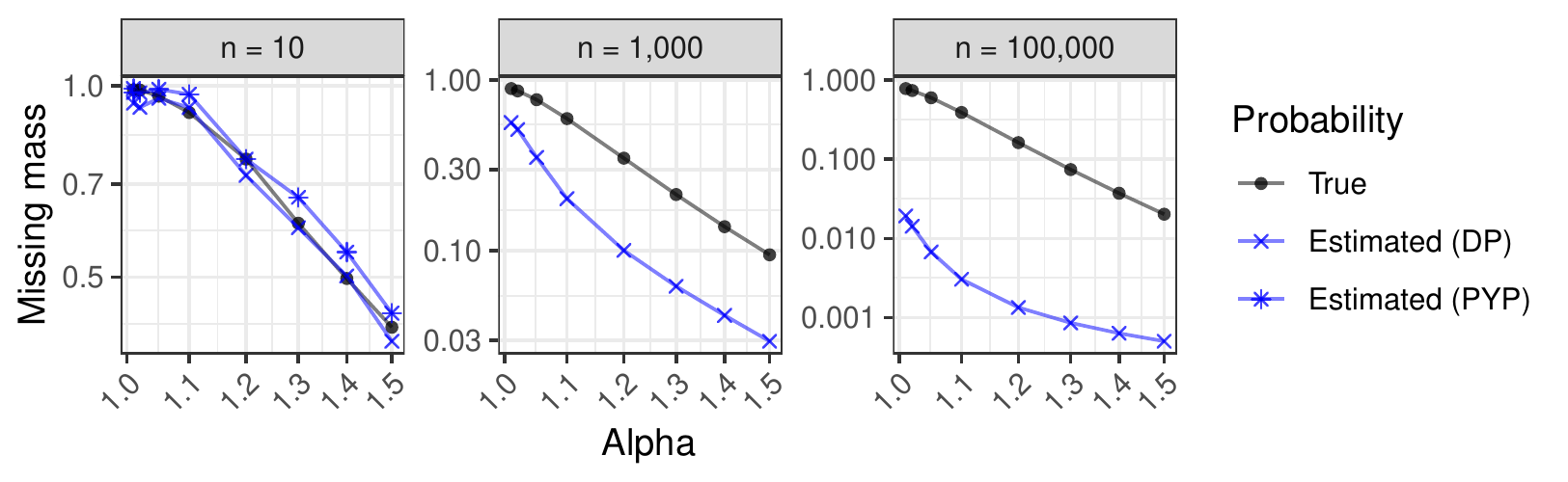}
\caption{True and estimated missing mass for data sampled from a Zipf distribution, as a function of the tail parameter $\alpha$. The estimated missing mass is calculated either assuming a mis-specified Dirichlet process prior with parameter estimated via maximum marginal likelihood, or assuming a PYP prior model with $\alpha=0.5$ and $\theta$ estimated empirically via Monte Carlo. The latter approach is only applied when $n=10$. Other details are as in Figure~\ref{fig:exp-dp-pyp}.}
\label{fig:exp-dp-zipf}
\end{figure}

\begin{figure}[H]
\centering
\includegraphics[width=0.9\linewidth]{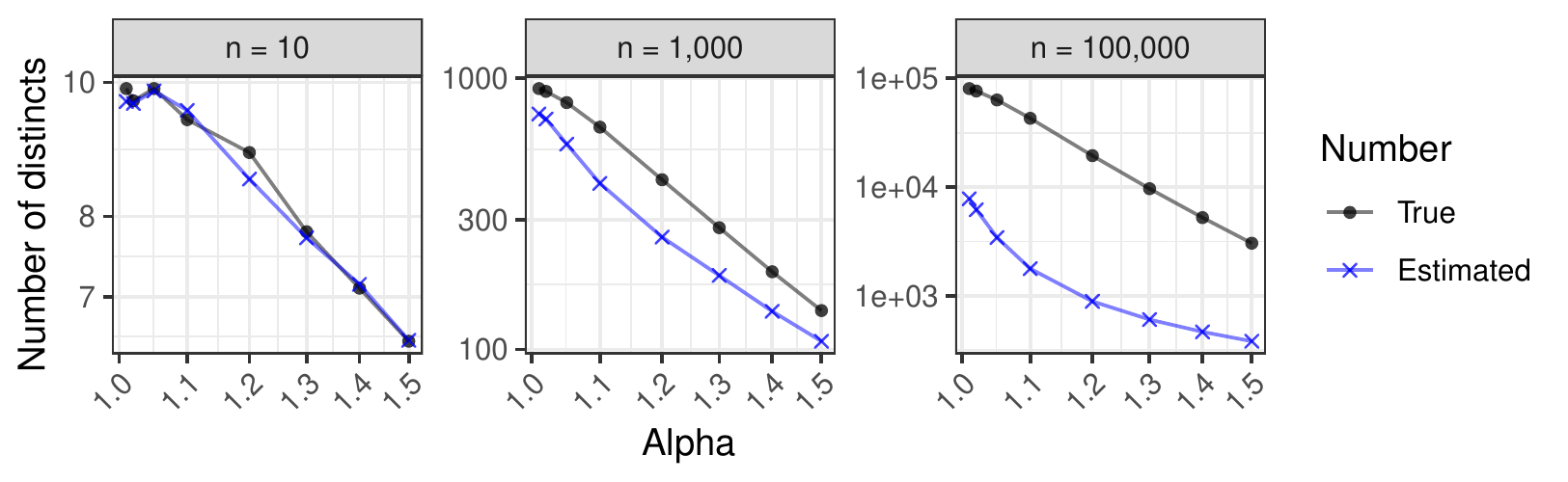}
\caption{True and estimated number of distinct species for data sampled from a Zipf distribution, as a function of the tail parameter $\alpha$. The estimated missing mass is calculated assuming a mis-specified Dirichlet process prior with parameter estimated via maximum marginal likelihood. Other details are as in Figure~\ref{fig:exp-dp-zipf}.}
\label{fig:exp-dp-zipf-distinct}
\end{figure}

\end{appendices}


\section*{Acknowledgement}

S.~F.~is also affiliated to IMATI-CNR ``Enrico  Magenes" (Milan, Italy), and he received funding from the European Research Council (ERC) under the European Union's Horizon 2020 research and innovation programme under grant agreement No 817257. S.~F.~also gratefully acknowledges the financial support from the Italian Ministry of Education, University and Research (MIUR), ``Dipartimenti di Eccellenza" grant 2018-2022. 



\begin{thebibliography}{9}


\bibitem[Anevski et al.(2017)]{Ane(17)}
\textsc{Anevski, D., Gill, R.D., and Zohren, S.} (2017). Estimating a probability mass function with unknown labels. \textit{Annals of Statistics} \textbf{45}, 2708--2735.

\bibitem[Arbel et al.(2017)]{Arb(17)}
\textsc{Arbel, J., Favaro, S., Nipoti, B., and Teh, Y.W.} (2017). Bayesian nonparametric inference for discovery probabilities: credible intervals and large sample asymptotics. \textit{Statistica Sinica} \textbf{27}, 839--858.

\bibitem[Ayed et al.(2018)]{Fad(18)}
\textsc{Ayed, F., Battiston, M., Camerlenghi, F., and Favaro, S.} (2018). On consistent and rate optimal estimation of the missing mass. \textit{Annales de l'Institut Henri Poincar\'e - Probabilit\'es et Statistques} \textbf{57}, 1476--1494.  

\bibitem[Bacallado et al.(2017)]{Bac(17)} 
\textsc{Bacallado, S., Battiston, M., Favaro, S., and Trippa, L.} (2015). Sufficientness postulates for Gibbs-type priors and hierarchial generalizations. \textit{Statistical Science} \textbf{32}, 487--500.

\bibitem[Baldi(2017)]{Bal(17)}
\textsc{Baldi, P.} (2017) \textit{Stochastic calculus.} Springer.

\bibitem[Balocchi et al.(2022)]{Bal(22)} 
\textsc{Balocchi, C., Favaro, S., and Naulet, Z.} (2022). Bayesian nonparametric inference for ``species-sampling" problems. \textit{Preprint arXiv:2203.06076}.

\bibitem[Bassily et al.(2017)]{Bas(17)} 
\textsc{Bassily, R., Nissim, K., Stemmer, U., and Guha Thakurta, A.} (2017). Practical locally private heavy hitters. \textit{Advances in Neural Information Processing Systems} \textbf{30}.

\bibitem[Ben-Hamou et al.(2017)]{Ben(17)} 
\textsc{Ben-Hamou, A., Boucheron, S., and Ohannessian, M.I.} (2017). Concentration inequalities in the infinite urn scheme for occupancy counts and the missing mass, with applications. \textit{Bernoulli} \textbf{23}, 249--287.

\bibitem[Ben-Hamou et al.(2018)]{Ben(18)}
\textsc{Ben-Hamou, A., Boucheron, S., and Gassiat, E.} (2018). Pattern coding meets censoring: (almost) adaptive coding on countable alphabets. \textit{Preprint arXiv:1608.08367}.

\bibitem[Berger et al.(2018)]{Ber(18)}
\textsc{Berger, B., Daniels, N.M., and Yu, Y.W.} (2016). Computational biology in the 21st century: scaling with compressive algorithms. \textit{Communication of the ACM} \textbf{59}, 72.

\bibitem[Bernton et al.(2019)]{Ber(19)}
\textsc{Bernton, E., Jacob, P.E., Gerber, M., and Robert, C.P.} (2019). On parameter estimation with the Wasserstein distance. \textit{Information and Inference} \textbf{8}, 657--676.

\bibitem[Bird et al.(2009)]{bird2009natural}
\textsc{Bird, S., Klein, E., and Loper, E.} (2009). \textit{Natural language processing with Python: analyzing text with the natural language toolkit}. O'Reilly Media, Inc.

\bibitem[Bradley et al.(2013)]{Bra(19)}
\textsc{Bradley, P., Den Bakker, H. C., Rocha, E. P., McVean, G., and Iqbal, Z.} (2019). Ultrafast search of all deposited bacterial and viral genomic data. \textit{Nature Biotechnology} \textbf{37}(2), 152--159.

\bibitem[Bubeck et al.(2013)]{Bub(13)}
\textsc{Bubeck, S., Ernst, D., and Garivier, A.} (2013). Optimal discovery with probabilistic  expert  advice:  finite  time  analysis  and  macroscopic optimality. \textit{Journal of Machine Learning Research} \textbf{14}, 601--623.

\bibitem[Bunge and Fitzpatrick(1993)]{Bun(93)}
\textsc{Bunge, J. and Fitzpatrick, M.} (1993) Estimating the number of species: a review. \textit{Journal of the American Statistical Association} \textbf{88}, 364-373.

\bibitem[Cai et al.(2018)]{Cai(18)}
\textsc{Cai, D., Mitzenmacher, M., and Adams, R. P.} (2018) A Bayesian nonparametric view on count-min sketch. \textit{Advances in neural information processing systems} \textbf{31}.

\bibitem[Cereda(2017)]{Cer(17)}
\textsc{Cereda, G.} (2017) Impact of model choice on LR assessment in case of rare haplotype match (frequentist approach). \textit{Scandinavian Journal of Statistics} \textbf{44}, 230--248.

\bibitem[Charalambides(2005)]{Cha(05)}
\textsc{Charalambides, C.} (2005) \textit{Combinatorial methods in discrete distributions}. Wiley.

\bibitem[Chung et al.(2013)]{Chu(13)}
\textsc{Chung, K., Mitzenmacher, M., and Vadhan, S.P.} (2013). Why simple hash functions work: exploiting the entropy in a data stream. \textit{Theory of Computing} \textbf{9}, 897--945.

\bibitem[Cormode(2017)]{Cor(17)}
\textsc{Cormode, G.} (2017). Data sketching. \textit{Communications of the ACM} \textbf{60}, 48--55.

\bibitem[Cormode et al.(2012)]{Cor(12)}
\textsc{Cormode, G., Garofalakis, M., and Haas, P.J.} (2012). \textit{Synopses for massive data: samples, histograms, wavelets, sketches}. Foundations and Trends in Databases.

\bibitem[Cormode et al.(2018)]{Cor(18)}
\textsc{Cormode, G., Jha, S., Kulkarni, T., Li, N., Srivastava, D., and Wang, T.} (2018). Privacy at scale: Local differential privacy in practice. \textit{Proceedings of the International Conference on Management of Data}, 1655--1658.

\bibitem[Cormode and Muthukrishnan(2005)]{Cor(05)}
\textsc{Cormode, G. and Muthukrishnan, S.} (2005). An improved data stream summary: the count-min sketch and its applications. \textit{Journal of Algorithms} \textbf{55}, 58--75.

\bibitem[Cormode and Yi(2020)]{Cor(20)}
\textsc{Cormode, G. and Yi, K.} (2020). \textit{Small summaries for big data}. Cambridge University Press.

\bibitem[Daley and Smith(2013)]{Dal(13)}
\textsc{Daley, T. and Smith, A.D.} (2013). Predicting the molecular complexity of sequencing libraries. \textit{Nature Methods} \textbf{10}, 325--327.

\bibitem[Deng et al.(2019)]{Den(19)}
\textsc{Deng, C. Daley, T., De Sena Brandine, G., and Smith, A.D.} (2019). Molecular heterogeneity in large-scale biological data: techniques and applications. \textit{Annual Review of Biomedical Data Science} \textbf{2}, 39--67.

\bibitem[Devroye(2009)]{Dev(09)}
\textsc{Devroye, L.} (2009). Random variate generation for exponentially and polynomially tilted stable distributions. \textit{ACM Transactions on Modeling and Computer Simulation} \textbf{19}, 4.

\bibitem[Ding et al.(2017)]{Din(17)}
\textsc{Ding, B., Kulkarni, J., and Yekhanin, S.} (2017). Collecting telemetry data privately. \textit{Advances in Neural Information Processing Systems} \textbf{30}.

\bibitem[Dolera and Favaro(2020)]{DF(20)}
\textsc{Dolera, E. and Favaro, S.} (2020). A Berry--Esseen theorem for Pitman's $\alpha$--diversity. \textit{Annals of Applied Probability} \textbf{30}, 847--869.

\bibitem[Dolera et al.(2022)]{Dol(22)}
\textsc{Dolera, E., Favaro, S., and Peluchetti, S.} (2022). Learning-augmented count-min sketches via Bayesian nonparametrics. \textit{Preprint arXiv:2102.04462}.

\bibitem[Efron and Thisted(1976)]{Efr(76)}  
\textsc{Efron, B. and Thisted, R.} (1976). Estimating the number of unseen species: How many words did Shakespeare know? \textit{Biometrika} \textbf{63}, 435--447.

\bibitem[Efron(2003)]{Efr(03)}  
\textsc{Efron, B.} (2003). Robbins, empirical Bayes and micorarrays \textit{Annals of Statistics} \textbf{31}, 366--378.

\bibitem[Erlingsson et al.(2014)]{Erl(14)}
\textsc{Erlingsson, U., Pihur, V., and Korolova, A.} (1972). Rappor: Randomized aggregatable privacy-preserving ordinal response. \textit{Proceedings of the ACM SIGSAC Conference on Computer and Communications Security}, 1054--1067.

\bibitem[Farahat and Bailey(2012)]{Fah(12)}  
\textsc{Farahat, A. and Bailey, M. C.} (2012). How effective is targeted advertising? \textit{Proceedings of the International Conference on World Wide Web} 111--120.

\bibitem[Favaro et al.(2009)]{Fav(09)}  
\textsc{Favaro, S., Lijoi, A., Mena, R.H., and Pr\"unster, I.} (2009). Bayesian nonparametric inference for species variety with a two parameter Poisson-Dirichlet process prior. \textit{Journal of the Royal Statistical Society Series B} \textbf{71}, 992--1008.

\bibitem[Favaro et al.(2012)]{Fav(12)}  
\textsc{Favaro, S., Lijoi, A., and Pr\"unster, I.} (2012). A new estimator of the discovery probability. \textit{Biometrics} \textbf{68}, 1188--1196.

\bibitem[Favaro et al.(2016)]{Fav(16)}  
\textsc{Favaro, S., Nipoti, B., and Teh, Y.W.} (2016). Rediscovery of Good-Turing estimators via Bayesian nonparametrics. \textit{Biometrics} \textbf{72}, 136--145.

\bibitem[Ferguson(1973)]{Fer(73)}
\textsc{Ferguson, T.S.} (1973). A Bayesian analysis of some nonparametric problems. \textit{Annals of Statistics} \textbf{1}, 209--230.

\bibitem[Gale and Sampson(1995)]{Gal(95)}
\textsc{Gale, W.A. and Sampson, G.} (1995). Good-Turing frequency estimation without tears. \textit{Journal of Quantitative Linguistics} \textbf{2}, 217--237.

\bibitem[Gao et al.(2007)]{Gao(07)}
\textsc{Gao, Z., Tseng, C.H., Pei, Z. an Blaser, M.J.} (2007). Molecular analysis of human forearm superficial skin bacterial biota. \textit{Proceedings of the National Academy of Sciences of USA} \textbf{104}, 2927--2932.

\bibitem[Ghosal and van der Vaart(2017)]{Gho(17)}
\textsc{Ghosal, S. and van der Vaart, A.} (2017) \textit{Fundamentals of Nonparametric Bayesian Inference.} Cambridge University Press. 

\bibitem[Good(1953)]{Goo(53)}
\textsc{Good, I.J.}(1953). The population frequencies of species and the estimation of population parameters. \textit{Biometrika} \textbf{40}, 237--264.

\bibitem[Good and Toulmin(1956)]{Goo(56)}  
\textsc{Good, I.J. and Toulmin, G.H.} (1956). The number of new species, and the increase in population coverage, when a sample is increased. \textit{Biometrika} \textbf{43}, 45--63.

\bibitem[Hatcher et al.(2017)]{hatcher2017virus}   
\textsc{Hatcher, E.L., Zhdanov, S.A., Bao, Y., Blinkova, O., Nawrocki, E.P., Ostapchuck, Y., Sch{\"a}ffer, A.A., and Brister, J.R.} (2017). Virus Variation Resource-improved response to emergent viral outbreaks. \textit{Nucleic acids research} \textbf{45}, D482--D490.

\bibitem[Heule et al.(2013)]{Heu(13)}
  \textsc{Heule, S., Nunkesser, M., and Hall, A.} (2013). Hyperloglog in practice: Algorithmic engineering of a state of the art cardinality estimation algorithm. \textit{Proceedings of the 16th International Conference on Extending Database Technology}.

\bibitem[Ionita-Laza et al.(2009)]{Ion(09)}
\textsc{Ionita-Laza, I., Lange, C., and Laird, N.M.} (2009). Estimating the number of unseen variants in the human genome. \textit{Proceedings of the National Academy of Sciences of USA} \textbf{106}, 5008--5013.

\bibitem[Kockan et al.(2020)]{Koc(20)}
\textsc{Kockan, C., Zhu, K., Dokmai, N., Karpov, N., Kulekci, M. O., Woodruff, D. P., and Sahinalp, S. C.} (2020). Sketching algorithms for genomic data analysis and querying in a secure enclave. \textit{Nature Methods} \textbf{17}(3), 295--301.

\bibitem[Leo Elworth et al.(2020)]{Leo(20)}
\textsc{Leo Elworth, R.A., Wang, Q., Kota, P.K., Barberan, C.J., Coleman, B., Balaji, A., Gupta, G., Baraniuk, R.G., Shrivastava, A., and Treangen, T.J.} (2020). To petabytes and beyond: recent advances in probabilistic and signal processing algorithms and their application to metagenomics. \textit{Nucleic Acids Research} \textbf{48}, 5217--5234.

\bibitem[Lijoi et al.(2007)]{Lij(07)}   
\textsc{Lijoi, A., Mena, R.H. and Pr\"unster, I.} (2007). Bayesian nonparametric estimation of the probability of discovering new species. \textit{Biometrika} \textbf{94}, 769--786.

\bibitem[Mao and Lindsay(2002)]{Mao(02)}   
\textsc{Mao, C.X. and Lindsay, B.G.} (2004). A Poisson model for the coverage problem with a genomic application. \textit{Biometrika} \textbf{89}, 669--682.

\bibitem[Mar\c{c}ais et al.(2019)]{Mar(19)}   
\textsc{Mar\c{c}ais, G., Solomon, B., Patro, R., and Kingsford, C.} (2019). Sketching and sublinear data structures in genomics. \textit{Annual Review of Biomedical Data Science} \textbf{89}, 669--682.

\bibitem[McAllester and Ortiz(2003)]{McA(03)}
\textsc{McAllester, D. and Ortiz, L.} (2003). Concentration inequalities for the missing mass and for histogram rule error. \textit{Journal of Machine Learning Research} \textbf{4}, 895--911.

\bibitem[Melis et al.(2016)]{Mel(16)}
\textsc{Melis, L., Danezis, G., and Cristofaro, ED.} (2016). Efficient Private Statistics with Succinct Sketches. \textit{Proceedings of the NDSS Symposium}.

\bibitem[Mitzenmacher and Upfal (2017)]{Mit(17)}
\textsc{Mitzenmacher, M. and Upfal, E.} (2017). \textit{Probability and computing: randomization and probabilistic techniques in algorithms and data analysis}. Cambridge University Press.

\bibitem[Mossel and Ohannessian(2019)]{Mos(15)}
\textsc{Mossel, E. and  Ohannessian, M.I.} (2019). On the impossibility of learning the missing mass. \textit{Entropy} \textbf{21}, 28.

\bibitem[Motwani and Vassilvitskii(2006)]{Mot(06)}
\textsc{Motwani, S. and Vassilvitskii, S.} (2006) Distinct value estimators in power law distributions. In \textit{Proceedings of the Workshop on Analytic Algorithms and Combinatorics}.

\bibitem[Ohannessian and Dahleh(2012)]{Oha(12)}  
\textsc{Ohannessian, M.I. and Dahleh, M.A.} (2012). Rare probability estimation under regularly varying heavy tails. In \textit{Proceedings of the Conference on Learning Theory}.

\bibitem[Orlitsky et al.(2003)]{Orl(03)}
\textsc{Orlitsky, A., Santhanam, N.P., and Zhang, J.} (2003). Always Good-Turing: asymptotically optimal probability estimation. \textit{Science} \textbf{302}, 427--431.

\bibitem[Orlitsky et al.(2004)]{Orl(04)}
\textsc{Orlitsky, A., Santhanam, N.P., and Zhang, J.} (2004). 
Universal compression of memoryless sources over unknown alphabets. \textit{IEEE Transaction on Information Theory} \textbf{50}, 1469--1481.

\bibitem[Orlitsky et al.(2016)]{Orl(17)}  
\textsc{Orlitsky, A., Suresh, A.T. and Wu, Y.} (2017). Optimal prediction of the number of unseen species. \textit{Proceeding of the National Academy of Sciences of USA} \textbf{113}, 13283--13288.

\bibitem[Perman et al.(1992)]{Per(92)}
\textsc{Perman, M., Pitman, J., and Yor, M.} (1992). Size-biased sampling of Poisson point processes and excursions. \textit{Probability Theory and Related Fields} \textbf{92}, 21--39.

\bibitem[Pitman(1995)]{Pit(95)}
\textsc{Pitman, J.} (1995). Exchangeable and partially exchangeable random partitions. \textit{Probability Theory and Related Fields} \textbf{102}, 145--158.

\bibitem[Pitman(2006)]{Pit(06)}
\textsc{Pitman, J.} (2006). \textit{Combinatorial stochastic processes}. Lecture Notes in Mathematics, Springer Verlag.

\bibitem[Pitman and Yor(1997)]{Pit(97)}
\textsc{Pitman, J. and Yor, M.} (1997). The two parameter Poisson-Dirichlet distribution derived from a stable subordinator. \textit{Annals of Probability} \textbf{25}, 855--900.

\bibitem[Project Gutenberg(2022)]{Gutenberg}
\textsc{Project Gutenberg} (accessed: 2022-02-05). www.gutenberg.org.

\bibitem[Quenouille(1956)]{quenouille1956notes}
\textsc{Quenouille, M.H.} (1956). Notes on bias in estimation. \textit{Biometrika} \textbf{43}, 353--360.

\bibitem[Regazzini(1978)]{Reg(78)} 
\textsc{Regazzini, E.} (1978). Intorno ad alcune questioni relative alla definizione del premio secondo la teoria della credibili\`a. \textit{Giornale dell'Istituto Italiano degli Attuari} \textbf{41}, 77--89.

\bibitem[Regazzini(2001)]{Reg(01)}
\textsc{Regazzini, E.} (2001). \textit{Foundations of Bayesian statistics and some theory of Bayesian nonparametric methods.} Lecture Notes, Stanford University.

\bibitem[Robbins(1956)]{Rob(56)}  
\textsc{Robbins, H.E.} (1956). An empirical Bayes approach to statistics. \textit{Proceedings of the Berkeley Symposium} \textbf{1}, 157--163.

\bibitem[Robbins(1968)]{Rob(68)}  
\textsc{Robbins, H.E.} (1968). Estimating the total probability of the unobserved outcomes of an experiment. \textit{Annals of Mathematical Statistics} \textbf{39}, 256--257.

\bibitem[Rojas et al.(2018)]{rojas2018personalized}  
\textsc{Rojas, J.S., Gall{\'o}n, A.R. and Corrales, J.C.} (2018). Personalized service degradation policies on OTT applications based on the consumption behavior of users. \textit{International Conference on Computational Science and Its Applications}, 543--557.

\bibitem[Rothchild et al.(2020)]{Rot(20)}  
\textsc{Rothchild, D., Panda, A., Ullah, E., Ivkin, N., Stoica, I., Braverman, V., Gonzalez, J. and Arora, R.} (2020). Fetchsgd: Communication-efficient federated learning with sketching. \textit{International Conference on Machine Learning}, 8253--8265.

\bibitem[Sangalli(2006)]{San(06)}
\textsc{Sangalli, M.L.} (2006). Some developments of the normalized random measures with independent increments. \textit{Sankhya A} \textbf{68}, 461--487.

\bibitem[Sesia and Favaro(2022)]{sesia2022conformalized}
\textsc{Sesia, M. and Favaro, S.} (2022). Conformalized frequency estimation from sketched data. \textit{Preprint arXiv:2204.04270}.

\bibitem[Solomon and Kingsford(2016)]{Sol(16)}
\textsc{Solomon, B. and Kingsford, C.} (2016). Fast search of thousands of short-read sequencing experiments. \textit{Nature Biotechnology} \textbf{34}, 300--302.

\bibitem[Tin(1965)]{tin1965comparison}  
\textsc{Tin M.} (1965). Comparison of some ratio estimators. \textit{Journal of the American Statistical Association} \textbf{60}, 294--307.

\bibitem[Toubiana et al.(2010)]{Tou(10)}  
\textsc{Toubiana, V., Narayanan, A., Boneh, D., Nissenbaum, H., and Barocas, S.} (2010). Adnostic: Privacy preserving targeted advertising. \textit{Proceedings Network and Distributed System Symposium}, 294--307.

\bibitem[Zabell(1982)]{Zab(82)}
\textsc{Zabell, S.L.}  (1982). W. E. Johnson's ``sufficientness" postulate. \textit{Annals of Statistics} \textbf{10}, 1090--1099.

\bibitem[Zhang(2005)]{Zha(05)}
\textsc{Zhang, C.H.} (2005). Estimation of sums of random variables: examples and information bound. \textit{Annals of Statistics} \textbf{33}, 2022--2041.

\bibitem[Zhang and Zhang(2009)]{Zha(09)}
\textsc{Zhang, C.H. and Zhang, Z.} (2009). Asymptotic normality of a nonparametric estimator of sample coverage. \textit{Annals of Statistics} \textbf{37}, 2582--2595.

\bibitem[Zhang et al.(2014)]{Zha(14)}
\textsc{Zhang, Q., Pell, J., Canino-Koning, R., Howe, A. C., and Brown, C. T.} (2014). These are not the k-mers you are looking for: efficient online k-mer counting using a probabilistic data structure. \textit{PloS one} \textbf{9}(7), e101271.

\end{thebibliography}
\end{document}